 \documentclass[iop]{emulateapj}
 \usepackage{natbib}

\newcommand{\kms}{km~s$^{-1}$}

\newcommand{\lam}{$\lambda$}
\newcommand{\rgc}{R$_{\rm gc}$}

\slugcomment{Published In The Astronomical Journal, 2011, 142, 59}

\shorttitle{Abundances in \rgc$\sim$9-13 kpc Open Clusters}
\shortauthors{Jacobson, Pilachowski \& Friel}

\begin{document}

\title{A Chemical Abundance Study of 10 Open Clusters Based on WIYN\altaffilmark{1}-Hydra Spectroscopy}

\author{Heather R. Jacobson\altaffilmark{2}}
\affil{Department of Physics \& Astronomy, Michigan State University, East Lansing, MI 48823;
jacob189@msu.edu}

\author{Catherine A. Pilachowski}
\affil{Department of Astronomy, Indiana University, Bloomington, IN 47405;
catyp@astro.indiana.edu}

\and 

\author{Eileen D. Friel}
\affil{Department of Astronomy, Boston University, Boston, MA 02215; edfriel@mac.com}

\altaffiltext{1}{The WIYN Observatory is a joint facility of the University of Wisconsin-Madison, Indiana University, Yale University, and the National Optical Astronomy Observatory.}
\altaffiltext{2}{National Science Foundation Astronomy and Astrophysics Postdoctoral Fellow.}

\begin{abstract}
We present a detailed chemical abundance study of evolved stars in 10 open clusters based on Hydra multi-object echelle spectra obtained with the WIYN 3.5m telescope.  From an analysis of both equivalent widths and spectrum synthesis, abundances have been determined for the elements Fe, Na, O, Mg, Si, Ca, Ti, Ni, Zr, and for two of the 10 clusters, Al and Cr.  To our knowledge, this is the first detailed abundance analysis for clusters NGC 1245, NGC 2194, NGC 2355 and NGC 2425.  These 10 clusters were selected for analysis because they span a Galactocentric distance range \rgc$\sim$9--13 kpc, the approximate location of the transition between the inner and outer disk.  Combined with cluster samples from our previous work and those of other studies in the literature, we explore abundance trends as a function of cluster \rgc, age, and [Fe/H].  As found previously by us and other studies, the [Fe/H] distribution appears to decrease with increasing \rgc\ to a distance of $\sim$12 kpc, and then flattens to a roughly constant value in the outer disk.  Cluster average element [X/Fe] ratios appear to be independent of \rgc, although the picture for [O/Fe] is more more complicated by a clear trend of [O/Fe] with [Fe/H] and sample incompleteness.  Other than oxygen, no other element [X/Fe] exhibits a clear trend with [Fe/H]; likewise, there does not appear to be any strong correlation between abundance and cluster age.  We divided clusters into different age bins to explore temporal variations in the radial element distributions.  The radial metallicity gradient appears to have flattened slightly as a function of time, as found by other studies.  There is also some indication that the transition from the inner disk metallicity gradient to the $\sim$constant [Fe/H] distribution of the outer disk occurs at different Galactocentric radii for different age bins.  However, interpretation of the time evolution of radial abundance distributions is complicated by the unequal \rgc\ and [Fe/H] ranges spanned by clusters in different age bins.
\end{abstract}


\section{Introduction}
Until relatively recently, our understanding of the chemical abundance distributions of the Milky Way disk has been shaped by objects predominantly located in the Galactocentric distance range of \rgc$\sim$7--16 kpc\footnote{Note R$_{gc,\odot}$, depending on the study, is generally taken to be $\sim$8--8.5 kpc.}.  Such distributions have been determined using objects for which distances can be determined, such as open clusters, Cepheids, H II regions, planetary nebulae and young OB stars, predominantly in the Galactic anticenter ({\it l} $\sim$180$^{\circ}$), where the line of sight is relatively less obscured by dust and overcrowding.  The abundance distributions (e.g., of [Fe/H], log N(O), etc.) shown in this Galactocentric distance range have been generally characterized as linearly-dependent on \rgc, with the more distant objects being more metal-poor than objects closer to the Galactic center.  Astronomers have more often than not chosen to draw linear gradients due to the relatively small sample sizes they were working with.  That said, Twarog et al.\ (1997) proposed an alternative view for their sample of 76 open clusters within the range \rgc$\sim$7--16 kpc.  They found that a bimodal distribution, with clusters inside \rgc = 10 kpc having [Fe/H] = 0.00, and clusters outside \rgc = 10 kpc having [Fe/H] = $-$0.30, better fit the metallicity distribution than a simple linear gradient.

\begin{deluxetable*}{lcrclccccll}
\tabletypesize{\footnotesize}
\tablewidth{0pt}
\tablenum{1}
\tablecaption{Clusters Observed\label{cluster_info}}
\tablehead{
\colhead{} & \colhead{\it l} & \colhead{\it b} & \colhead{}  & \colhead{} & \colhead{\it d} & \colhead{R$_{gc}$\tablenotemark{b}}
& \colhead{\it z} &  \colhead{} & \colhead{Age\tablenotemark{c}} & \colhead{} \\ 
\colhead{Cluster} & \colhead{(deg.)} 
& \colhead{(deg.)} & 
\colhead{E(B$-$V)} & \colhead{Ref.\tablenotemark{a}} & \colhead{(kpc)} &
\colhead{(kpc)} &
\colhead{(pc)} & \colhead{Ref.\tablenotemark{a}} &
\colhead{(Gyr)} 
 & \colhead{\# Stars}}
\startdata
M 67 & 215.6 & $+$31.7 & 0.04$\pm$0.02 & 11 & 0.83$\pm$0.02 & 9.09$\pm$0.02 & $+$440$\pm$10 & 10 & 4.3  & 31\\
NGC 188 & 122.8 & $+$22.5 & 0.09$\pm$0.02 & 8 & 1.70$\pm$0.07 & 9.4$\pm$0.1 & $+$650$\pm$30 & 8 &6.3 & 51 \\
NGC 1245 & 146.6 & $-$8.9 & 0.21$\pm$0.03 & 3 & 3.0$\pm$0.3 & 11.1$\pm$0.3 & $-$460$\pm$50 & 10 & 1.1 & 24 \\
NGC 1817 & 186.1 & $-$13.1 & 0.27$\pm$0.05 & 2 & 1.5$\pm$0.5 & 10.0$\pm$0.4 & $-$340$\pm$100 & 2 & 1.1 & 72 \\
NGC 2158 & 186.6 & $+$1.8 & 0.430$\pm$0.013 & 5 & 4.0$\pm$0.1 & 12.5$\pm$0.1 & $+$130$\pm$5 & 10 & 1.9 & 36  \\
NGC 2194 & 197.3 & $-$2.3 & 0.45$\pm$0.02 & 7 & 1.9$\pm$0.1 & 10.3$\pm$0.2 & $-$80$\pm$5& 10 & 0.9 & 18 \\
NGC 2355 & 203.4 & $+$11.8 & 0.16 & 9 & 1.9$\pm$0.1 & 10.3$\pm$0.1 & $+$400$\pm$20 & 10 & 0.8 & 12 \\
NGC 2420 & 198.1 & $+$19.6 & 0.050$\pm$0.004 & 1 & 2.5$\pm$0.3 & 10.7$\pm$0.3 & $+$830$\pm$100 & 10 & 2.2 & 22 \\
NGC 2425 & 231.5 & $+$3.3 & 0.29$\pm$0.05 & 6 & 3.3$\pm$0.2 & 10.6$\pm$0.1 & 190$\pm$10 & 10 & 2.5\tablenotemark{d} & 31\\
NGC 7789 & 115.5 & $-$5.4 & 0.28$\pm$0.03 & 4 & 2.2$\pm$0.2 & 9.6$\pm$0.2 & $-$200$\pm$20 & 10 & 1.8 & 44 \\
\enddata
\tablenotetext{a}{References for cluster reddening and/or distance: (1) Anthony-Twarog et al.\ 2006; (2) Balaguer-N\'{u}\~{n}ez et al.\ 2004b; (3) Burke et al.\ 2004; (4) Gim et al.\ 1998b; (5) Grochalski \& Sarajedini 2002; (6) Pietrukowicz et al.\ 2006; (7) Sanner et al.\ 2000; (8) Sarajedini et al.\ 1999; (9) Soubiran et al.\ 2000; (10) This Study; (11) Twarog et al.\ 1997.}
\tablenotetext{b}{R$_{\odot}$ = 8.5 kpc}
\tablenotetext{c}{Adopted from Salaris et al.\ (2004)}
\tablenotetext{d}{Age from Pietrukowicz et al.\ (2006)}
\end{deluxetable*}
%1

Within the past ten years or so, a number of chemical abundance studies have extended the Galactocentric distance range of several populations, and rather unsurprisingly, the resulting element abundance distributions have shown themselves to be more complicated than the early linear gradients.  For example, Andrievsky et al.\ (2002b) determined chemical abundances for five Cepheids in the \rgc\ range 4--6 kpc (where R$_{gc, \odot}$ = 7.9 kpc).  The combination of these results with a sample of Cepheids with \rgc$\sim$6--11 kpc (Andrievsky et al.\ 2002a) showed the Cepheid metallicity gradient sharply increased inside \rgc$\sim$6.5 kpc.    

In the opposite direction, high resolution spectroscopic studies of very distant open clusters have extended our reach nearly 23 kpc from the center of the Galaxy.  Our understanding of the Galactic disk beyond \rgc$\sim$13 kpc rests on roughly a dozen open clusters.  
  The first studies of these outer disk clusters (Carraro et al.\ 2004, Yong et al.\ 2005) lent support to the view of Twarog et al.\ (1997) that the metallicity distribution differs between the outer disk and the inner disk.  The increasingly large number of detailed abundance studies of open clusters have made it clear that the abundance gradients of the inner disk flatten out at \rgc$\sim$9--14 kpc, and remain relatively independent of Galactocentric distance for some 10 kpc in the outer disk (e.g, Carraro et al.\ 2007).  This trend is seen not only in the traditional metallicity gradient (i.e., [Fe/H]) for open clusters, but in trends of other element [X/H] ratios with \rgc\ (e.g., Magrini et al.\ 2009).  However, the nature of the transition from gradient to plateau is poorly constrained.  Is there a sharp break, or is there a large dispersion in abundances in the transition region?  How does the abundance dispersion in the transition region compare to those in the inner and outer disks?

In the effort to address these questions, we have assembled a sample of 20 open clusters located in the transition region.  A robust determination of their elemental abundances based on observations of as many confirmed cluster member stars as possible in a self-consistent and homogeneous fashion will allow us to better trace the dispersions in cluster element abundances in the transition area of the disk.  A little more than half our sample has been presented in previous papers.  These include detailed abundances of old open clusters based on single star, multi-order echelle spectra obtained with 4m class telescopes (Friel et al.\ 2005, 2010, Jacobson et al.\ 2008, 2009), and of two Southern hemisphere clusters observed with the Hydra multi-object spectrograph on the CTIO 4m telescope (Jacobson et al.\ 2011).  In this paper, we present results for the remainder of the clusters in our sample observed with the WIYN 3.5m telescope.  Results for four of these clusters have been presented previously in Jacobson et al.\ (2009; NGC 1817, NGC 2158) and Friel et al.\ (2010; M 67, NGC 188); the analysis presented here both complements and adds to that work, allowing us not only to directly compare to previous results but to increase the number of stars used to determine mean cluster abundances.

\section{Target Selection and Observations}
Assembly of our cluster sample was constrained by the efficiency and throughput of the Hydra spectrograph on the WIYN 3.5m telescope.  To obtain a signal-to-noise ratio (S/N) of $\sim$70 per pixel generally requires $\sim$6 hours' integration time for stars with V$\sim$15 observed in conditions of moderate seeing, light cirrus and bright moon.  We selected clusters in the direction of the Galactic anticenter that contained a large enough sample (preferably a dozen or more) of potential cluster members with V$\sim$15 or brighter.
Table~\ref{cluster_info} lists the clusters for which stellar spectra of sufficient S/N ratio ($\sim$70 or higher) allowed for detailed abundance analysis.  Of these clusters, M 67 and NGC 188 have been studied in great detail, and as already mentioned, they and NGC 1817 and NGC 2158 were studied previously by us.  Of the remaining clusters, NGC 2194, NGC 2425 and to a lesser extent, NGC 2355 are relatively little-studied to date, while NGC 1245 has been the subject of planet searches but not chemical abundance analysis (e.g., Burke et al.\ 2004, 2006; Pepper \& Burke 2006).  To our knowledge, only NGC 2355 has been subject to spectroscopic study (Soubiran et al.\ 2000), but in this analysis stellar [Fe/H] values were determined by comparison of spectra to a library of stars with known parameters, rather than via a ``classical" detailed abundance analysis based on measurement of individual spectral lines.  NGC 7789 is another classic, old open cluster; Tautvai\v{s}ien\.{e} et al.\ (2005) and Pancino et al.\ (2010) presented detailed abundance analyses of several evolved stars in NGC 7789, which will serve as a useful comparison to our work.  Brief descriptions and summaries of previous studies of individual clusters are given in Appendix A.  

\begin{deluxetable*}{llccc}
\tabletypesize{\footnotesize}
\tablewidth{0pt}
\tablenum{2}
\tablecaption{Observing Log\label{obs_log}}
\tablehead{
\colhead{Cluster} & \colhead{UT Date} & \colhead{Wavelength (\AA)} & \colhead{Exposure (s)} & \colhead{Detector}
}
\startdata
M 67         & 2006 April 16 & 6590 & 4x1800 & T2KA\\
                  & 2007 Feb 27, Mar 1 & 6270 & 6x1800 & T2KA\\
NGC 188 &  2006 Dec 3  & 6280 & 4x2700 & T2KA\\
                  & 2006 Dec 27 & 6275 & 2x2700 & T2KA\\
                  & 2007 Feb 25,26 & 6280 & 5x2700, 2x1800 & T2KA\\
NGC 1245 &  2006 Dec 7 & 6280 & 7x1800, 4x2700 & T2KA\\
                     & 2007 Dec 18,19,20 & 6265 & 6x2700 & T2KA\\
NGC 1817 &  2006 Dec 27 & 6275 & 10x1800 &T2KA \\
                    & 2008 Dec 6,7 & 6245 & 3x3600 & STA-1 \\
NGC 2158 & 2006 Feb 14 & 6700 &  5x3600 & T2KA \\
                     & 2007 Nov 21,23 & 6260 & 6x3600, 1x1800 & T2KA\\
                     & 2008 Feb 25 & 6280 & 2x3600, 1x2700 & T2KA \\
                    & 2008 Dec 6 & 6245 & 5x3600 & STA-1\\
NGC 2194 & 2006 Dec 2,3,5,6 & 6280 & 12x1800, 1x1200 & T2KA\\
NGC 2355 & 2006 Dec 5 & 6280 & 6x1800 & T2KA \\
                     & 2006 Dec 27 & 6280 & 4x1800 & T2KA \\
NGC 2420 & 2008 Dec 7 & 6245 & 3x3600, 1x1800 & STA-1  \\
NGC 2425 & 2007 Feb 25 & 6280 & 5x3600, 1x1800 & T2KA\\
                     & 2007 Nov 20,23 & 6260 & 6x3600, 1x1800, 1x1200 & T2KA\\
                     & 2008 Dec 6,7 & 6245 & 4x3600 & STA-1 \\
NGC 7789 & 2006 Dec 2 & 6280 &  7x1800, 4x1200 & T2KA\\
\enddata
\end{deluxetable*}

Primary stellar targets were red clump stars, with brighter red giant stars serving as secondary targets.   NGC 188 has no identifiable red clump, so all evolved stars were selected for observation.  Given the relatively small angular diameter of some of these clusters on the sky ($\sim$5--10 arcminutes), they filled up only a small portion of Hydra's field of view.  This, combined with constraints on fiber placement, often greatly limited the number of targets it was possible to observe.  As there was generally only time enough to observe one field configuration per cluster, the number of stars observed was sometimes limited to $\sim$12 or so per field.  As will be seen, examination of radial velocities often showed only a much smaller number of stars were actually cluster members.  

The majority of the observations were obtained with the Hydra fiber positioner and bench spectrograph on the WIYN 3.5m telescope from 2006 December through 2008 December.  Table~\ref{obs_log} gives a summary of the observations, including UT date of observations, central wavelength, and the number and length of exposures.  Most clusters were observed in more than one observing run, and each is listed separately.  

We used the bench spectrograph camera and the red-optimized 200 $\mu$m fiber cables.
The echelle 613 lines mm$^{-1}$ grating along with the X18 filter at 9th order were used to obtain single order $\sim$300 \AA\ spectra centered at $\sim$6280 \AA.  This wavelength range was selected because it includes the Na \lam6154/6160 \AA\ doublet, the [O I] \lam6300 \AA\ feature, and a handful of Mg, Si, Ca, Ti and Ni lines, as well as $\sim$20 Fe I lines.  
Comparison of cluster $\alpha$ element abundances to metallicity will yield information about the relative contributions of Type II and Type Ia supernovae, and thus any trends of [$\alpha$/Fe] with \rgc\ would indicate differences in chemical evolutionary history.  Na abundances are also interesting, as many open clusters have been found to have enhanced light element abundances relative to field stars (Friel 2006).

Most data were taken with the T2KA CCD detector (2048x2048 pixels, 24 $\mu$m in size), with the spectra generally having a resolution (\lam/$\Delta$\lam) R$\sim$18\,000, based on measurement of the FWHM of ThAr lines.  In the weeks prior to the 2008 December observing run, the bench spectrograph was updated with a new collimator and new CCD, STA-1.  The new CCD has 2600x4000 12 $\mu$m pixels, and we observed in binned 2x2 mode.  The new collimator increased the throughput, allowing us to achieve minimum desired S/N ratios in fewer integrations.  What is more, the spectra obtained with STA-1 have a spectral resolution of R$\sim$21\,000 due to its higher dispersion (0.07 \AA\ pix$^{-1}$ compared to 0.15 \AA\ pix$^{-1}$ for T2KA).  Given the better quality of these 2008 December spectra, both in terms of S/N and spectral resolution, only these data for NGC 2425 and NGC 2158 were used in the abundance analysis (though all spectra were used for radial velocity determination).  The field configuration of NGC 1817 obtained in 2008 December was different from that obtained in 2006 December, so both data sets were analyzed.
Calibration data obtained each evening of each observing run included 3--10 flat field images, 2--3 twilight sky exposures, 9--15  bias frames, and one 10 minute ThAr spectrum for wavelength calibration.  Spectra of hot, rapidly-rotating early-type stars near each cluster's position on the sky were obtained either before or after target observations.  These were used for removal of telluric absorption features, particularly those near the [O I] \lam6300 \AA\ feature.  %Information about individual stellar targets is given in Table~\ref{star_info}.  

Targets in M 67 and NGC 2158 were observed at redder wavelengths earlier in 2006 as part of a program to study Na and Al abundances in open clusters.  These Hydra echelle spectra, centered at roughly \lam6600 \AA, were obtained using the 613 lines mm$^{-1}$ echelle grating with the X19 filter at 8th order, and have a resolution of R$\sim$14\,500.  The field configuration for M 67 used in these observations is essentially identical to that of the 2007 observations, so all stars analyzed in this study have spectra in both wavelength ranges.  
The field configurations for NGC 2158 changed from one observing run to another, as radial velocity nonmembers were replaced with other potential member stars.  As a result, the field configuration used in 2006 differed enough from the 2008 December configuration that only a few stars were observed in both wavelength regimes.
  Of these that turned out to be radial velocity members, only one star (3216) had S/N ratios large enough in both wavelength ranges for abundance analysis.  Two confirmed radial velocity members (see below), 7773 and 7866, were only observed at \lam6600 \AA.

Information for each observed star, including J2000 coordinates and available optical and 2MASS\footnote{See http://irsa.ipac.caltech.edu/applications/Gator/} photometry (Cutri et al.\ 2003) are given in Table~\ref{star_info}.  The sources for the optical photometry and identification numbers are listed in Appendix A for each cluster.

\section{Data Reduction}
\begin{figure*}
\epsscale{0.8}
\plotone{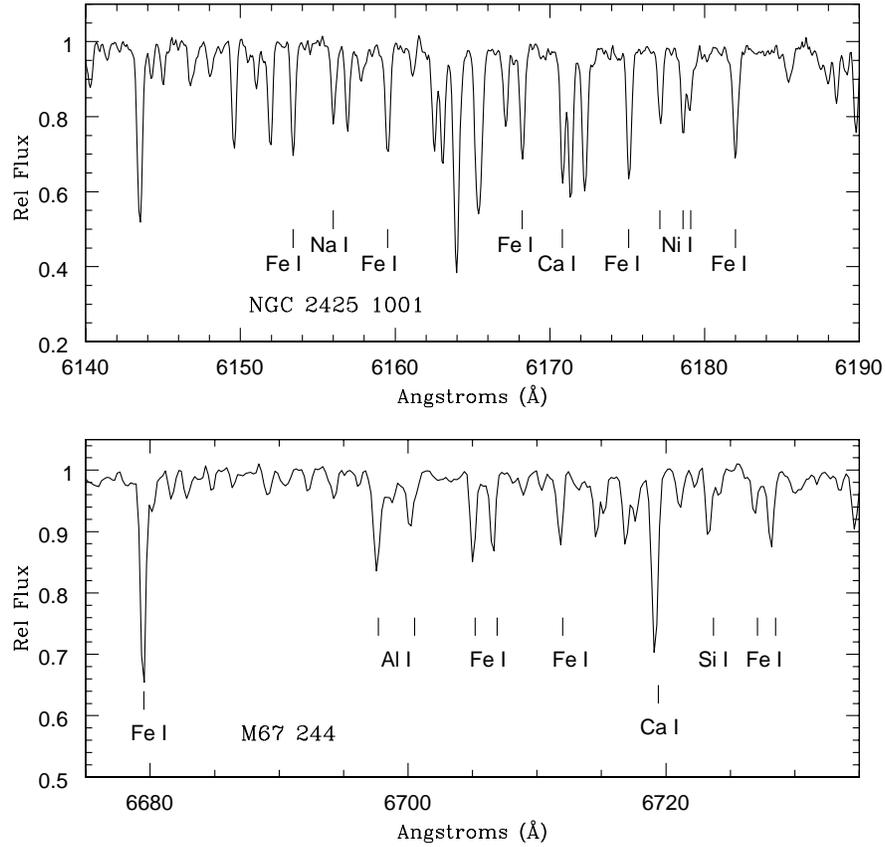}
\caption{Sample portions of the WIYN-Hydra spectra with some lines identified.  The top panel shows the region around the Na \lam6154--6160 doublet in the spectrum of NGC 2425 1001; the bottom panel shows the Al \lam6696--6698 doublet and surrounding features in the spectrum of M 67 244.  Only stars in M 67 and NGC 2158 have spectra in this wavelength range.  Note the spectra are not radial velocity corrected.}
\label{spec}
\end{figure*}

%\documentclass[manuscript]{aastex}
%\begin{document}
\begin{deluxetable*}{llllllllrcc}
\tabletypesize{\scriptsize}
\tablecolumns{11}
\tablewidth{0pt}
\tablecaption{Stellar Targets\label{star_info}}
\tablenum{3}
\tablehead{
\colhead{ID\tablenotemark{a}} & \colhead{ID\tablenotemark{b}} & \colhead{$\alpha$\tablenotemark{c}} & \colhead{$\delta$\tablenotemark{d}}
& \colhead{V} & 
\colhead{B$-$V\tablenotemark{4}} & \colhead{V$-$K} &
\colhead{J$-$K} &
\colhead{S/N} &
\colhead{V$_{rad}$} & \colhead{Member?}\\
\colhead{} & \colhead{} & \colhead{(J2000)} & \colhead{(J2000)} & \colhead{} & \colhead{} &
\colhead{} & \colhead{} & \colhead{} & \colhead{(km s$^{-1}$)} & \colhead{}}
\startdata
M67    4 & 5059 & 08 50 36.14 & $+$11 43 18.07 & 12.71 & 0.913 & 2.160 & 0.579 & 35  & $+$33.5 & M\\
M67   37 & 5318 & 08 50 58.16 & $+$11 52 22.15 & 12.86 & 0.941 & 2.236 & 0.571 & 60  & $+$33.4 & M\\
M67   79 &\nodata & 08 51 09.51 & $+$11 41 44.97 & 12.81 & 0.733 & 1.815 & 0.448 & 55 & $+$33.3 & M\\
M67   84 & 6492 & 08 51 12.69 & $+$11 52 42.31 & 10.59 & 1.120 & 2.614 & 0.674 & 185  & $+$33.8 & M\\
M67   86 & 5542 & 08 51 13.36 & $+$11 51 39.99 & 13.79 & 1.051 & 2.832 & 0.699 & 35  & $+$15.7 & M, SB\\
M67   96 & 5580 & 08 51 15.64 & $+$11 50 56.03 & 13.01 & 0.854 & 2.002 & 0.472 & 45  & $+$32.7 & M\\
%M67  104 & & 08 51 17.05 & $+$11 50 46.26 & 11.20 & 1.080 &  &    &  & $+$ & M\\
M67  105 & 6486 & 08 51 17.10 & $+$11 48 16.01 & 10.30 & 1.260 & 2.915 & 0.755 & 190  & $+$33.7 & M\\
M67  108 & 6482 & 08 51 17.49 & $+$11 45 22.61 &  9.72 & 1.370 & 3.226 & 0.831 & 250  & $+$33.9 & M\\
\enddata
\tablenotetext{a}{Star ID used in this study; see individual cluster summaries in Appendix A for sources.}
\tablenotetext{b}{Alternate ID; see individual cluster summaries in Appendix A for sources.} 
\tablenotetext{c}{For NGC 7789 stars, this column contains 1950B epoch coordinates.}
\tablenotetext{d}{For NGC 1817 stars, this column is ({\it b$-$y}) color.}
\tablecomments{This table is available in its entirety in electronic format.  A portion is shown here for
guidance regarding its form and content.}
\end{deluxetable*}
%\end{document}

The data were reduced using the usual routines in IRAF\footnote{IRAF is 
distributed by the National Optical Astronomy Observatories, which is
operated by the Association of Universities for Research in Astronomy,
Inc., under cooperative agreement with the National Science Foundation.}.  Briefly, the individual frames were trimmed and overscan-subtracted.  Bias frames were combined into a master bias that was then subtracted from the rest of the images.  The object frames were run through L. A. Cosmic, an iraf script developed by P. van Dokkum (2001) to remove cosmic rays.  After aperture extraction, the one-dimensional spectra were flat fielded and then dispersion corrected.  Lastly, sky spectra from 10--30 sky fibers uniformly distributed throughout the field of view were combined to create a single sky spectrum that was subtracted from the stellar spectra.  This effectively removed the telluric emission lines that plague the \lam$\sim$6250--6350 \AA\ region of the electromagnetic spectrum.  The individual spectra for each star were then summed to form a single, high S/N ratio spectrum, which was used for both preliminary radial velocity measurements and for the abundance analysis.  The individual spectra were used to measure the final stellar radial velocities, as discussed in Section 4.2.  Figure~\ref{spec} shows examples of the 
combined, high S/N spectra used for abundance analysis.

\section{Analysis}
\subsection{Distance Determination}

As in our previous work, we have attempted to place all clusters in our sample on a uniform distance scale.  We followed the methods of Grochalski \& Sarajedini (2002; hereafter GS02) and Carney et al.\ (2005; hereafter CLD05) to determine a cluster's distance modulus and reddening based on the K-band magnitude and (J$-$K) color of red giant clump stars.  GS02 found M$_{\rm K}$ =  $-$1.6 for red clump stars over a wide range of cluster age and metallicity.  For clusters outside of this range, Figure 6 in GS02 can be used to determine the appropriate M$_{\rm K}$ value.  CLD05 determined an empirical relation between a cluster's intrinsic (J$-$K)$_{0}$ color and [Fe/H], which, combined with the cluster's observed color, measures the cluster's color excess.  This reddening value can then be used along with the cluster's K magnitude to determine the absolute distance modulus, (K$-$M$_{\rm K}$)$_{0}$.  

All open clusters in our sample exhibit identifiable red clumps save for NGC 188 (see, e.g., Sarajedini et al.\ 1999).  Three clusters, M 67, NGC 1817, and NGC 2420, were used by GS02 to calibrate their relation, so comparison of our distance and reddening determinations here with literature values serves as validation of this method.  Details of the red clump star identifications and distance and reddening calculations are offered for each cluster in Appendix A.  There the reader will find that in some cases identification of red clump stars proved difficult, and in all but two cases the interstellar extinction values are in disagreement with other determinations in the literature, sometimes by factors of two or more.  However, given the relatively small dependence of K-band magnitudes on reddening, the corresponding distance moduli are in general good agreement with literature values.
The cause of the discrepancy between these reddening determinations and those found by other methods is unclear, but at least in some cases it is possibly due to incorrect identification of red clump stars due to crowded CMDs and lack of membership information (see Appendix A for details).

It is important to note that the distances and reddenings determined by the red clump method have {\it not} been used to calculate atmospheric parameters for cluster stars as described in Section 4.4.  As the color excesses found from clump (J$-$K) colors were generally different from previous determinations, and also as the clump star intrinsic  (J$-$K)$_{0}$ given by the CLD05 relation is dependent upon metallicity, potentially large systematic uncertainties could be introduced into stellar effective temperatures.  As a brief example, consider the case for NGC 1817 (more details can be found in Appendix A).  Depending on the adopted reddening for the cluster, stellar effective temperatures can vary by 300 K, and the resulting cluster [Fe/H] by 0.3 dex.  As discussed below, comparison to the results for two NGC 1817 stars analyzed by us previously (Jacobson et al.\ 2009) found that the atmospheric parameters determined using a reddening and distance modulus from the literature were in much better agreement than those calculated using the reddening and distance modulus from the red clump method.  Therefore, we have chosen photometric studies from the literature that we have deemed reliable for each cluster for the determination of atmospheric parameters; the distances calculated by us using the cluster red clump morphology is used only in the discussion of cluster abundance trends with location in the Galaxy.  As already mentioned, and as can be seen in Appendix A, these distances are in general good agreement with literature values.

\subsection{Radial Velocity Determination}

Stellar radial velocities were determined using the {\it fxcor} task in IRAF.  The primary motivation for determining radial velocities was to verify cluster membership for each stellar target; as a result, we used a twilight sky spectrum obtained in the same run as the target spectra for the radial velocity template for each cross-correlation.  However, in the 2008 December observing run, spectra of the late-type radial velocity standard HD 212943 were obtained and used as velocity templates.  All individual spectra of stellar targets observed in previous observing runs were then cross-correlated against a spectrum of HD 212943 for final determination of radial velocities.  These velocities were corrected for heliocentric motion using IRAF's {\it rvcor} routine.  Final radial velocities for each star observed are given in Table~\ref{star_info}.  
Stars that were either known to be spectroscopic binaries or were discovered to be so are also indicated.

Agreement between our radial velocity values with those of previous studies for most clusters is generally quite good (see below) and errors are relatively small.  Velocity shifts from fiber to fiber as measured in twilight sky spectra are generally no larger than 0.2 \kms.  Shifts in velocities calculated for stars observed over multiple epochs are also generally small, within 0.4 \kms, with exceptions noted below.  The error in the zero point of our velocity scale can be estimated based on the radial velocity we determined for HD 212943 using a twilight sky spectrum as template: our value of 53.5 \kms\ agrees well with the value of 54.3 \kms\ listed in the SIMBAD\footnote{See http://simbad.u-strasbg.fr/simbad/} database for this star.  Comparison to literature radial velocities for certain clusters below confirms that our velocities are on a common scale within 1 \kms.

\begin{figure*}
\epsscale{0.75}
\plotone{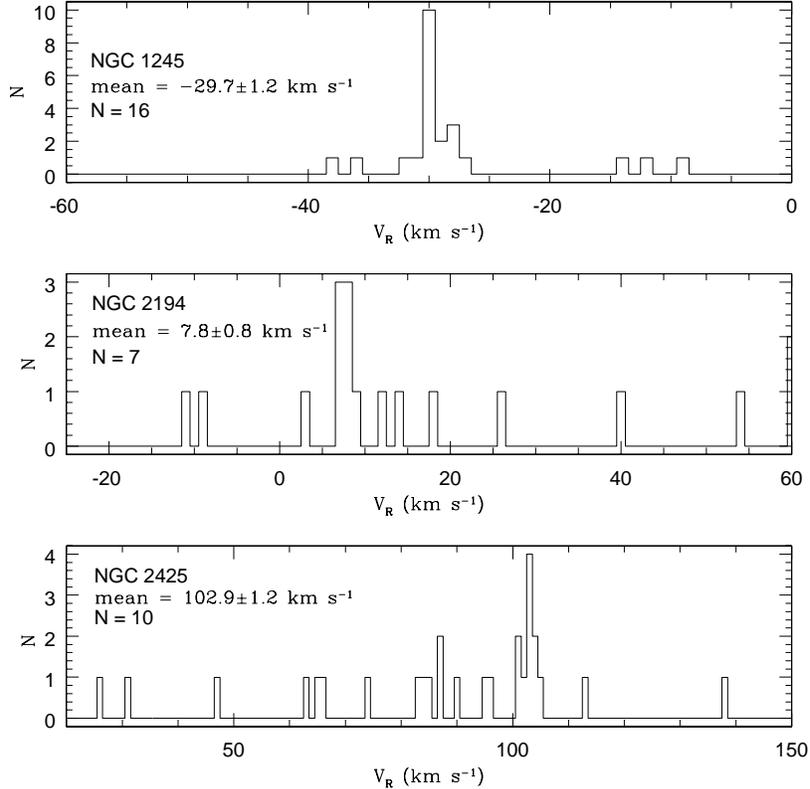}
\caption{Radial velocity histograms for stars observed in open clusters NGC 1245 (top), NGC 2194 (middle) and NGC 2425 (bottom).  The cluster mean velocity, standard deviation, and number of stars used in the calculation are indicated in each panel.  Each histogram bin size is 1 \kms. }
\label{rvhist}
\end{figure*}

Extensive studies of proper motions and radial velocities exist in the literature for the clusters M 67, NGC 188, NGC 7789 and NGC 1817.  Consequently, comparison of our radial velocities to the literature values serves as another estimation of errors in our measurements.  For the less well-studied clusters,  membership was determined by examination of histograms of stellar radial velocities, such as in Figure~\ref{rvhist}.  In most cases, a clear peak in the histogram made membership identification straightforward.  Stars with velocities within 15 \kms\ of the adopted cluster mean velocity are tentatively identified as possible (binary star) cluster members (noted with {\it M?} in Table~\ref{star_info}).  We chose this value because $\sim$20\%\ of binary systems can have semi-amplitudes as large as 20 \kms\ (Mermilliod \& Mayor 2007, Mermilliod et al.\ 2008).

To our knowledge, no previous radial velocity studies for NGC 1245, NGC 2194 and NGC 2425 exist in the literature.  For these clusters, we have found mean radial velocities of $-$29.5$\pm$1.0 \kms\ (N=15), $+$7.5$\pm$0.8 (N=7), and $+$102.9$\pm$1.2 (N=10) \kms, respectively (Figure ~\ref{rvhist}).  More details about radial velocity and membership determination are given below.  For the more well-studied clusters, we also offer a brief comparison of our results to literature values.

\subsubsection{M 67}
As we used the radial velocity study of Mathieu et al.\ (1986) and the proper motion study of Sanders (1977) to identify targets for observation, 
all our targets in M 67 save one star were included in Mathieu et al.\ (1986), who found a cluster mean V$_{r}$ = $+$33.5$\pm$2.8 \kms\ (65 stars).  Excluding binary systems, the average difference between our velocities and those of Mathieu et al.\ is 0.5$\pm$0.3 (s.d.) \kms, with the Mathieu et al.\ values being larger.  Star 2152, which was not included in the Mathieu et al.\ study, is a cluster member, as shown by its radial velocity.  
We have detected radial velocity variations of several known spectroscopic binaries in the cluster; these are indicated in Table 3.  Star 286 also exhibited velocity variations; it has been shown to exhibit variability (Stello et al.\ 2007), as described in the WEBDA database.  The mean radial velocity as determined by our cluster sample (excluding known binaries and variable stars) is $+$33.3$\pm$0.6 \kms\ (N=22).  This is in excellent agreement with the values found by, e.g., Friel et al.\ (2002), who found a mean value of $+$33$\pm$8 (s.d.; 25 stars) \kms\ for the cluster, and Mermilliod et al.\ (2007), who reported $+$33.7$\pm$0.7 \kms\ based on 23 stars.

\subsubsection{NGC 188}
NGC 188 has been studied as part of the WIYN Open Cluster Survey (WOCS; Mathieu 2000); consequently, membership has been established via both proper motions (Platais et al.\ 2003) and radial velocities (Geller et al.\ 2008).  We used the proper motion survey of Platais et al.\ (2003) to identify stars for observation, and therefore the majority of our targets are included in the radial velocity survey of Geller et al.\ (2008).
Excluding spectroscopic binaries, the difference between our measured radial velocities and those of Geller et al.\ is 0.7$\pm$0.5 (s.d.) \kms, with our values being larger.  As with M 67, we confirmed the radial velocity variability of several stars previously identified as spectroscopic binaries (Table~\ref{star_info}).  Excluding binary systems, we have found a mean cluster radial velocity of $-$41.9$\pm$0.7 (s.d) \kms\ (N=40) for NGC 188.  This compares well with the values found by Geller et al. ($-$42.36$\pm$0.04 \kms\ using single member stars) and by Friel et al.\ (2002; $-$45$\pm$10 \kms, 21 stars).  Stars 5212 and 6712 have radial velocities consistent with cluster membership, but have proper motion membership probabilities of 26\%\ and 6\%, respectively (Platais et al.\ 2003).  Therefore, they are given as possible members in Table~\ref{star_info}.

\subsubsection{NGC 1245}
Of the 24 stars observed in this cluster field, the majority appear to have a common velocity.  The radial velocity histogram for this cluster is highly peaked at $-$30 \kms\ (see Figure~\ref{rvhist}).  Based on visual examination of this peak, we have identified all stars with radial velocities between $-$35 \kms\ and $-$25 \kms\ to be cluster members; this results in a cluster mean velocity of $-$29.7$\pm$1.2 \kms\ (N=16).  Two other stars have velocities within 15 \kms\ of this value; they are considered as possible binary star members.
NGC 1245 495 and 902 are the only stars declared definite nonmembers, while for stars 406, 528,  and 536 the fibers were not well centered on the stars and velocities could not be measured.  Star 28 appears to be a rapidly rotating early type star, for which no radial velocity could be calculated using the twilight sky or HD 212943 templates.  We tentatively identify stars 10, 185 and 534 as spectroscopic binaries, as the standard deviations of their velocities calculated from multiple epochs were $\sim$2-3 \kms, much larger than the $\sim$0.4 \kms\ shown by the other stars.

\subsubsection{NGC 1817}
As the target stars in NGC 1817 were relatively bright, we were able to obtain data using two different Hydra field configurations, resulting in radial velocity determination for over seventy stars.  Only four stars were common to  the two configurations, which were observed two years apart.  The radial velocities determined from the different epochs for three of the four stars differed between 0.2 and 0.6 \kms, and for the fourth star (185) by 11.4 \kms.  Although this latter star could be a binary, its 2006 epoch spectrum had S/N$\sim$3; therefore, we favor the velocity determined from the 2008 epoch spectrum, which indicates it is a cluster member (Table 3).  The velocity histogram for all the stars observed exhibits a narrow peak at $\sim$66 \kms.  From 36 stars with velocities between 62 and 68 \kms, we have found a mean radial velocity of $+$65.2$\pm$1.0 \kms\ for NGC 1817.

We were not able to determine velocities for two stars, 285 and 1474.  The former is an early-type star, while the spectrum of the latter was too noisy to determine a velocity.  Mermilliod et al.\ (2003) determined high-precision radial velocities for several red giants in the field of NGC 1817; their cluster average velocity $+$65.33$\pm$0.52 (standard error; 29 stars) \kms\ is in excellent agreement with our value.  Of the 17 stars common to our two samples, three are identified as binary systems: 1420, 1424 and 1467.  The latter two stars are non-members according to their radial velocities and are identified as binaries based on the differences between our values and Mermilliod et al.\ (2003) being $-$29.2 and $+$4.2 \kms, respectively.  Considering only the fourteen stars identified as single star systems common to our samples, the average difference between our determined radial velocities is $-$0.05$\pm$0.24 (s.d.) \kms, with our values being smaller.  Lastly, we note that star 1502, which has a velocity $\sim$5 \kms\ from the cluster mean, is identified by Mermilliod et al.\ (2003) as a nonmember based on its 0\%\ membership probability assigned by Balaguer-N\'{u}\~{n}ez et al.\ (1998).  Balaguer-N\'{u}\~{n}ez et al.\ (2004a) redetermined the proper motion probability for this star (among others) using both parametric and non-parametric methods and found membership probabilities of 65\%\ or 48\%, identifying it as a probable non-member.  We have listed it as a possible member, based on its radial velocity.
Based on KPNO 4m echelle spectra, Jacobson et al.\ (2009) determined velocities of $+$65.9 and $+$66.1 \kms\ for stars 73 and 79, respectively.  These values are in good agreement with the results presented here: $+$65.1 and $+$65.8 \kms.

\subsubsection{NGC 2158}
NGC 2158 is an important, rich, old cluster, but its location in a rich stellar field makes determination of cluster members in photometric studies difficult.  Relatively few radial velocity studies exist for this cluster, most of which are based on low-resolution spectroscopy.  We have determined radial velocities for 36 stars in the cluster field from WIYN-Hydra data.  In addition to the three epochs of observations centered at $\sim$$\lambda$6280 \AA, observations centered at $\lambda$6600 \AA\ were obtained in 2006 February.  Radial velocities were measured using a twilight sky spectrum as a template, and then were shifted to the same scale as the 2008 December measures based on 12 stars common to each field configuration.  (The average difference in radial velocity between the two epochs of observation was $+$3.9$\pm$0.5 \kms (s.d.).)  Radial velocities based on the 2006 Feb spectra are given in Table~\ref{star_info} for eight stars.

The histogram of radial velocities for NGC 2158 stars presents a fairly broad peak around $+$27 \kms.  From 27 stars, we have calculated an average radial velocity of $+$26.9$\pm$1.9 (s.d.) \kms\ for the cluster.  Two more stars with velocities within 15 \kms\ are identified as possible members.  Our average value for the cluster is in good agreement with that of Friel et al.\ (2002), who found $+$28$\pm$10 \kms\ from moderate-resolution spectra of seven stars.  Minniti (1995) found a much lower mean velocity based on 20 stars: $+$14$\pm$9 \kms, although with a possible 10 \kms\ uncertainty in the zero point.  Geisler (1988) determined velocities for two stars from medium-resolution spectra, of which only 4230 is common to our study.  His value of $+$9.8 \kms\ for this star is much lower than the value we found here ($+$25.3 \kms) as well as the value determined from KPNO 4m echelle spectra of this star (Jacobson et al.\ 2009; $+$24.3 \kms).  There is no indication in our data that star 4230 is a binary, although more epochs of observations would be needed to confirm this.  In the meantime, our radial velocity measurements indicate it is a cluster member.  We have identified stars 4305 and 5102 as likely binary systems.

\subsubsection{NGC 2194}
NGC 2194 has the appearance of a rich cluster, though it is located in a crowded field (Piatti et al.\ 2003).  A CMD of the cluster shows relatively few evolved stars to target for observation.  Of the 20 or so stars observed in the field, seven cluster around 7-8 \kms\ in the velocity histogram (Figure~\ref{rvhist}).  These are taken to be cluster members, yielding an average velocity of $+$7.5$\pm$0.8 (s.d.) \kms.  Four more stars with velocities within 15 \kms\ of this value are taken to be possible cluster members.  No velocity variations were detected for any stars over the six day period in which the data were taken.  The low value of this cluster's radial velocity meant that the telluric emission line at $\lambda$6300 \AA\ was exactly superposed onto the [O I] forbidden feature in the stellar spectra, and a clean removal of the sky line was not possible.  Therefore, no oxygen abundances for stars in this cluster were determined.

\subsubsection{NGC 2355}
Only a dozen stars in the field of NGC 2355 were targeted for observation.  Fortunately, a clear cluster locus was readily identified in the radial velocity histogram.  Based on six stars identified as members, we have found an average radial velocity of $+$35.4$\pm$0.5 \kms\ for NGC 2355.  This value is in excellent agreement with those found by Soubiran et al.\ (2000; $+$35.13$\pm$0.39 \kms\ from 9 stars) and Mermilliod et al.\ (2008; $+$35.02$\pm$0.42 \kms\ from 7 stars).  Our measures for individual stars are also in general excellent agreement with these two studies, save for star 548.  Soubiran et al.\ (2000) identified it as a nonmember based on a radial velocity of $-$32.43 \kms, however, both we and Mermilliod et al.\ (2008) found it to have a radial velocity consistent with the cluster mean.  Apart from this star, average differences between our values and those of Soubiran et al.\ and Mermilliod et al.\ for stars in common are 0.4 \kms\ or smaller.

Soubiran et al.\ (2000) remarked on the curious nature of star 398, a cluster member two magnitudes brighter than the red clump, with the same color as red clump stars.  We confirm its velocity is consistent with that of the cluster (as does Mermilliod et al.\ 2008), although as Soubiran et al.\ pointed out, it could be a less distant field star with similar radial velocity.  Comparison of this star's chemical abundances to those of the red clump stars may shed light on this issue.  Soubiran et al.\ also noted that the FWHM of its cross correlation function was larger than typical for other cluster members, possibly indicating binarity or fast rotation.  The radial velocities we determined for this star based on spectra from two different epochs did not show any variation, nor did Mermilliod et al.\ (2008) detect velocity variations in their observations.  Likewise, the FWHM of individual absorption lines in its spectrum are comparable to those of other members of NGC 2355.

\subsubsection{NGC 2420}
Of 22 stars observed in the field of NGC 2420, twelve occupied a narrow peak in the velocity histogram.  The average radial velocity based on these twelve stars is $+$73.6$\pm$0.6 \kms.  One star, 8010, is a possible cluster member, as its velocity is within 15 \kms\ of the cluster mean.  All other stars in the field are clearly nonmembers.  Our measured radial velocity for the cluster is in good agreement with that of Mermilliod \& Mayor (2007), who found $+$73.18$\pm$0.94 from 14 single cluster members.  For the 13 stars common to our two studies, the average difference between our measures is smaller than 1 \kms.  Pancino et al.\ (2010) determined radial velocities of three cluster members from high resolution echelle spectra.  The average velocity of these three stars is $+$74.5$\pm$0.9 \kms.  Star 174 is a known spectroscopic binary (e.g., Smith \& Suntzeff 1987), while Mermilliod \& Mayor (2007) identified stars 76, 111 and 174 as possible binaries based on differences between their velocities and literature values.  Our measures for these stars agree with those of Mermilliod \& Mayor (2007) to better than 0.6 \kms, though we note that both our values for star 76 differ from that of Pancino et al.\ (2010) by more than 1 \kms.  These stars are identified as possible binaries in Table~\ref{star_info}.

\subsubsection{NGC 2425}
As can be seen in the bottom panel of Figure~\ref{rvhist}, stars in the field of NGC 2425 have relatively large radial velocities: ten stars clump in the histogram at an average velocity of $+$102.9$\pm$1.2 \kms, which we adopt as the cluster mean.  Four other stars with velocities within 15 \kms\ of this value are identified as possible cluster members. Star 815 is a binary star, as its radial velocity varied across all three epochs of observations, with its 2007 November spectra showing a velocity consistent with the cluster mean.

\subsubsection{NGC 7789}
The radial velocity histogram for NGC 7789 is strongly peaked at $-$54 \kms\ but is also broader compared to other open clusters, with several stars having velocities within 2 \kms\ of this value.  A similar broad distribution can also be seen in the radial velocity study of Gim et al.\ (1998a).  
They calculated radial velocities for 112 stars in the field of NGC 7789 and found a binary fraction of 32\%, which can explain the broad distribution in the radial velocity histogram.  Based on 50 single star members, Gim et al.\ determined an average radial velocity of $-$54.9$\pm$0.86 (s.d.) \kms.  As we used their measurements for our target selection, it is encouraging that all but one observed star are at least possible cluster members.  Comparison of our radial velocities to those of Gim et al.\ finds very good agreement, with the average difference between our values (excluding binary systems) being 0.1$\pm$0.7 \kms\, with our values larger.  The average radial velocity for NGC 7789 determined from our targets (26 single stars) is $-$54.7$\pm$1.3 \kms (s.d.).  Gim et al.\ (1998a) identified several binary systems in the cluster, these are indicated in Table~\ref{star_info}.  They remarked on the unknown membership status of several stars in the field based on large differences between their measured velocities and the cluster mean despite their having proper motions consistent with membership (McNamara \& Solomon 1981).  Of these uncertain stars, 9728 is most surely a binary, based on the 26.6 \kms\ difference between our measures and that of Gim et al.  As our two measures of this star bracket the cluster mean velocity, we identify it as a possible cluster member.  The membership of star 7840 is also uncertain; our measure of its velocity agrees within 1 \kms\ with that of Gim et al., and with a velocity $\sim$5 \kms\ within the cluster mean, we identify it as a possible cluster member.  Nine stars in our sample were not targeted by Gim et al.  Five have velocities within 1 standard deviation of the mean and are identified as members; the remaining four have velocities within 15 \kms\ of the cluster mean and are taken as possible members.
Pancino et al.\ (2010) presented radial velocities for three stars in NGC 7789, only one of which (7840) is common to our study.  Our velocity for this star agrees with that of Pancino et al.\ within 1 \kms.  The average velocity of all three stars in their sample is $-$51.2$\pm$2.2 \kms, in good agreement with our value.

\subsection{Line list and Equivalenth Width Measurements}

Although the development of our line list and comparison of it to others in the literature have been described 
in our previous work (particularly Friel et al.\ 2003, 2010 and Jacobson et al.\ 2008), it is worthwhile to include a 
brief review here.  Our line list is composed of absorption lines selected to be mostly free of blending in cool red 
giant star spectra at echelle resolutions.  These lines were selected from various sources that included both 
laboratory and inverse-solar log {\it gf}'s (see Friel et al.\ 2003 for details).  Rather than assemble a line list with 
heterogeneous {\it gf}-values, we opted to calculate {\it gf}-values differentially relative to the mildly metal-poor 
red giant Arcturus, a star more similar to those in our study than is the Sun.  Equivalent widths were measured in the 
Hinkle et al.\ (2000) Arcturus atlas, and the log {\it gf} values were altered until 
each line in our list reproduced the element abundances of Arcturus as found by Fulbright et al. (2006, 2007) and 
Peterson et al.\ (1993).  Specifically, we reproduced the [X/H] values for Arcturus found by Fulbright et al.\, and used 
them along with the solar abundances of Anders \& Grevesse (1989) to calculate log N(X) values for Arcturus.  For 
the elements not analyzed by Fulbright et al., we used the [X/Fe] ratios found by Peterson et al. (1993) to determine 
log N(X) values.

Uncertainties and/or systematics in our log {\it gf} values therefore arise from errors in the equivalent width 
measurements and our choice of Arcturus element abundances.  To estimate the former, we measured lines in our 
list multiple times and then redetermined the log {\it gf} values using the largest and smallest measurements for 
each line.  The average change in  log {\it gf} was 0.03 dex for Fe I, 0.02 dex for Fe II, and no more than 0.05 dex 
for other elements.  Such uncertainties have a much smaller impact on abundance uncertainties than do other 
parameters, such as effective temperature (see next section).  We have made careful comparisons of our 
abundance scale to those of others in the literature, not only based on comparison of log {\it gf} values but of 
abundance results for stars studied in common.  We refer the interested reader to discussions in Jacobson et al.\ 
(2008) and Friel et al.\ (2010) for more details.  Where necessary, we identify known or suspected systematic 
differences between our scale and others in later sections of this paper.

Equivalent widths (EWs) of absorption lines were measured in the spectra of all cluster single star radial velocity members with S/N ratios higher than $\sim$70.  In some cases, spectra with S/N$\sim$60 were analyzed, but these measurements are more uncertain, especially given that many of the features are much more blended in these R$\sim$18--21K spectra.  All EWs were fitted with Gaussians using interactive routines in IRAF's {\it splot} task, with great care taken to deblend features.  A few lines in our line list were deemed too blended for accurate measurement; these lines were discarded.  Some lines, particularly of Mg and Ti, were blended with telluric absorption features near $\lambda$6300 \AA, depending on a star's radial velocity.  In such cases, these lines were also discarded.  To aid in continuum placement, the high-resolution spectrum of Arcturus (Hinkle et al.\ 2000) and occasionally KPNO 4m (R$\sim$28K) echelle spectra of cool red giants were consulted.  The Mg $\lambda$6318 \AA\ line is affected by a Ca autoionization feature.  This line was measured, treating the Ca autoionization feature as the continuum. 

\begin{deluxetable}{lllcccc}
\tabletypesize{\scriptsize}
\tablenum{4}
\tablewidth{0pt}
\tablecaption{Equivalent Width Measurements\label{EW}}
\tablehead{ \colhead{Cluster} & \colhead{Star} & 
\colhead{$\lambda$} & \colhead{El.} & \colhead{E. P.} & \colhead{log {\it gf}} & \colhead{EW (m\AA)}
} 
\startdata
M67 &    84 & 6154.23 & 11.0 & 2.100 & -1.58 &  82 \\
M67 &    84 & 6160.75 & 11.0 & 2.100 & -1.20 & 106 \\
M67 &    84 & 6319.24 & 12.0 & 5.110 & -2.25 &  63 \\
M67 &    84 & 6696.03 & 13.0 & 3.140 & -1.45 & 120 \\
M67 &    84 & 6698.67 & 13.0 & 3.130 & -1.87 &  64 \\
M67 &    84 & 6142.49 & 14.0 & 5.620 & -1.64 &  50 \\
M67 &    84 & 6145.02 & 14.0 & 5.610 & -1.52 &  54 \\
M67 &    84 & 6155.13 & 14.0 & 5.620 & -0.82 &  87 \\
\enddata
\tablecomments{This table is published in its entirety in electronic format.  A portion is shown here for
guidance regarding its form and content.}
\end{deluxetable}

All EWs were measured multiple ($\sim$3) times to assess the uncertainties in continuum placement and fitting of blended features.  Measurement uncertainties naturally depended upon the quality of the spectra: for good S/N (70 or higher) spectra, measurement uncertainties were typically 2--5 m\AA, but were 5--8 m\AA\ for poorer-quality spectra.  For some of the more blended, strong features, EW uncertainties could be as high as 10--15 m\AA, due to both blending and difficulty in identifying local continua.  Fortunately, these instances were mainly confined to one or two strong Fe I lines, and so the effect on the abundance analysis of their large measurement error was mitigated by the presence of several other more reliably-measured Fe I lines.  Measured EWs for all stars are given in Table~\ref{EW}, fully available in electronic format.

\subsection{Atmospheric Parameter Determination}
\subsubsection{Photometric Parameters and Uncertainties}
Atmospheric parameters were determined from available optical and 2MASS photometry for each cluster (see Appendix A for information on individual clusters).  Stellar effective temperatures were determined using extinction-corrected magnitudes and colors and the relations of Alonso et al.\ (1999).  In general, we adopted the average of the (B$-$V), (V$-$K), and (J$-$K) temperatures, unless any appeared suspicious depending on the quality of the available photometry.
Surface gravities were calculated using the well-known relation: 
 \begin{equation}
log\ g = log\ (m/m_{\odot}) -0.4(M_{bol,\odot} - M_{bol,*}) + 4log\ (T/T_{\odot}) + log\ g_{\odot},
\end{equation}
where $M_{bol,\odot}$ = 4.72, $T_{\odot}$ = 5770 K, and log $g_{\odot}$ = 4.44 (Allen 1976).
Bolometric corrections were calculated using the Alonso et al.\ (1999) relations, and  turn-off masses
appropriate for the ages of the clusters.  As the number of Fe I lines suitable for analysis was small ($\sim$11--17), and weak Fe I lines were more difficult to measure, we chose not to determine microturbulent velocities by removing iron abundance trends with line strength.  Rather, 
we adopted a microturbulent velocity of 1.5 \kms\ for all stars.  We found this value to be adequate for the majority of stars in our previous work based on multi-order spectra, and it is within the range of values generally adopted for giant stars by different spectroscopic studies in the literature ($\sim$1.3--1.8 \kms; but see discussion below).
Atmospheric parameters for all stars are given in Table~\ref{star_atmparam_mod2}.

%checked for errors 3 May 2011
\begin{deluxetable*}{rrcccccccccc}
\tabletypesize{\scriptsize}
\tablecolumns{12}
\tablenum{5}
\tablewidth{0pt}
\tablecaption{Atmospheric Parameters and Fe abundances\label{star_atmparam_mod2}}
\tablehead{ \colhead{Cluster} &
\colhead{Star} & \colhead{T$_{eff}$} & \colhead{log \it g} & \colhead{v$_{t}$} & \colhead{log N(Fe I)}  & \colhead{$\sigma$} &\colhead{\# lines} & \colhead{[Fe/H]} & \colhead{log N(Fe II)} & \colhead{$\sigma$} &\colhead{\# lines}\\
\colhead{} & \colhead{} & \colhead{(K)} & \colhead{(dex)} & \colhead{(km s$^{-1}$)} & \colhead{} & \colhead{} & 
\colhead{} & \colhead{(dex)} & \colhead{} & \colhead{} & \colhead{}}
\startdata
M67 & 84   & 4650 & 2.5 & 1.5 & 7.52 & 0.20 & 29 & $+$0.00 & 7.83 & 0.25 & 7\\
M67 & 105  & 4400 & 2.2 & 1.5 & 7.60 & 0.21 & 30 & $+$0.08 & 7.74 & 0.19 & 7\\
M67 & 108  & 4200 & 1.8 & 1.5 & 7.54 & 0.22 & 30 & $+$0.02 & 7.80 & 0.27 & 7\\
M67 & 135  & 4700 & 2.8 & 1.5 & 7.50 & 0.20 & 30 & $-$0.02 & 7.83 & 0.32 & 7\\
M67 & 141  & 4700 & 2.4 & 1.5 & 7.60 & 0.18 & 30 & $+$0.08 & 7.75 & 0.21 & 7\\
M67 & 143  & 5100 & 3.0 & 1.5 & 7.41 & 0.20 & 29 & $-$0.11 & 7.62 & 0.19 & 7\\
M67 & 151  & 4700 & 2.4 & 1.5 & 7.50 & 0.19 & 29 & $-$0.02 & 7.69 & 0.25 & 7\\
M67 & 170  & 4200 & 1.8 & 1.5 & 7.55 & 0.19 & 29 & $+$0.03 & 7.87 & 0.26 & 7\\
\enddata
\tablecomments{This table is published in its entirety in electronic format.  A portion is shown here for
guidance regarding its form and content.}
\end{deluxetable*}

For the same reason, we opted not to refine atmospheric parameters spectroscopically (Jacobson et al.\ 2011).  Therefore, uncertainties in T$_{\rm eff}$ and log {\it g} are predominantly due to uncertainties in the adopted cluster distances and reddening values.  To estimate the magnitude of these uncertainties, we determined effective temperatures for stars in the clusters NGC 2158, NGC 2194, NGC 2355 and NGC 7789 using different reddening values.  The values chosen were $\sim$0.1--0.2 magnitudes larger than those shown in Table 1, and were generally selected because they were the more extreme values found in the literature.  Effective temperatures changed by $\sim$100--200 K, with the larger differences for NGC 2158, which is one of the more reddened clusters.  Surface gravities only changed 0.1--0.2 dex.  Therefore, we adopted conservative uncertainties of 200 K and 0.2 dex in T$_{\rm eff}$ and log {\it g}, respectively.  As for our analysis of clusters NGC 2204 and NGC 2243 (Jacobson et al.\ 2011), we adopted an uncertainty 0.2 \kms\ for microturbulent velocity.  
Table~\ref{tab_unc} shows the sensitivities of element abundances to uncertainties in the atmospheric parameters for a hotter and cooler star in our sample.  Fortunately, changes of 0.2 dex in gravity have little effect on most element abundances, save oxygen (see Section 4.5).

 \begin{deluxetable}{llccc}
\tabletypesize{\scriptsize}
\tablewidth{0pt}
\tablenum{6}
\tablecaption{Abundance Uncertainties due to Atmospheric Parameters\label{tab_unc}}
\tablehead{
\colhead{} &
\colhead{} &
\colhead{T$_{eff}$} &
\colhead{log g} &
\colhead{v$_{t}$} \\%& \colhead{[M/H]} \\
\colhead{Star} & \colhead{[X/H]} &
\colhead{$+$200 K} &
\colhead{$+$0.2 dex} &
\colhead{$+$0.2 km s$^{-1}$}} %& \colhead{$+$0.5 dex}}
\startdata
N1817$-$164 & Fe I & $+$0.21 & $+$0.02 & $-$0.07 \\%& $+$0.01 \\
 & Fe II & $-$0.01 & $+$0.13 & $-$0.03 \\%& $+$0.20 \\
& Na I & $+$0.12 & $+$0.00 & $-$0.02 \\%& $+$0.01  \\
& Mg I & $+$0.09 & $+$0.01 & $-$0.01 \\%& \nodata \\
%& Al I & $+$0.05 & $-$0.01 & $-$0.02 \\%& $+$0.00 \\
& Si I & $+$0.08 & $+$0.02 & $-$0.02 \\%& $+$0.09  \\
& Ca I & $+$0.17 & $+$0.01 & $-$0.05 \\%& $-$0.01  \\
 & Ti I & $+$0.18 & $-$0.02 & $-$0.01 \\%& $-$0.02 \\
% & Cr I & $+$0.10 & $+$0.00 & $-$0.01 \\%& $-$0.01  \\
  & Ni I & $+$0.14 & $+$0.01 & $-$0.02 \\%& $+$0.06 \\
 & Zr I &  $+$0.23 & $-$0.02 & $+$0.00 \\%& \nodata  \\
\hline
N188$-$6712 & Fe I & $-$0.01 & $+$0.03 & $-$0.14 \\%& $+$0.14\\
& Fe II & $-$0.42 & $+$0.14 & $-$0.07 \\%& $+$0.23\\
& Na I & $+$0.25 & $-$0.03 & $-$0.08 \\%& $+$0.01\\
& Mg I & $-$0.07 & $+$0.04 & $-$0.04 \\%& $+$0.07\\
%& Al I & $+$0.07 & $-$0.01 & $-$0.05 \\%& $+$0.01\\
& Si I & $-$0.21 & $+$0.06 & $-$0.03 \\%& $+$0.14\\
& Ca I & $+$0.29 & $-$0.04 & $-$0.14 \\%& $+$0.04\\
& Ti I & $+$0.32 & $+$0.03 & $-$0.16 \\%& $+$0.05\\
%& Cr I & $+$0.15 & $+$0.03 & $-$0.06 \\%& $+$0.07\\
& Ni I & $-$0.09 & $+$0.07 & $-$0.09 \\%& $+$0.14\\
& Zr I & $+$0.46 & $+$0.04 & $-$0.18 \\%& $+$0.08\\
\enddata

\end{deluxetable}

\subsubsection{Comparison to Literature Results}
In general our atmospheric parameters agree well with literature values for clusters previously studied, particularly high resolution spectroscopic studies.  For most cases, our effective temperatures differ from spectroscopically-determined values in the literature by no more than 100 K, while log {\it g}'s typically agree within 0.2 dex, though occasionally within 0.4 dex.  The high resolution spectroscopy study with which we agree less well is that of Pancino et al.\ (2010) for M 67 stars and some NGC 2420 stars common to our studies.  Differences are on order 150--200 K and 0.1 to 0.4 dex for temperature and gravity.  For all clusters, differences between our assumed microturbulence value of 1.5 \kms\ and literature values are typically 0.1--0.2 \kms, and no more than 0.3 \kms, with our values being sometimes lower and sometimes higher.  

Our effective temperatures are also in general good agreement with those determined from Washington photometry for stars in NGC 1817 (Parisi et al.\ 2005) and NGC 2158 (Geisler 1987).  For NGC 1817, our values are 100-300 K lower than those of Parisi et al.\ (2005), while for NGC 2158 our values are 100-200 K higher than those of Geisler (1987).  It is likely that choice of adopted reddening to the cluster can explain at least some of these differences (differences between our adopted reddening and those of the above studies are 0.03 magnitudes).  Soubiran et al.\ (2000) determined effective temperatures and gravities for several stars in NGC 2355 by comparing low S/N ratio ELODIE spectra to the TGMET library.  Our temperatures agree to within 100 K with their values, while gravities agree to within 0.2--0.4 dex.

As already mentioned, we performed an abundance analysis of stars in the clusters M 67, NGC 188, NGC 1817, and NGC 2158 based on KPNO 4m echelle spectra.  Effective temperatures and surface gravities were determined spectroscopically, by forcing ionization and excitation equilibrium in Fe abundances.  The resulting stellar parameters are generally in good agreement  with those determined photometrically here for all clusters: effective temperatures within 100 K, and surface gravities within 0.4 dex, with the spectroscopically-determined gravities being lower.  

\subsubsection{Microturbulent Velocity}
It is well known that stellar microturblent velocity can vary greatly with position in the HR diagram, with more evolved stars having higher values ($\sim$2--3 \kms), and unevolved stars having lower values ($\lesssim$1 \kms).  Though all stars in this study are evolved stars, they vary widely in evolutionary state, as indicated by the nearly 2 dex range of surface gravities seen in Table~\ref{star_atmparam_mod2}.  Assignment of a single microturbulent velocity across the board can therefore introduce a systematic error to the resulting abundances for individual stars on the order of the values seen in Table~\ref{tab_unc}, or even larger.

Several relations between v$_{t}$ and log {\it g} and/or T$_{\rm eff}$ exist in the literature, and many studies use these relations to determine microturblent velocities for their program stars (and then often go on to modify them to remove any abundance trends with line strength).  However, it is difficult to select a particular relation to use, because each is derived for stars of a particular range of temperature, surface gravity and/or metallicity.  Furthermore, some relations are determined empirically from measured EWs for a sample of stars (e.g., Gratton et al.\ 1996), while others (e.g., Carretta et al.\ 2004) are determined using {\it theoretical} EWs to remove what Pancino et al.\ (2010) have dubbed the ``Magain effect" -- the fact that uncertainties in EW measurements drive any determination of v$_{t}$ by making abundance independent of line strength to a value superficially larger than is correct (Magain 1984).   As a result, we have been reluctant to select a relation or relations in the literature to apply to our sample.  We can, however, investigate the effect the adoption of a single microturbulent velocity has on our abundance results by selecting a few relations from the literature, calculating new abundances, and comparing them to our results.

We selected five relations from the literature: v$_{t}$ = 1.5 $-$ 0.13 log {\it g} (Carretta et al.\ 2004; C04); v$_{t}$ = 4.08 $-$ 5.01$\times$10$^{-4}$ T$_{\rm eff}$ (Ram\'irez \& Cohen 2003; RC); v$_{t}$ = $-$0.254 log {\it g} + 1.930 (Marino et al.\ 2008; M08); v$_{t}$ = 2.22 $-$ 0.322 log {\it g} (Gratton et al.\ 1996; G96); and v$_{t}$ = $-$0.0011 T$_{\rm eff}$ + 6.66 (Johnson et al.\ 2008; J08).  While all these relations are appropriate for red giant stars, the majority of them were determined based on metal-poor stars: the RC, M08 and J08 relations were determined for globular cluster stars, while the metallicity range of the bulk of stars used in the G96 relation is [Fe/H]$\sim$ $-$1.0 to 0.0.  Only the C04 relation is based on open cluster stars.  

We used these relations to calculate microturbulent velocities for stars in the clusters M 67, NGC 1817, NGC 2420 and NGC 7789.  These clusters were selected for this analysis because either their element abundances are very well known (M 67), they have a sample of stars ranging 1.8 dex in log {\it g} (NGC 7789), they are relatively metal-poor (NGC 2420), or conversely to NGC 7789, the sample of stars analyzed have log {\it g} $\sim$2.0--3.0, for which v$_{t}$ = 1.5 \kms\ may be systematically incorrect for all stars (NGC 1817).
The values for individual stars ranged as much as 0.5 \kms\ from all the different relations used.    In general, the C04 relation resulted in the lowest microturbulent velocities (median values 1.1--1.2 \kms), and the M08 (1.2-1.3 \kms), G96 (1.4 \kms) and RC (1.5--1.7 \kms) relations in successively larger values for all four clusters considered.  Median microturbulent values from the J08 relation varied from 1.1 \kms\ for NGC 1817 to 1.5 \kms\ for M 67.  As mentioned by Pancino et al.\ (2010), the systematically low v$_{t}$ values from the C04 relation are likely due to the relation being based on theoretical EWs, rather than measured EWs.  

\begin{deluxetable*}{lcccccccccccccccc}
\tabletypesize{\scriptsize}
%\rotate
\tablewidth{0pt}
\tablenum{7}
\tablecaption{Abundance Differences for Different Microturbulent Velocities\label{vtabund}}
\tablehead{ \colhead{} & \colhead{} & \multicolumn{3}{c}{M 67} & \colhead{} & \multicolumn{3}{c}{NGC 1817} & \colhead{} & \multicolumn{3}{c}{NGC 2420} & \colhead{} & 
\multicolumn{3}{c}{NGC 7789} \\
\cline{3-5} \cline{7-9} \cline{11-13} \cline{15-17}\\
\colhead{El.} & \colhead{} & \colhead{G96} & \colhead{C04} & \colhead{J08} & \colhead{} & \colhead{G96} & \colhead{C04} & \colhead{J08} & \colhead{} &
\colhead{G96} & \colhead{C04} & \colhead{J08} & \colhead{} & \colhead{G96} & \colhead{C04} & \colhead{J08}
} 
\startdata
Fe I & & $-$0.05 & $-$0.16 & $-$0.01 & & $-$0.07 & $-$0.18 & $-$0.15 & & $-$0.03 & $-$0.16 & $-$0.10 & & $-$0.05 & $-$0.20 & $-$0.03 \\
 Na & & $-$0.01 & $-$0.05 & $+$0.01 & & $-$0.02 & $-$0.05 & $-$0.04 & & $-$0.01 & $-$0.01 & $-$0.03 & & $-$0.02 & $+$0.00 & $-$0.05 \\
 Mg & & $-$0.02 & $-$0.05 & $+$0.00 & & $-$0.01 & $-$0.03 & $-$0.03 & & $-$0.01 & $-$0.02 & $-$0.01 & & $-$0.02 & $-$0.06 & $-$0.01 \\
  Al & & $-$0.05 & $-$0.10 & $+$0.01 & & \nodata & \nodata & \nodata & & \nodata & \nodata & \nodata & & \nodata & \nodata & \nodata \\
  Si & & $-$0.02 & $-$0.04 & $+$0.01 & & $-$0.04 & $-$0.07 & $-$0.04 & & $-$0.01 & $-$0.04 & $-$0.03 & & $-$0.06 & $-$0.12 & $-$0.05 \\
Ca & & $-$0.05 & $-$0.16 & $+$0.00 & & $-$0.08 & $-$0.15 & $-$0.16 & & $-$0.05 & $-$0.15 & $-$0.11 & & $-$0.05 & $-$0.18 & $-$0.01 \\
  Ti & & $-$0.02 & $-$0.08 & $+$0.01 & & $-$0.01 & $-$0.02 & $-$0.02 & & $+$0.00 & $-$0.03 & $-$0.02 & & $+$0.03 & $-$0.09 & $+$0.02 \\
  Ni & & $-$0.03 & $-$0.12 & $+$0.01 & & $-$0.03 & $-$0.06 & $-$0.05 & & $-$0.02 & $-$0.08 & $-$0.05 & & $-$0.01 & $-$0.03 & $+$0.02 \\
  Zr & & $+$0.00 & $-$0.14 & $-$0.09 & & $-$0.08 & $-$0.09 & $-$0.09 & & $+$0.00 & $-$0.01 & $+$0.00 & & $+$0.10 & $-$0.09 & $+$0.13 \\
\enddata
\tablecomments{Shown here are differences between log N(X) abundances calculated adopting v$_{t}$ = 1.5 \kms\ for each star and abundances calculated using the v$_{t}$ relations from G96, C04 and J08.  See text for more information.}
\end{deluxetable*}

To illustrate the effect of the adopted microturbulent velocity on resulting abundances, we highlight only three relations: C04 and G96 because they are often used by other studies in the literature (e.g., Sestito et al.\ 2008), and J08.  This latter relation was determined from a sample of 180 giant stars in ${\omega}$ Cen observed with the Hydra multi-object spectrograph on the CTIO 4m telescope, most similar to the data analyzed here (though note that the log {\it g} range of ${\omega}$ Cen giants is smaller than considered here: $\sim$0.4 to 1.5, and its [Fe/H] range of $-$2.2 to $-$0.7 is also lower).  We determined abundances for stars in all four clusters using our determined temperatures and gravities and the microturbulent velocities from these three relations and then calculated weighted mean cluster abundances and standard deviations, as described in the next section.  Table~\ref{vtabund} shows the differences between weighted mean cluster abundances (log N(X)) when 1.5 \kms\ is assigned to all stars and when microturbulent velocities are determined by the above relations.  

%checked for errors 3 May 2011
\begin{deluxetable*}{rrcccccccccc}
\tabletypesize{\scriptsize}
\tablecolumns{12}
\tablenum{8}
\tablecaption{$\alpha$ Element Abundances\label{logN_alpha_abund}}
\tablehead{ \colhead{Cluster} &
\colhead{Star} & 
 \colhead{log N(Mg)} & \colhead{log N(Si)} & \colhead{$\sigma$$_{Si}$} & \colhead{\#$_{Si}$} & \colhead{log N(Ca)\tablenotemark{1}} & 
\colhead{$\sigma$$_{Ca}$} & \colhead{\#$_{Ca}$} & \colhead{log N(Ti)} & \colhead{$\sigma$$_{Ti}$} &
\colhead{\#$_{Ti}$} %& \colhead{$\sigma$[Ni/H]} & \colhead{[Zr/H]} & \colhead{$\sigma$[Zr/H]}
}
\startdata
M 67 &   84 & 7.73 & 7.78 & 0.11 & 4 & 6.29 & 0.17 & 5 & 4.85 & 0.06 & 3 \\
M 67 &  105 & 7.94 & 7.75 & 0.29 & 4 & 6.21* & 0.11 & 5 & 4.89 & 0.12 & 3 \\
M 67 &  108 & 7.90 & 8.01 & 0.27 & 4 & 6.11* & 0.27 & 5 & 4.79 & 0.24 & 3 \\
M 67 &  135 & 7.82 & 7.74 & 0.08 & 4 & 6.19 & 0.15 & 5 & 4.91 & 0.07 & 3\\
M 67 &  141 & 7.88 & 7.76 & 0.11 & 4 & 6.37 & 0.16 & 5 & 4.94 & 0.10 & 3\\
M 67 &  143 & 7.74 & 7.64 & 0.08 & 4 & 6.18 & 0.27 & 5 & 4.84 & 0.13 & 3 \\
M 67 &  151 & 7.69 & 7.80 & 0.18 & 4 & 6.26 & 0.28 & 5 & 4.84 & 0.09 & 3 \\
M 67 &  170 & 7.85 & 8.00 & 0.20 & 4 & 6.13* & 0.23 & 5 & 4.72 & 0.22 & 3 \\
\enddata
\tablenotetext{1}{Asterisk indicates abundance determination includes lines $>$150 m\AA.}
\tablecomments{This table is published in its entirety in electronic format.  A portion is shown here for
guidance regarding its form and content.}
\end{deluxetable*}%8
% values checked 4 May 2011
 \begin{deluxetable*}{rrccccccccc}
\tabletypesize{\scriptsize}
\tablecolumns{10}
\tablenum{9}
\tablecaption{Other Element Abundances\label{logN_other_abund}}
\tablehead{ \colhead{Cluster} &
\colhead{Star} & 
\colhead{log N(Na)} & \colhead{$\sigma$$_{Na}$} & \colhead{\#$_{Na}$} & \colhead{log N(Ni)} & 
\colhead{$\sigma$$_{Ni}$} & \colhead{\#$_{Ni}$} & \colhead{log N(Zr)} & \colhead{$\sigma$$_{Zr}$} &
\colhead{\#$_{Zr}$} %& \colhead{$\sigma$[Ni/H]} & \colhead{[Zr/H]} & \colhead{$\sigma$[Zr/H]}
}
\startdata
M 67 &   84 & 6.27 & 0.03 & 2 & 6.25 & 0.14 & 6 & 2.42 & 0.18 & 2 \\
M 67 &  105 & 6.29 & 0.17 & 2 & 6.29 & 0.18 & 6 & 2.60 & 0.01 & 2 \\
M 67 &  108 & 6.34 & 0.06 & 2 & 6.24 & 0.14 & 6 & 2.39 & 0.21 & 2 \\
M 67 &  135 & 6.31 & 0.17 & 2 & 6.24 & 0.20 & 6 & 2.48 & 0.12 & 2 \\
M 67 &  141 & 6.56 & 0.22 & 2 & 6.31 & 0.10 & 6 & 2.52 & 0.14 & 2 \\
M 67 &  143 & 6.34 & 0.12 & 2 & 6.11 & 0.13 & 6 & 2.77 & 0.27 & 2 \\
M 67 &  151 & 6.41 & 0.18 & 2 & 6.22 & 0.20 & 6 & 2.48 & 0.37 & 2 \\
M 67 &  170 & 6.38 & 0.28 & 2 & 6.20 & 0.20 & 6 & 2.29 & 0.13 & 2 \\
\enddata
\tablecomments{This table is published in its entirety in electronic format.  A portion is shown here for
guidance regarding its form and content.}
\end{deluxetable*}%9

As can be seen, the differences in log N(X) abundances for non-Fe elements are generally less than 0.1 dex, save for Ca.
Not surprisingly, the effect on Fe abundances can be large. 
 Regarding the abundances of individual stars, we found that the line-by-line dispersion in Fe abundance is independent of the adopted microturbulent velocity.  Likewise, the standard deviation of cluster weighted mean abundances is independent of adopted microturbulent velocities, which implies that the choice of microturbulent velocity does not impact clusters with different ranges of log {\it g} in different ways.  The smallest abundance differences occur for the G96 relation, which of all the relations produces median velocities near 1.5 \kms.

Although Table~\ref{vtabund} shows that cluster weighted mean abundances {\it are} impacted by the adoption of a single mictroturbulent velocity for all stars, the magnitude of the effect appears comparable to that seen when comparing abundances determined using velocities from different relations in the literature.  Furthermore, the majority of high resolution spectroscopic studies in the literature refine microturbulent velocities in the abundance analysis, and as already mentioned, the final adopted velocities often fall within 1.3--1.7 \kms\ for the studies we often compare our results to (see, e.g., Yong et al.\ 2005, Pancino et al.\ 2010).  Comparisons of abundance results in our previous work to the literature for clusters such as M 67 have shown general good agreement in spite of our adoption of v$_{t}$ = 1.5 \kms\ for the majority of our program stars.  Therefore, we conclude that no major systematics are introduced into our results as a result of our treatment of microturbulence.  It is important to note, though, that the adoption of different relations by different studies may introduce systematic differences between studies: a 0.4 \kms\ difference in v$_{t}$ corresponds to $\sim$0.18-0.20 dex difference in [Fe/H], as shown here.  However, given that the spectroscopic determination of T$_{\rm eff}$ and v$_{t}$ are linked, this effect may be at least partly compensated for by choice of effective temperature.

\subsection{Abundance Analysis}
Element abundances were determined using the 2002 version of the LTE analysis code MOOG (Sneden 1973).  Model atmospheres were interpolated from a grid of plane parallel MARCS models (Bell et al.\ 1976) with appropriate effective temperatures and surface gravities for the majority of program stars (exceptions noted below).  Initial values of [M/H] were chosen based on previous estimations of cluster metallicity in the literature.  [M/H] values were then altered depending on the Fe abundance determined in the first iteration.  The metallicity step-size of our atmosphere grid is 0.25 dex.  Only Fe I lines with equivalent widths 150 m\AA\ or smaller were used to calculate stellar metallicity; given the smaller number of lines available for non-Fe elements, all lines were used to determine abundances regardless of strength.  Such cases where strong line abundances were used are noted in the tables.

Several stars in NGC 188 have log {\it g} values greater than 3.0, beyond the extent of our grid of plane parallel MARCS models.  Rather than extrapolate to gravities as high as 3.7, we opted to use a grid of the newer spherical MARCS models\footnote{See http://marcs.astro.uu.se/} (Gustafsson et al.\ 2008).  We compared the
plane parallel and spherical models to be sure there were no gross differences, and we also compared the 
abundances determined for sample stars (with log {\it g} $<$ 3.0) using both models.  For all elements, the abundance differences 
amounted to no more than 0.03 dex, so we are confident that no major systematic uncertainties are introduced by
the adoption of the spherical MARCS models for a subset of our stars. 

%\begin{landscape}
 \begin{deluxetable*}{rrcccccccccccc}
\tabletypesize{\scriptsize}
\tablecolumns{14}
\tablenum{10}
\tablecaption{$\alpha$ Element Abundance Ratios\label{alpha_abund_ratios}}
\tablehead{ \colhead{Cluster} &
\colhead{Star} & 
 \colhead{[Mg/H]} & \colhead{[Mg/Fe]} & \colhead{$\sigma$$_{[Mg/Fe]}$} & 
 \colhead{[Si/H]} & \colhead{[Si/Fe]} & \colhead{$\sigma$$_{[Si/Fe]}$} 
& \colhead{[Ca/H]} & \colhead{[Ca/Fe]} & 
\colhead{$\sigma$$_{[Ca/Fe]}$} & \colhead{[Ti/H]} & \colhead{[Ti/Fe]} & \colhead{$\sigma$$_{[Ti/Fe]}$} %&
%\colhead{\#$_{Ti}$} %& \colhead{$\sigma$[Ni/H]} & \colhead{[Zr/H]} & \colhead{$\sigma$[Zr/H]}
}
\startdata
M 67 &   84 & $+$0.15 & $+$0.15 & 0.28 & $+$0.23 & $+$0.23 & 0.23 & $-$0.07 & $-$0.07 & 0.26 & $-$0.14 & $-$0.14 & 0.21 \\
M 67 &  105 & $+$0.36 & $+$0.28 & 0.29 & $+$0.20 & $+$0.12 & 0.36 & $-$0.15 & $-$0.23 & 0.24 & $-$0.10 & $-$0.18 & 0.24 \\
M 67 &  108 & $+$0.32 & $+$0.30 & 0.30 & $+$0.46 & $+$0.44 & 0.35 & $-$0.25 & $-$0.27 & 0.35 & $-$0.20 & $-$0.22 & 0.33 \\
M 67 &  135 & $+$0.24 & $+$0.26 & 0.28 & $+$0.19 & $+$0.21 & 0.22 & $-$0.17 & $-$0.15 & 0.25 & $-$0.08 & $-$0.06 & 0.21 \\
M 67 &  141 & $+$0.30 & $+$0.22 & 0.27 & $+$0.21 & $+$0.13 & 0.21 & $+$0.01 & $-$0.07 & 0.24 & $-$0.05 & $-$0.13 & 0.21 \\
M 67 &  143 & $+$0.16 & $+$0.27 & 0.28 & $+$0.09 & $+$0.20 & 0.22 & $-$0.18 & $-$0.07 & 0.34 & $-$0.15 & $-$0.04 & 0.24 \\
M 67 &  151 & $+$0.11 & $+$0.13 & 0.28 & $+$0.25 & $+$0.27 & 0.26 & $-$0.10 & $-$0.08 & 0.34 & $-$0.15 & $-$0.13 & 0.21 \\
M 67 &  170 & $+$0.27 & $+$0.24 & 0.28 & $+$0.45 & $+$0.42 & 0.28 & $-$0.23 & $-$0.26 & 0.30 & $-$0.27 & $-$0.30 & 0.29 \\
\enddata
\tablecomments{This table is published in its entirety in electronic format.  A portion is shown here for
guidance regarding its form and content.}
\end{deluxetable*}
%\end{landscape}
%10

Individual star iron abundances based on Fe I and Fe II lines are given in Table~\ref{star_atmparam_mod2},  in the form log N(Fe), along with standard deviations of the mean and the number of lines measured.  Also listed are [Fe/H] ratios calculated from the Fe I abundances only, which we adopt as the stellar metallicities, relative to a solar log N(Fe) = 7.52, the default used in MOOG.  Table~\ref{logN_alpha_abund} presents stellar abundances of Mg, Si, Ca and Ti in the form log N(X), along with standard deviations of the mean and the number of lines used to calculate the averages.  Table~\ref{logN_other_abund} presents abundances of Na, Ni and Zr in the same format.  Tables~\ref{alpha_abund_ratios} and ~\ref{other_abund_ratios} present abundance ratios [X/H] and [X/Fe] for the $\alpha$ elements and Na, Ni and Zr, respectively, along with standard deviations for [X/Fe].  [X/Fe] ratios are calculated using each individual star's [Fe/H] value, and $\sigma$$_{[X/Fe]}$ is calculated by adding the standard deviations of [X/H] and [Fe/H] in quadrature.  Stellar [X/H] and [X/Fe] ratios are all calculated relative to the solar abundances of Anders \& Grevesse (1989), the default solar abundances used in MOOG.

As in our previous analysis of single-order Hydra echelle spectra (Jacobson et al.\ 2011), we have chosen to calculate weighted cluster mean abundances (Taylor 1982) in order to minimize the contribution of individual star abundances with relatively large uncertainties.  Individual stellar abundances are weighted by their standard deviations in the calculation of the cluster average.  For the majority of elements, the difference between weighted and unweighted cluster means is generally within 0.05 dex, save for elements such as Zr, the lines of which are often very weak or not measured in some stars.  The cluster exhibiting the largest difference between weighted and unweighted mean abundances is NGC 2355, for which element abundances vary rather largely ($\sim$0.3 dex) among its five stars.
Weighted cluster mean [X/H] ratios along with standard deviations and the number of stars used, are presented in Table~\ref{cluster_wxh}, while Table~\ref{cluster_wxfe} presents the same for [X/Fe].

% Values doublechecked 5 May 2011
%\begin{landscape}
 \begin{deluxetable*}{rrccccccccc}
\tabletypesize{\scriptsize}
\tablecolumns{11}
\tablenum{11}
\tablewidth{0pt}
\tablecaption{Other Element Abundance Ratios\label{other_abund_ratios}}
\tablehead{ \colhead{Cluster} &
\colhead{Star} & 
 \colhead{[Na/H]} & \colhead{[Na/Fe]} & \colhead{$\sigma$$_{[Na/Fe]}$} & 
 \colhead{[Ni/H]} & \colhead{[Ni/Fe]} & \colhead{$\sigma$$_{[Ni/Fe]}$} 
& \colhead{[Zr/H]} & \colhead{[Zr/Fe]} & 
\colhead{$\sigma$$_{[Zr/Fe]}$}% & \colhead{[Ti/H]} & \colhead{[Ti/Fe]} & \colhead{$\sigma$$_{[Ti/Fe]}$} &
%\colhead{\#$_{Ti}$} %& \colhead{$\sigma$[Ni/H]} & \colhead{[Zr/H]} & \colhead{$\sigma$[Zr/H]}
}
\startdata
M 67 &   84 & $-$0.06 & $-$0.06 & 0.20 & $+$0.00 & $+$0.00 & 0.24 & $-$0.18 & $-$0.18 & 0.27  \\
M 67 &  105 & $-$0.04 & $-$0.12 & 0.27 & $+$0.04 & $-$0.04 & 0.28 & $+$0.00 & $-$0.08 & 0.21  \\
M 67 &  108 & $+$0.01 & $-$0.01 & 0.23 & $-$0.01 & $-$0.03 & 0.26 & $-$0.21 & $-$0.23 & 0.30  \\
M 67 &  135 & $-$0.02 & $+$0.00 & 0.26 & $-$0.01 & $+$0.01 & 0.28 & $-$0.12 & $-$0.10 & 0.23 \\
M 67 &  141 & $+$0.23 & $+$0.15 & 0.28 & $+$0.06 & $-$0.02 & 0.21 & $-$0.08 & $-$0.16 & 0.23 \\
M 67 &  143 & $+$0.01 & $+$0.12 & 0.23 & $-$0.14 & $-$0.03 & 0.24 & $+$0.17 & $+$0.28 & 0.34  \\
M 67 &  151 & $+$0.08 & $+$0.10 & 0.26 & $-$0.03 & $-$0.01 & 0.28 & $-$0.12 & $-$0.10 & 0.42  \\
M 67 &  170 & $+$0.05 & $+$0.02 & 0.34 & $-$0.05 & $-$0.08 & 0.28 & $-$0.31 & $-$0.34 & 0.23   \\
\enddata
\tablecomments{This table is published in its entirety in electronic format.  A portion is shown here for
guidance regarding its form and content.}
\end{deluxetable*}
%11

We stress that the magnesium abundances are based on measurement of the $\lambda$6318 \AA\ feature only.  This feature is greatly blended at this spectral resolution, and therefore the resulting Mg abundances are very uncertain and likely overestimates.  To calculate $\sigma$$_{[Mg/Fe]}$ for each star, we assigned an uncertainty of 0.2 dex to its [Mg/H] ratio.  Likewise, we adopted 0.2 dex uncertainty for Zr abundances based on measurement of only a single line.  This is approximately the line-by-line dispersion in Fe abundance for individual stars.  Examination of the Si abundances showed that the \lam6243 \AA\ line produced systematically larger abundances for nearly every star in our sample.  Given that this line is quite blended in these Hydra data, we excluded it from calculation of stellar Si abundances.  
The Na abundances presented here are LTE abundances; we make no corrections for NLTE effects.  Given that most of our sample stars are relatively metal-rich and warm, and that we use the \lam\lam6154-6160 \AA\ Na doublet, the NLTE corrections are not expected to be large (Mashonkina et al.\ 2000).

%\begin{landscape}
%\input{tab12}%12
%\end{landscape}
%\input{tab13}%13
%\end{landscape}

Abundances of Al and Cr were also determined for all M 67 stars and three NGC 2158 stars from Hydra \lam$\sim$6600 \AA\ spectra.  These abundances are shown in Table~\ref{abund_6600}.  To estimate the uncertainties in [Cr/Fe], which rests on the measurement of only one line, we assigned an uncertainty of 0.2 dex to log N(Cr).  Similarly, we assigned 0.2 dex uncertainty to the Si, Ti and Ni log N(X) values of NGC 2158 stars 7773 and 7866, since they are also determined from a single line only.  Given that non-Fe element abundances for these two stars are based on only one line, they are excluded from the calculation of the cluster average abundances shown in Tables~\ref{cluster_wxh} and \ref{cluster_wxfe}.  The average [Fe/H] for NGC 2158 changes only by 0.01 dex with no change in standard deviation if these stars are included in the calculation, therefore their exclusion does not affect the cluster's mean metallicity measurement.  Weighted cluster mean Al and Cr abundance ratios for M 67 and NGC 2158 are given in Table~\ref{cluster_wAlCr}.

% Values checked 5 May 2011
\begin{deluxetable}{rrccccc}
\tabletypesize{\scriptsize}
\tablecolumns{7}
\tablenum{14}
\tablewidth{0pt}
\tablecaption{Al and Cr Abundances For M 67 and NGC 2158 Stars\label{abund_6600}}
\tablehead{ \colhead{Cluster} &
\colhead{Star} & 
 \colhead{log N(Al)} & \colhead{$\sigma$$_{Al}$} & \colhead{[Al/H]} & \colhead{[Al/Fe]} &  \colhead{$\sigma$$_{[Al/Fe]}$} 
}
\startdata
 M 67 &   84 & 6.79 & 0.30 & $+$0.32 & $+$0.32 & 0.36 \\
 M 67 &  105 & 6.91 & 0.33 & $+$0.44 & $+$0.36 & 0.39 \\
 M 67 &  108 & 6.88 & 0.40 & $+$0.41 & $+$0.39 & 0.46 \\
 M 67 &  135 & 6.74 & 0.31 & $+$0.27 & $+$0.29 & 0.37 \\
 M 67 &  141 & 6.86 & 0.35 & $+$0.39 & $+$0.31 & 0.39 \\
 M 67 &  143 & 6.65 & 0.24 & $+$0.18 & $+$0.29 & 0.31 \\
 M 67 &  151 & 6.82 & 0.29 & $+$0.35 & $+$0.37 & 0.35 \\
 M 67 &  170 & 6.82 & 0.36 & $+$0.35 & $+$0.32 & 0.41 \\
 M 67 &  173 & 6.71 & 0.10 & $+$0.24 & $+$0.27 & 0.23 \\
 M 67 &  217 & 7.04 & 0.18 & $+$0.57 & $+$0.63 & 0.28 \\
 M 67 &  218 & 6.88 & 0.46 & $+$0.41 & $+$0.48 & 0.51 \\
 M 67 &  223 & 6.89 & 0.31 & $+$0.42 & $+$0.41 & 0.36 \\
 M 67 &  224 & 6.95 & 0.45 & $+$0.48 & $+$0.43 & 0.51 \\
 M 67 &  244 & 6.77 & 0.21 & $+$0.30 & $+$0.39 & 0.28 \\
 M 67 &  266 & 6.81 & 0.38 & $+$0.34 & $+$0.30 & 0.42 \\
 M 67 &  286 & 6.84 & 0.28 & $+$0.37 & $+$0.25 & 0.40 \\
 M 67 & 2152 & 6.89 & 0.29 & $+$0.42 & $+$0.40 & 0.35 \\
 M 67 & 3035 & 6.78 & 0.34 & $+$0.31 & $+$0.37 & 0.38 \\
 M 67 & 4169 & 6.73 & 0.41 & $+$0.26 & $+$0.32 & 0.45 \\
N2158 & 3216 & 6.22 & 0.38 & $-$0.25 & $-$0.09 & 0.44 \\
N2158 & 7773 & 6.65 & 0.05 & $+$0.18 & $+$0.39 & 0.20 \\
N2158 & 7866 & 6.74 & 0.09 & $+$0.27 & $+$0.49 & 0.24 \\
\hline
Cluster & Star & log N(Cr) & [Cr/H] & [Cr/Fe] & $\sigma$$_{[Cr/Fe]}$ & \\
\hline
 M 67 &   84 & 5.63 & $-$0.04 & $-$0.04 & 0.28  & \\
 M 67 &  105 & 5.76 & $+$0.09 & $+$0.01 & 0.29  & \\
 M 67 &  108 & 4.89 & $-$0.78 & $-$0.80 & 0.30  & \\
 M 67 &  135 & 5.59 & $-$0.08 & $-$0.06 & 0.28  & \\
 M 67 &  141 & 5.77 & $+$0.10 & $+$0.02 & 0.27  & \\
 M 67 &  143 & 5.66 & $-$0.01 & $+$0.10 & 0.28  & \\
 M 67 &  151 & 5.65 & $-$0.02 & $+$0.00 & 0.28  & \\
 M 67 &  170 & 5.69 & $+$0.02 & $-$0.01 & 0.28  & \\
 M 67 &  173 & 5.65 & $-$0.02 & $+$0.01 & 0.29  & \\
 M 67 &  217 & 5.75 & $+$0.08 & $+$0.14 & 0.30  & \\
 M 67 &  218 & 5.98 & $+$0.31 & $+$0.38 & 0.29  & \\
 M 67 &  223 & 5.68 & $+$0.01 & $+$0.00 & 0.27  & \\
 M 67 &  224 & 5.93 & $+$0.26 & $+$0.21 & 0.31  & \\
 M 67 &  244 & 5.61 & $-$0.06 & $+$0.03 & 0.27  & \\
 M 67 &  266 & 5.72 & $+$0.05 & $+$0.01 & 0.27  & \\
 M 67 &  286 & 5.66 & $-$0.01 & $-$0.13 & 0.35 & \\
 M 67 & 2152 & 5.81 & $+$0.14 & $+$0.12 & 0.28  & \\
 M 67 & 3035 & 5.71 & $+$0.04 & $+$0.10 & 0.27  & \\
 M 67 & 4169 & 5.52 & $-$0.15 & $-$0.09 & 0.28  & \\
N2158 & 3216 & 5.06 & $-$0.61 & $-$0.45 & 0.37  & \\
N2158 & 7773 & 5.40 & $-$0.27 & $-$0.06 & 0.36  & \\
N2158 & 7866 & 5.42 & $-$0.25 & $-$0.03 & 0.37  & \\
\enddata
\end{deluxetable}
% Values checked 5 May 2011
 \begin{deluxetable*}{lccccccccc}
\tabletypesize{\footnotesize}
\tablewidth{0pt}
\tablenum{15}
\tablecaption{Cluster Average Al and Cr Abundance Ratios\label{cluster_wAlCr}}
\tablehead{\colhead{Cluster} & \colhead{\# Stars} & 
\colhead{[Al/H]} & \colhead{$\sigma$[Al/H]} & \colhead{[Al/Fe]} & \colhead{$\sigma$[Al/Fe]} &
\colhead{[Cr/H]} & \colhead{$\sigma$[Cr/H]} & \colhead{[Cr/Fe]} & \colhead{$\sigma$[Cr/Fe]}
}
\startdata
%Best Fit & $-$0.32$\pm$0.04 & $+$0.01$\pm$0.16 & \nodata\\
M67   & 19 & $+$0.33 & 0.06 & $+$0.36 & 0.08 & $+$0.00 & 0.05 & $+$0.00 & 0.07\\
N2158 &  3 & $+$0.20 & 0.04 & $+$0.38 & 0.14 & $-$0.38 & 0.12 & $-$0.18 & 0.21 \\

\enddata
\end{deluxetable*}

\subsection{Spectrum Synthesis of [O I] $\lambda$6300 \AA}
\begin{figure*}
\epsscale{0.85}
\plotone{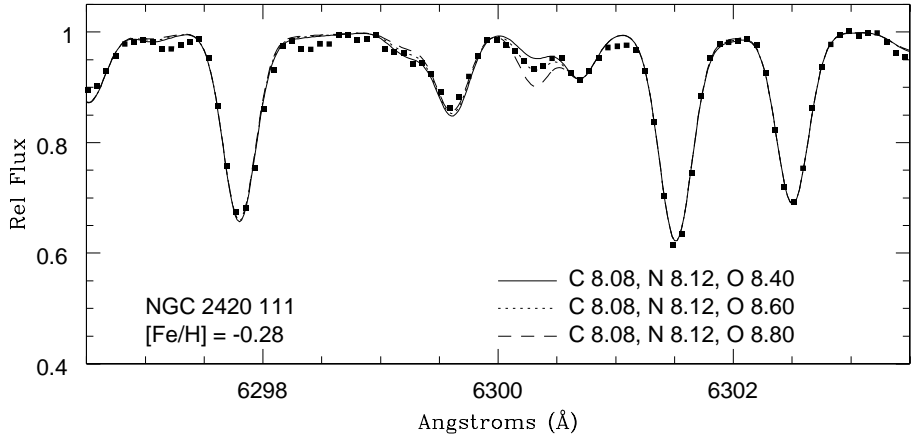}
\caption{Spectrum synthesis of the [O I] \lam6300 \AA\ feature in a program star.  Three synthetic spectra with different oxygen abundances are compared to the observed spectrum and the best match determined by eye.}
\label{Osynth}
\end{figure*}

As in all our previous work, oxygen abundances were determined via spectrum synthesis of the [O I] \lam6300.3 \AA\ feature. The spectra of stars in clusters NGC 188, NGC 1245, NGC 1817, NGC 2425 and NGC 7789 required correction for telluric contamination, and no oxygen abundances were determined for stars in NGC 2194.  Readers are referred to, e.g., Jacobson et al.\ (2011) or Friel et al.\ (2010) for details of our technique.  Briefly, spectrum synthesis was performed using MOOG, with a line list provided by C.\ Sneden (2003, private communication).  The abundances of Fe, Ti, and Ni features near the oxygen line were set to equal those found in the EW analysis.  The FWHM of the Gaussian used to match the resolution of the synthetic spectra to the observed one for each star was found by synthesizing a $\sim$10 \AA\ region around the [O I] line and minimizing the rms of the difference between the synthetic and observed spectra.  Three synthetic spectra were generated at a time, typically in 0.1--0.25 dex steps in oxygen abundance, and the best fit spectrum was determined by eye (see Figure~\ref{Osynth}).  The uncertainty of this fit was then found by decreasing the step size in abundance until no single best fit abundance could be identified.  This uncertainty was typically $\sim$0.07 dex.  Table~\ref{Oabund_mod} shows the log N(O), fitting uncertainty, [X/H]  and [O/Fe] value for each star analyzed, relative to log N$_{\odot}$(O) = 8.93 set as the default in MOOG (Anders \& Grevesse 1989).  Note that oxygen abundances were not determined for all stars in each cluster.  In general, stars were excluded when the telluric emission line near the oxygen feature had not been cleanly removed in the data reduction process and the quality of the fit was judged too poor to determine a reliable oxygen abundance.  Cluster average [O/H] and [O/Fe] ratios along with standard deviations and the number of stars per cluster are shown in Table~\ref{oxygentab}.

% Values checked 6 May 2011, updated 12 May 2011 to include log N(O).
\begin{deluxetable}{rrcccc}
\tabletypesize{\scriptsize}
\tablecolumns{5}
\tablenum{16}
\tablewidth{0pt}
\tablecaption{Oxygen abundances for Hydra Sample\label{Oabund_mod}}
\tablehead{ \colhead{Cluster} & \colhead{Star} & \colhead{log N(O)} & \colhead{unc.} & \colhead{[O/H]} &  \colhead{[O/Fe]}}
\startdata
M 67 &   84 & 8.88 & 0.06  & $-$0.05 &  $-$0.05 \\
M 67 &  105 & 8.93 & 0.07  & $+$0.00 &  $-$0.08 \\
M 67 &  108 & 8.78 & 0.08  & $-$0.15 &  $-$0.17 \\
M 67 &  135 & 8.83 & 0.08  & $-$0.10 &  $-$0.08 \\
M 67 &  141 & 8.83 & 0.07  & $-$0.10 &  $-$0.18 \\
M 67 &  143 & 8.58 & 0.10  & $-$0.35 &  $-$0.24 \\
M 67 &  151 & 8.83 & 0.07  & $-$0.10 &  $-$0.08 \\
M 67 &  170 & 8.83 & 0.07  & $-$0.10 &  $-$0.13 \\
\enddata
\tablecomments{This table is published in its entirety in electronic format.  A portion is shown here for
guidance regarding its form and content.}
\end{deluxetable}%16
% Values checked 6 May 2011
\begin{deluxetable}{lccc}
\tabletypesize{\footnotesize}
\tablewidth{0pt}
\tablenum{17}
\tablecaption{Cluster Average Oxygen Abundances and Uncertainties\label{oxygentab}}
\tablehead{\colhead{Cluster} & \colhead{\# Stars} & \colhead{[O/H]} & \colhead{[O/Fe]}
}
\startdata
%Best Fit & $-$0.32$\pm$0.04 & $+$0.01$\pm$0.16 & \nodata\\
M67   & 19 & $-$0.11$\pm$0.02 & $-$0.11$\pm$0.05 \\
%N188  &  8 & $-$0.03$\pm$0.02 & $-$0.06$\pm$0.09 \\
N1245 & 12 & $+$0.04$\pm$0.02 & $+$0.08$\pm$0.06 \\
N1817 & 27 & $-$0.18$\pm$0.01 & $-$0.02$\pm$0.03 \\
N2158 & 10 & $-$0.24$\pm$0.02 & $+$0.03$\pm$0.06 \\
%N2355 &  5 & $-$0.19$\pm$0.04 & $-$0.05$\pm$0.08 \\
N2420 &  9 & $-$0.21$\pm$0.02 & $-$0.01$\pm$0.06 \\
%N2425 &  3 & $-$0.12$\pm$0.04 & $+$0.03$\pm$0.11\\
N7789 & 15 & $+$0.08$\pm$0.02 & $+$0.07$\pm$0.05 \\
\hline
  & N1817-164  & N188-6712 & \\         
  Unc.          & $\Delta$[O/H] & $\Delta$[O/H] & \\                  
T$_{eff}$$+$200 K & $+$0.08 &  $-$0.02 & \\
 log {\it g}$+$0.2 dex & $+$0.08 & $+$0.00 & \\
 v$_{t}$$+$0.2 \kms & $+$0.00 & $-$0.05 & \\
 smooth$+$0.03 & $+$0.03 & $+$0.00 & \\
 $[\rm{N/H}]$$+$0.3 & $-$0.04 & $-$0.05 & \\
 $[\rm{C/H}]$$+$0.3 & $-$0.02 & $+$0.05 & \\
\enddata
\end{deluxetable}%17

Overall uncertainties in the oxygen abundances can be attributed to uncertainties in atmospheric parameters, uncertainties in the fit of the syntheses as described above, uncertainties in the smoothing of the synthetic spectra, and uncertainties due to lack of any abundance information for carbon and nitrogen.  As before, we have adopted [C/Fe] and [N/Fe] ratios for each star according to its [Fe/H] value, based on Figures 4 and 5 in Tautvai\v{s}ien\.{e} et al.\ (2010).  Typically, these were [C/Fe] = $-$0.20, [N/Fe] = $+$0.30.  Table~\ref{oxygentab} shows uncertainties in oxygen abundance for a hotter (NGC 1817 164) and cooler (NGC 188 6712) star in our sample.  As can be seen, the uncertainties are generally smaller than the fitting uncertainties for most stars.  It is well-known that the [O I] \lam\ 6300.3 \AA\ feature is blended with a nickel line.  This line was included in the synthesis line list, and fit with the Ni abundance found for each star from the equivalent width analysis, as already mentioned.  The adopted log {\it gf} 
value of this Ni feature could impact the resulting oxygen abundance, especially as the value used here ($-$3.0) is 
smaller than that found by Johansson et al.\ (2003; $-$2.11).  To test this, we performed a spectrum synthesis using 
the Johansson et al.\ log {\it gf} value and found the oxygen abundance was unchanged.  
Therefore, the presence of this feature should have negligible effect on the determined oxygen abundance.  

\section{Comparison to previous results}
We offer in Appendix B a brief comparison of our abundance results to those of previous studies for each individual cluster.  Here, we confine the discussion to the comparison of the results for four clusters which we have analyzed previously,
based on R$\sim$28\,000 KPNO 4m echelle spectra: M 67, NGC 188 (Friel et al.\ 2010), NGC 1817 and NGC 2158 (Jacobson et al.\ 2009).  It is worthwhile to briefly compare the results from the WIYN Hydra spectra to this earlier work, since, although the Hydra data are lower resolution, single-order spectra, the number of stars analyzed per cluster should make the Hydra results much more statistically robust.  What is more, we have made a careful comparison of results found from KPNO 4m and WIYN Hydra echelle spectra for stars in the cluster NGC 7142 (Jacobson et al.\ 2008), and found them to be consistent within the (larger) uncertainties of the Hydra spectra.

In general, the abundances found from the Hydra spectra are slightly lower than those found from the 4m spectra.  Abundances of elements for clusters M 67, NGC 188 and NGC 1817 generally agree within $\sim$0.12 dex, which is better than some individual star uncertainties.  The Hydra abundances of Mg and Al for M 67 are 0.2 dex larger than those determined from the KPNO 4m spectra.  That said, agreement is much poorer for NGC 2158: the [Fe/H] = $-$0.28 found here is much lower than that found in our previous work ($-$0.03), which was based on an analysis of one star only (Jacobson et al.\ 2009).  In addition to iron, abundances of Na, Ca, Ti and Zr also differ by 0.20 dex or more.

The differences arise from the choice of temperature scale, which is dependent upon the choice of E(B$-$V) towards the cluster.  One can see this in the abundances determined for the star for which we have both KPNO 4m and WIYN Hydra spectra, star 4230.  The T$_{\rm eff}$ value we determined from the KPNO 4m spectrum based on excitation equilibrium was 4400 K, which is consistent with a photometric temperature found with the adoption of E(B$-$V) = 0.55 and (m$-$M)$_{0}$ = 13.0, roughly the median values found in the literature for NGC 2158 (see Appendix A).  These values were used to determine atmospheric parameters in the preliminary analysis of the Hydra spectra mentioned in Jacobson et al.\ (2009).  Using this temperature scale, which is on average 200 K hotter than that adopted here (using E(B$-$V) = 0.43, (m$-$M)$_{0}$ = 13.0), NGC 2158's mean metallicity becomes [Fe/H] = $-$0.12$\pm$0.10 (s.d.), and all element abundances found for star 4230 agree with that determined from the KPNO 4m spectrum to within 0.07 dex. 

The difficulties in determining the fundamental parameters for this cluster have been well documented in the literature: generally speaking, it is either near solar metallicity and rather heavily reddened, or less reddened and more metal-poor.  As already discussed here and in Jacobson et al.\ (2009), we consider the distance and reddening determination of Grochalski \& Sarajedini (2002), based on the K magnitude of NGC 2158's rich red clump, as the most reliable, since near-infrared photometry is much less sensitive to extinction.  Therefore, we consider the abundance results of this analysis to be more likely correct.

Soubiran et al.\ (2000) obtained high resolution (R$\sim$42\,000) spectra of 24 stars in the field of NGC 2355 with the ELODIE spectrograph at the Haute-Provence Observatory.  All of these are publicly available in the ELODIE Archive\footnote{http://atlas.obs-hp.fr/elodie/intro.html} (Moultaka et al.\ 2004), but the S/N ratios are too low for EW measurements.  However, a S/N = 42 spectrum of star 398 was obtained in 2002 (ELODIE archive identifier: 20021022/0024) and is available in the archive.  After visual examination, the spectrum was deemed good enough for EW measurement, despite its relatively low S/N ratio. 

The ELODIE spectrum spans \lam4000-6800 \AA, so it was possible to measure the majority of the lines in our list, including lines for the elements Al and Cr, which are not measurable in the Hydra spectra.  Given the larger number of Fe I and Fe II lines available, the atmospheric parameters for 398 were determined spectroscopically by forcing excitation and ionization equilibrium, and the resulting parameters are in good agreement with that found photometrically.  The resulting Fe abundance is also in decent agreement, with log N(Fe) = 7.52$\pm$0.15 (cf. 7.78$\pm$0.28).  Other elements agree within 0.2 dex, save for Mg, which is $\sim$0.3 dex larger from the Hydra spectrum.  In general, star 398 has similar element abundances to the other NGC 2355 stars, confirming it is indeed a cluster member, despite its unique location in the color-magnitude diagram.  However, given the large star-to-star abundance dispersion, this should be confirmed by higher resolution spectroscopic follow-up of these and other cluster members.  
Based on analysis of this one star in NGC 2355, [Al/H] = [Al/Fe] = $-$0.12$\pm$0.12 (s.d.) and [Cr/H] = [Cr/Fe] = $-$0.18$\pm$0.04 (s.d.).  Clearly, these abundances must be confirmed with the analysis of more cluster members.

\section{Discussion}
\subsection{Abundance trends with R$_{gc}$}
With chemical abundances determined for the current sample, we combine these with results from our previous work 
(Friel et al.\ 2005, 2010, Jacobson et al.\ 2008, 2009, 2011).  Keeping in mind the four clusters common to our previous and current work, our total sample is comprised of 19 open clusters with 
\rgc\ values between 9 and 14 kpc, which appears to be the transition region between the inner and outer disk.  All abundances have been determined in a consistent fashion using the same line list and methodology.  Table~\ref{prev_ocs} gives a summary of the clusters from our previous work.

\begin{figure*}
\epsscale{0.85}
\plotone{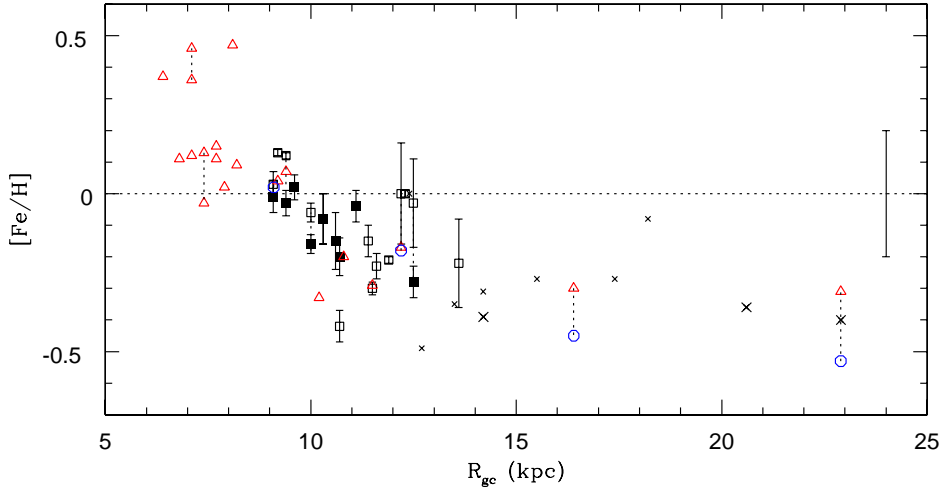}
\caption{Radial metallicity distribution of open clusters in the current sample (filled squares), our previous work (Friel et al.\ 2005, 2010; Jacobson et al.\ 2008, 2009, 2011; open squares). Circles represent the sample of Yong et al.\ (2005) and triangles represent clusters of Bragaglia et al.\ (2001, 2008), Carretta et al.\ (2004, 2005, 2007), Magrini et al.\ (2010), Mikolaitis et al.\ (2010), and Sestito et al.\ (2006, 2007, 2008).  Clusters studied by Carraro et al.\ (2004, 2007) and Villanova et al.\ (2005), placed on our abundance scale, are shown as crosses.  As described in Jacobson et al.\ (2010), the sizes of the crosses indicate the reliability of the measurement on our abundance scale (see that publication for details).  Error bars on the filled squares are weighted standard deviations; those on the open squares are standard errors of the mean.  A representative error bar for the abundance measurement of a {\it single star} in the current sample is shown to the right of the plot.  Clusters common to different studies are connected by dotted lines.}
\label{f:fe_rgc}
\end{figure*}

We note here that our previous determinations of distances to clusters in Friel et al.\ (2010) and Jacobson et al.\ (2009, 2011) via the red clump luminosity method did not take into account interstellar reddening in the K-band, A$_{\rm K}$, and so consequently are systematically too distant.  Revised values are given in Table~\ref{prev_ocs}.  Given the small amount of dust attenuation at near-infrared wavelengths, distances have changed only 0.1 kpc for all clusters save the more highly reddened cluster Be 17, for which the distance was decreased 0.3 kpc to \rgc=11.4 kpc.

In order to explore abundance trends across the full extent of the disk, we have combined our work with those of other groups from the literature, such as Yong et al.\ (2005), Bragaglia et al.\ (2001, 2008), Carretta et al.\ (2004, 2005, 2007) and Sestito et al.\ (2006, 2007, 2008).  These groups have studied clusters both in the inner and outer disk, and we have performed a detailed comparison of our results and methods to identify systematic differences (see Friel et al.\ 2010).  In this paper, we include the more recent work of Magrini et al.\ (2010) on three open clusters inside the solar circle and of Mikolaitis et al. (2010) on NGC 6134.  
As in Jacobson et al.\ (2011), we also include the outer disk clusters of  Carraro et al.\ (2004, 2007) and Villanova et al.\ (2005), which we have placed on our abundance scale.  (Of these latter clusters, indicated by crosses in all figures, we consider the abundances of Be 22, Be 29 and Be 66 to be more reliable on our abundance scale than the clusters of Carraro et al.\ (2007), and so these are indicated by larger symbols.  We also remind the reader that the oxygen abundances of Carraro et al.\ (2004) and Villanova et al.\ (2005) are {\it not} on our abundance scale, and consequently are indicated by small crosses as well.  See Jacobson et al.\ (2011) for details.) 
All combined, these samples comprise $\sim$42 open clusters spanning \rgc$\sim$6.5--23 kpc, with roughly half of the sample (our work) in the transition region between the inner and outer disk.  With the transition region now more populated and finely sampled, perhaps we can say something meaningful about the nature of the transition between the inner and outer disk.

Figure~\ref{f:fe_rgc} shows the cluster metallicity gradient from the cluster samples listed above.  In this and all
subsequent figures, the error bars shown are the standard errors of the mean abundance for each cluster.  Note that 
errors due to adoption of atmospheric parameters as discussed in Section 4.4 are not included here.  
%As T$_{\rm eff}$ and log {\it g} values are determined via photometry, the corresponding abundance errors
%are more likely than not to be systematic, affecting all stars in individual clusters and therefore 
Comparison of Figure~\ref{f:fe_rgc} to the radial metallicity distribution of a large heterogeneous cluster sample in Pancino et al.\ (2010; their Figure 9) shows them to be similar, though with our sample having smaller dispersion in [Fe/H] per unit \rgc.  This is likely due to the (hopefully) smaller systematic differences within our sample.  Our previous papers (open squares in Figure~\ref{f:fe_rgc}) showed the transition to be a smooth one, with no clear signs of a sharp discontinuity such as that seen at \rgc = 10 kpc in the sample of Twarog et al. (1997).  This is essentially unchanged by the inclusion of the new results presented here, yet we acknowledge that Figure~\ref{f:fe_rgc} leaves much room for interpretation.  Three features of Figure~\ref{f:fe_rgc} emphasize the ambiguity of the distribution.  First, there are the different [Fe/H] values for NGC 2158 found here ($-$0.28) and in Jacobson et al.\ (2009; $-$0.03).  At \rgc=12.5 kpc, NGC 2158 occupies an important place in the transition region, and the interpretation of the dispersion in [Fe/H] as a function of \rgc\ changes significantly if the results of the current analysis (of a larger stellar sample) for the cluster is more correct.  With an [Fe/H] $\sim$$-$0.3, NGC 2158 has a metallicity  more consistent with outer disk clusters, implying that the transition from the inner disk to the outer disk may be closer to \rgc$\sim$12 kpc instead of beyond $\sim$14 kpc as our earlier (smaller) samples suggested. 

Second, two other clusters in our sample with [Fe/H]$\sim$0 at \rgc$\sim$12 kpc also greatly influence the appearance of a large dispersion in metallicity in the transition region.  We call attention to the fact that the analysis of these two clusters (NGC 2141 and NGC 1883; Jacobson et al.\ 2009) rests on only 1--2 stars per cluster.  Given their crucial location in the transition region, these results should be confirmed.

\begin{figure*}
\epsscale{0.85}
\plotone{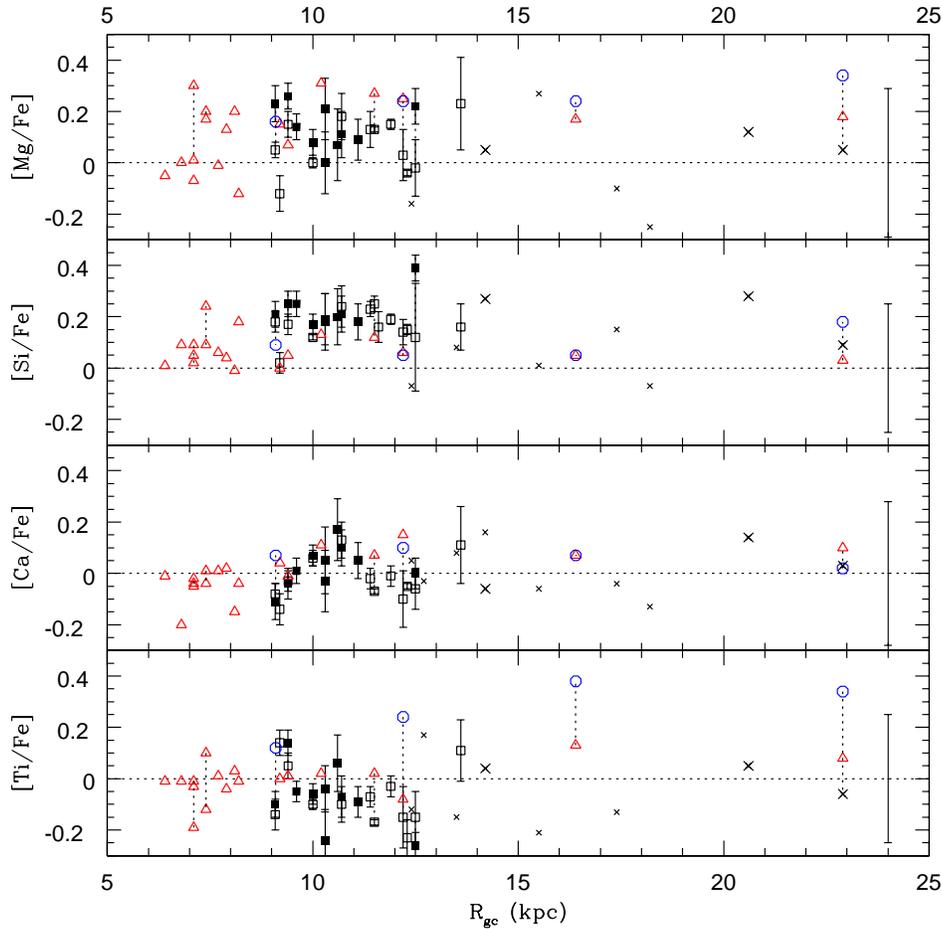}
\caption{The radial distribution of open cluster $\alpha$ element abundances.  Symbols same as in Figure~\ref{f:fe_rgc}. }
\label{f:a_rgc}
\end{figure*}

Third, the appearance of a slope in [Fe/H] interior to \rgc\ = 10 kpc is driven by the handful of supersolar metallicity clusters (NGC 6253, NGC 6791, NGC 6583).  One of these clusters, NGC 6791, has been considered an exceptional cluster for some time, considering its old age, extreme richness and interesting orbit (e.g., Carraro et al.\ 2006).  The recent work of Magrini et al.\ (2010) on the metallicity gradient in the inner disk showed NGC 6253 and NGC 6583 to be more metal-rich than other clusters and the vast majority of Cepheids in the inner disk. Calculations of these orbits showed that, at least in the case of NGC 6583, cluster perigalactic and apogalactic radii can differ by 3-4 kpc.  Depending on the choice of the galactic potential, the birth location of NGC 6253 can be as close to the Galactic center as \rgc$\sim$5 kpc.  Therefore, if these supersolar metallicity clusters are special cases, and not necessarily representative of inner disk clusters in general, then the metallicity distribution of clusters inside \rgc = 10 kpc is flat, with a mean [Fe/H] close to solar.  That said, one could argue that, supersolar metallicity clusters aside, there is a smooth decrease in [Fe/H] with \rgc\ between \rgc = 7 kpc and 13 kpc, albeit with increasing dispersion at larger \rgc.

\begin{figure*}
\epsscale{0.85}
\plotone{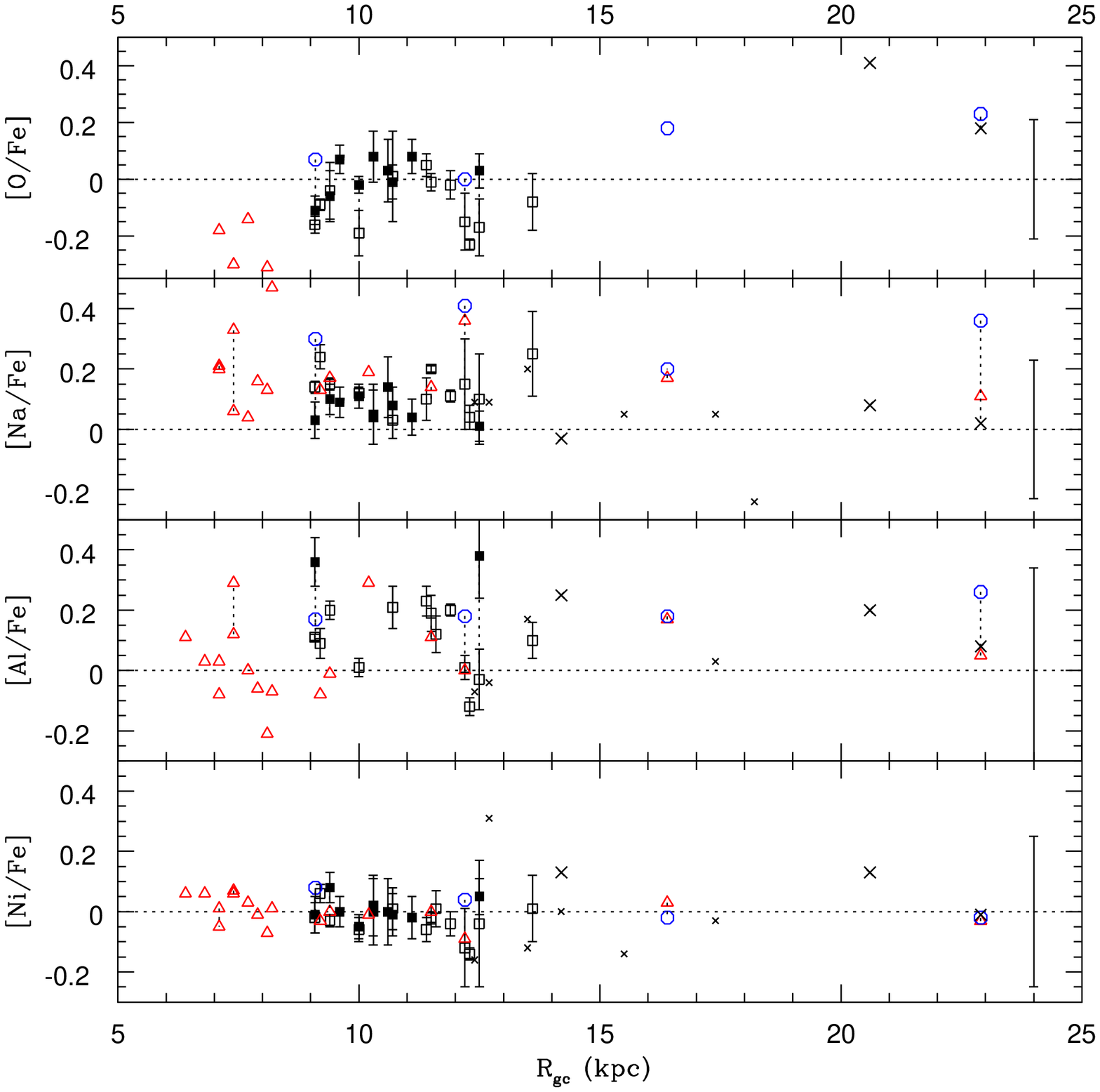}
\caption{The radial distribution of [O/Fe], [Na/Fe], [Al/Fe] and [Ni/Fe].  Symbols same as in Figure~\ref{f:fe_rgc}.  }
\label{f:el_rgc}
\end{figure*}

The radial dependence of element [X/Fe] ratios are shown in Figures~\ref{f:a_rgc} and \ref{f:el_rgc}.  
One is struck by the range in [X/Fe] found by different studies for the same cluster: variations can be as large as 0.3 dex, indicative of remaining uncertainties/systematics in the literature.  With the exception of [O/Fe], the range of element abundances appears roughly constant and independent of \rgc, with no evidence of a change from the inner to the outer disk.  The increase in [O/Fe] values from the inner disk through the transition region is a mirror image of the [Fe/H] trend seen in Figure~\ref{f:fe_rgc}.  This is due to a relationship between [O/Fe] and [Fe/H] that has been seen in numerous field star studies (see, e.g., Bensby et al.\ 2005), and which has been attributed to the increased contribution of Type Ia supernovae over time to the chemical enrichment of the disk.  However, the appearance of the trend is driven by the innermost (\rgc $\lesssim$8 kpc) and outermost (\rgc $\gtrsim$ 15 kpc) objects; the [O/Fe] distribution for \rgc$\sim$9--14 kpc clusters is quite flat, and within the errors, consistent with the solar ratio.  The enhanced [O/Fe] of the few studied-to-date outer disk clusters is intriguing, as it hints at a different chemical evolutionary history for the outer disk (e.g., Yong et al.\ 2005).  It is important to confirm the [O/Fe] values for clusters based on analyses of 1-2 stars and/or on the oxygen IR triplet, which is known to suffer from NLTE effects (e.g., Be 29, Saurer 1; Carraro et al.\ 2004).

\begin{figure*}
\epsscale{0.75}
\plotone{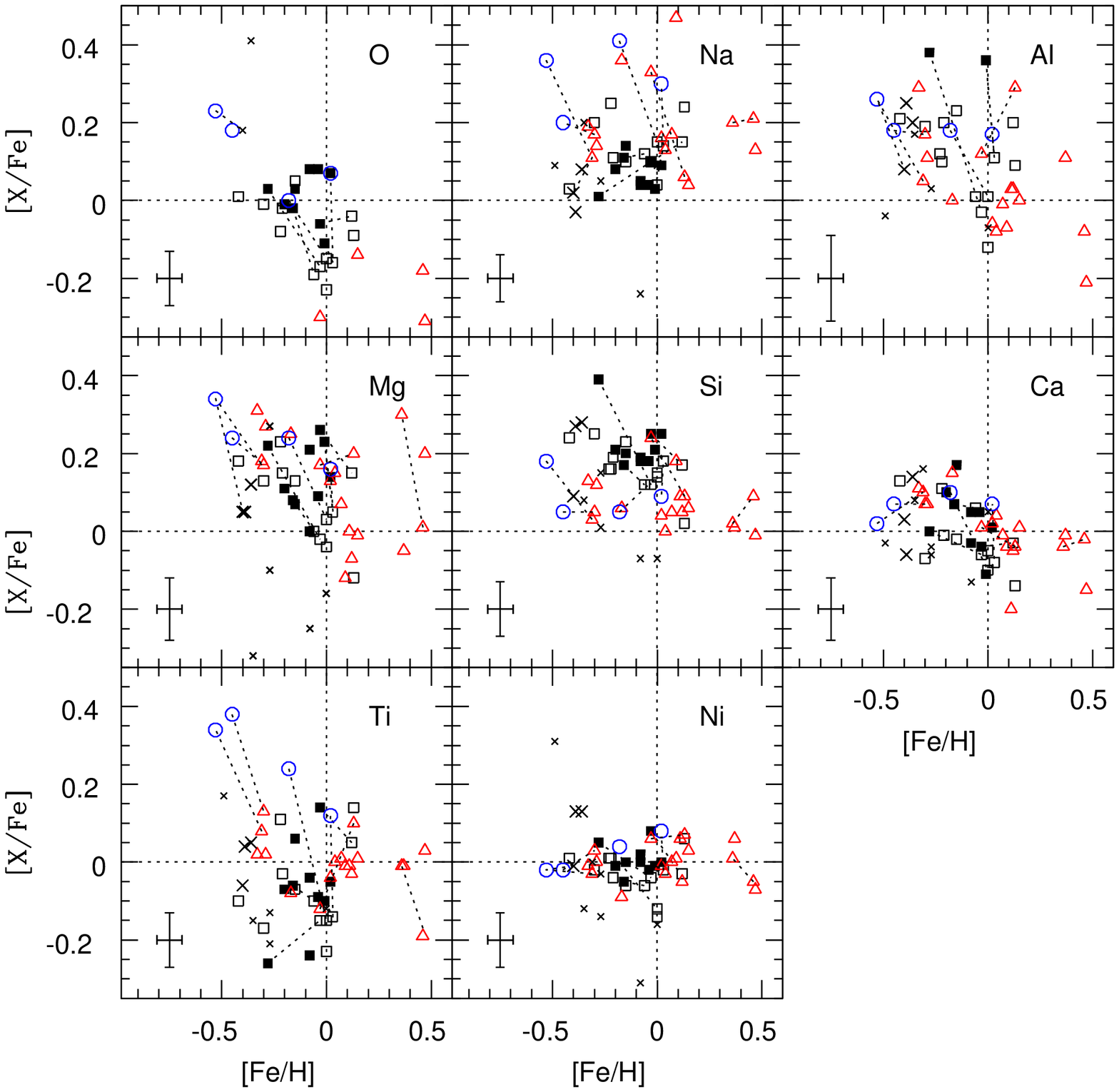}
\caption{Open cluster [X/Fe] as a function of [Fe/H].  Symbols same as in Figure~\ref{f:fe_rgc}.  Representative error bars for cluster mean abundance ratios (indicating typical standard errors of the mean) are shown in each panel.}
\label{f:xfe_fe}
\end{figure*}

\begin{figure*}
\epsscale{0.75}
\plotone{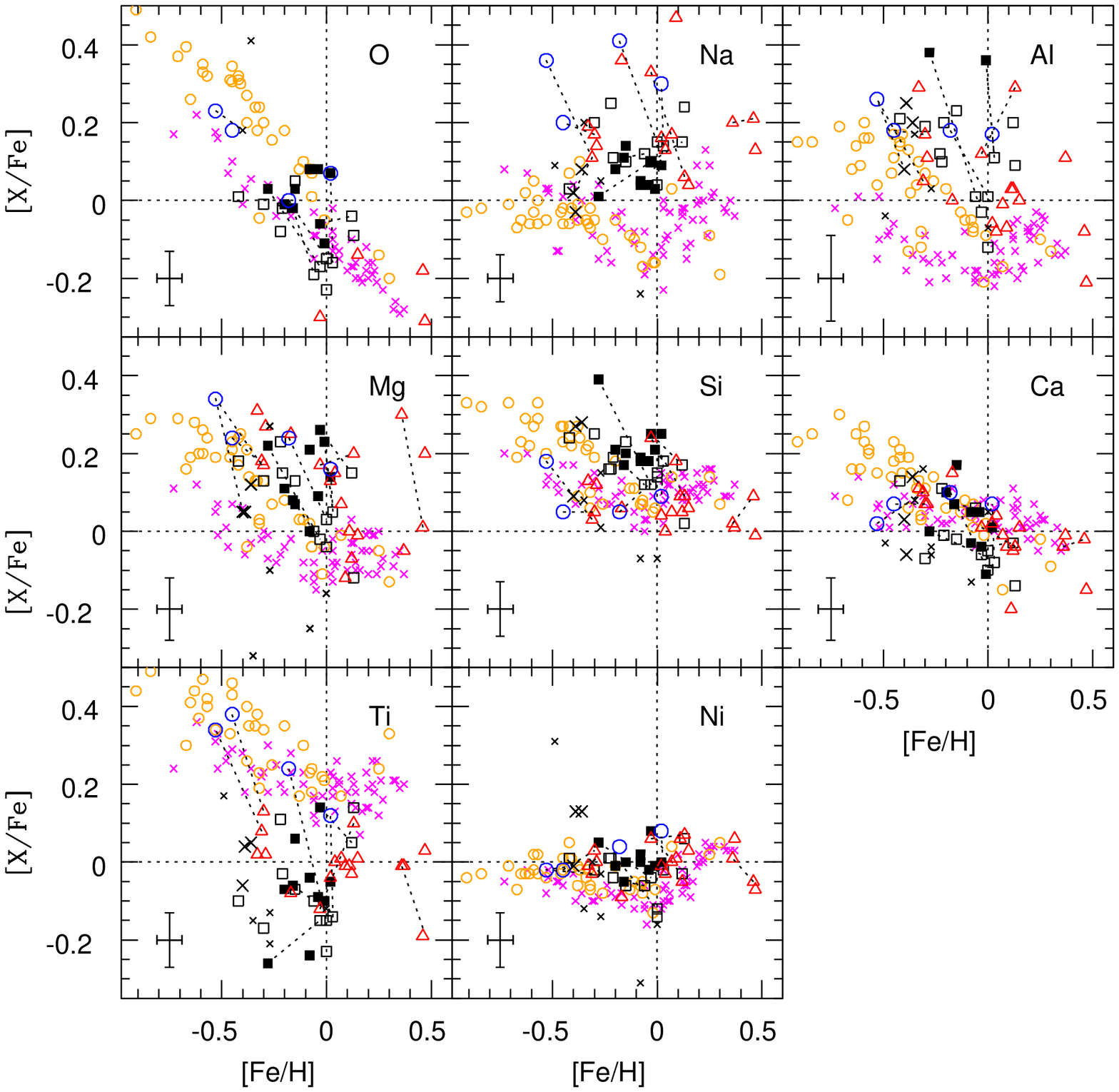}
\caption{Same as Figure~\ref{f:xfe_fe}, but with the addition of the thin (magenta crosses) and thick (orange open circles) disk field star samples from Bensby et al.\ (2003, 2005).} 
\label{f:bensxfe_fe}
\end{figure*}

The flat distributions of [X/Fe] with \rgc\ for the other $\alpha$ elements is consistent with that seen in the inhomogeneous cluster sample of Pancino et al.\ (2010).
They found that clusters beyond \rgc$\sim$15 kpc have slightly larger [$\alpha$/Fe] ratios than clusters inside this \rgc, but emphasized that the distribution is also consistent within the errors of having zero slope.  As seen in Figure~\ref{f:a_rgc}, the outer disk clusters have slightly enhanced [Mg/Fe] ratios compared to some inner disk clusters, but such is not the case for Si, Ca or Ti (accounting for the dispersion in individual cluster Ti abundances from different studies).

The magnitude of the abundance dispersions seen in Figures~\ref{f:a_rgc} and \ref{f:el_rgc} varies from element to element, with [Na/Fe] varying by $\sim$0.2 dex, and the $\alpha$ elements generally ranging 0.3--0.4 dex.  Many clusters are enhanced in elements Mg, Si, Na and Al, as has been known for some time (see, e.g., Friel 2006).  If cluster mean abundance ratios do vary with \rgc, the magnitude of such variation is smaller than the cluster abundance dispersions found in our study.  The small dispersion in [Ni/Fe] seen for clusters across the full \rgc\ range of the disk is reassuring: [Ni/H] generally follows [Fe/H] in the metallicity range spanned by open clusters (see discussion below), and so [Ni/Fe]$\sim$0 with small dispersion indicates robustness in the Fe analysis. 

\subsection{Abundance trends with [Fe/H]}
Trends of [X/Fe] with [Fe/H] are shown in Figure~\ref{f:xfe_fe}.  Of them, [O/Fe] exhibits a clear trend with metallicity, while a similar trend in [Ca/Fe] with [Fe/H] may also be present for our clusters especially (squares), though we note such trends are weakened when individual cluster abundance dispersions (shown by the error bars) are taken into account.       
The $\alpha$ elements Mg and Si show no noticeable trends in [X/Fe] with [Fe/H], both for our sample alone and when other groups' clusters are considered.  Clusters in general are predominantly enhanced in these elements at all metallicities, as already discussed.  Again, note the the very different [X/Fe] ratios found for individual clusters by different studies, connected by dashed lines here as in Figures~\ref{f:fe_rgc}-\ref{f:el_rgc}.  This contributes much to the scatter, especially for the element Ti. 

Figure~\ref{f:bensxfe_fe} again shows open cluster [X/Fe] versus [Fe/H], but with the addition of thin and thick disk field stars from Bensby et al.\ (2003, 2005).  As in our earlier work, we have shifted the abundance results of Bensby et al.\ on to our abundance scale in an attempt to minimize systematic differences (see Jacobson et al.\ 2011).  As can be seen, the clusters generally follow the field star trends with [Fe/H], especially for elements O, Mg, Si, Ca, and Ni.  
Even in cases where their abundances are systematically enhanced (Na, Si), the cluster locus is not far from the field star distributions.  This does not hold in the case of titanium, however, as can be seen in Figure~\ref{f:bensxfe_fe}.  The titanium abundances on our scale appear systematically low (squares and black crosses), though we note that some other clusters have been found to have [Ti/Fe] $\sim$ $-$0.1 to $-$0.2 dex, (e.g., NGC 2324 -- Bragaglia et al.\ 2008; NGC 2099 -- Pancino et al.\ 2010; Melotte 71 -- Brown et al.\ 1996).  (Note that the Bensby et al.\ sample on our scale is shifted up by $\sim$0.2 dex; compare to Figure 13 of Bensby et al.\ 2003.)

The difference between cluster and field star abundance distributions of the light elements Na and Al is in a sense systematic as well.  The field star sample of Bensby et al.\ is composed of solar-type dwarfs, while the cluster abundances here have been determined exclusively from giant stars.  The range of cluster [Na/Fe] ratios is in much better agreement with that of field red clump stars, which also show generally enhanced light element abundances with no obvious trend with metallicity (see, e.g., Mishenina et al.\ 2006, their Figure 14).  There has been debate in the literature whether or not differences in light element abundances between dwarfs and giants is real or merely a systematic error, namely due to the incorrect assumption of LTE.  Mishenina et al.\ (2006) determined NLTE Na abundances for their sample of red clump stars, and as already stated the resulting distribution is consistent with that of the cluster LTE Na abundances.  Some studies have found differences between dwarf and giant light element abundances for stars in the same cluster that cannot be the result of systematic effects alone (e.g., Schuler et al.\ 2009, Pace et al.\ 2010).  Clearly, it would be worthwhile to include both dwarfs and giant stars in cluster element abundance studies to search for further evidence of abundance differences.  At the very least, the different patterns of light element abundances shown by dwarfs and giants should be taken into consideration when comparing two different disk populations, as in Figure~\ref{f:bensxfe_fe}.

\subsection{Abundance trends with Age}
\begin{figure*}
\epsscale{0.85}
\plotone{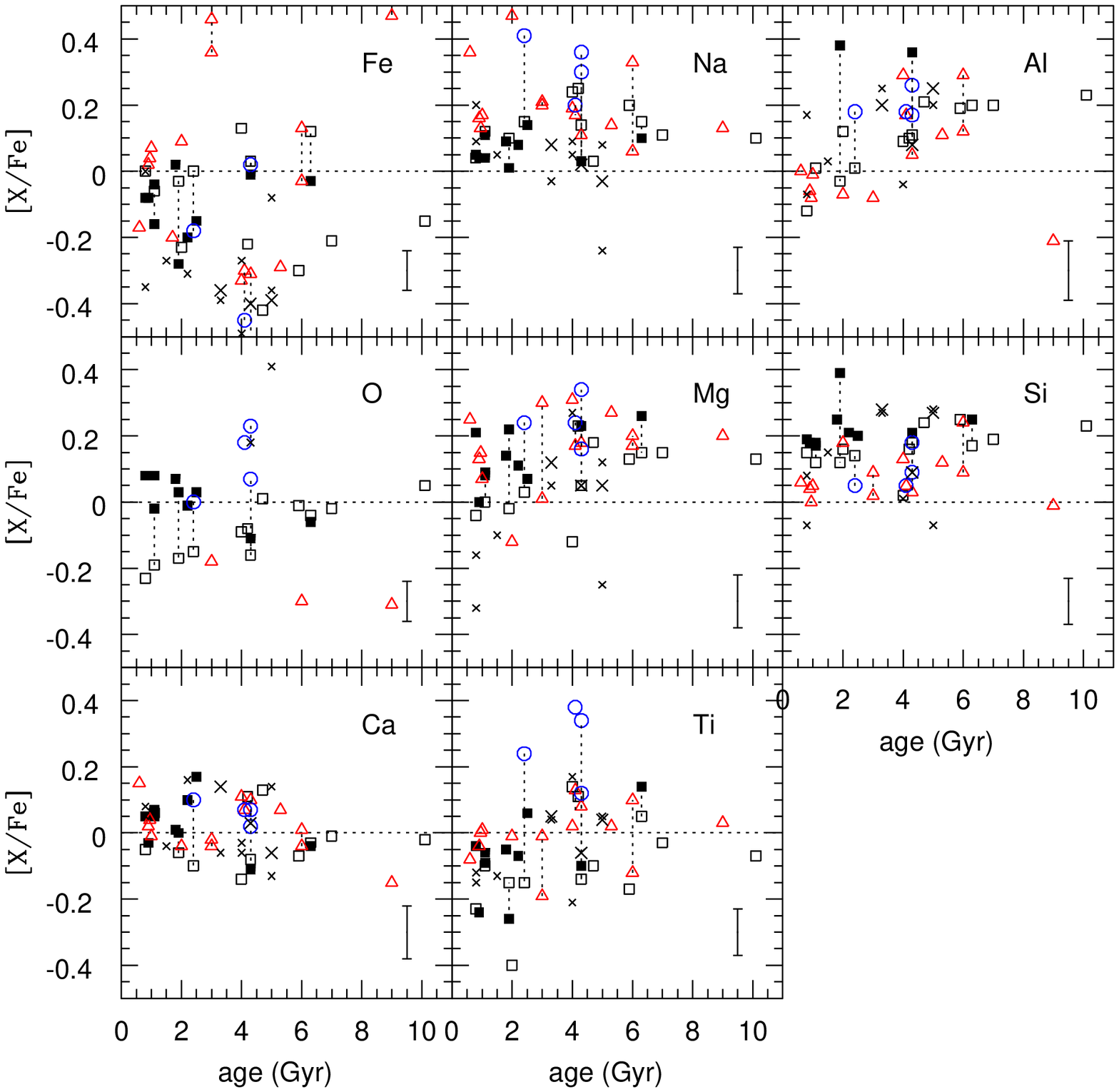}
\caption{Element abundance ratios as a function of cluster age.  Symbols same as in Figure~\ref{f:fe_rgc}.  The typical standard deviation of the weighted mean [X/Fe] for our cluster  sample (squares) is indicated in the bottom right corner of each panel.}
\label{f:age_el}
\end{figure*}

The clusters in our sample (including those of our previous work) span a wide range in age, from $\sim$700 Myr to 10 Gyr.  Figure~\ref{f:age_el} shows cluster trends of [Fe/H] and [X/Fe] with age.  As has been well remarked upon in the literature (e.g., Magrini et al.\ 2009, Pancino et al.\ 2010), clusters exhibit no statistically significant trend of metallicity with age, and instead show a large dispersion ($\sim$0.3-0.4 dex) at nearly all ages.  This would imply that the disk was relatively inhomogeneous when these clusters were formed, which is borne out by the lack of clear trends with age for other elements as well.  In Friel et al.\ (2010), we remarked on the trends of [O/Fe] and [Al/Fe] with age, shown by the twelve clusters in that study (open squares in Figure~\ref{f:age_el}).  The addition of the clusters in the current study has effectively erased the trend in the case of oxygen, while the enhanced [Al/Fe] shown in M 67 and NGC 2158 (note that the latter value has a large uncertainty) also weaken the trend with age for that element.  As with \rgc, the dispersion in cluster [X/Fe] as a function age varies from element to element, ranging from 0.2 dex for Na, to 0.3--0.4 dex for, e.g., O and Ti.  

\subsection{Chemical Evolution of the Milky Way Disk}
There are two general classes of Milky Way disk chemical evolution models: one predicts abundance gradients should flatten with time (e.g., Hou et al.\ 2000), while the other predicts a steepening with time (e.g., Chiappini et al.\ 2001).  These contrasting predictions are the result of differing treatments of gas infall and the star formation rate (see, e.g., Magrini et al.\ 2009 for a succinct summary).  Because open clusters span a wide range in age and location in the disk, their abundance patterns are crucial to constraining chemical evolution models.

The temporal variations in open cluster element abundance distributions have been investigated by many researchers, more recent examples being Magrini et al.\ (2009) and Andreuzzi et al.\ (2010; hereafter A10).  Both these studies used at least in part inhomogeneous cluster samples with chemical abundances collected from the literature, though it is worth remembering that much of the literature results are now comprised of increasingly large, internally-consistent cluster samples analyzed by individual groups.  Some such samples are incorporated in the analysis presented here, and as already stated, much work has been done to either place abundance results on a common abundance scale or else to identify sources of systematic error between studies.  Therefore, it is hoped that the picture presented here is less effected by systematic uncertainties than larger, heterogeneous literature samples.

\begin{figure*}
\epsscale{0.85}
\plotone{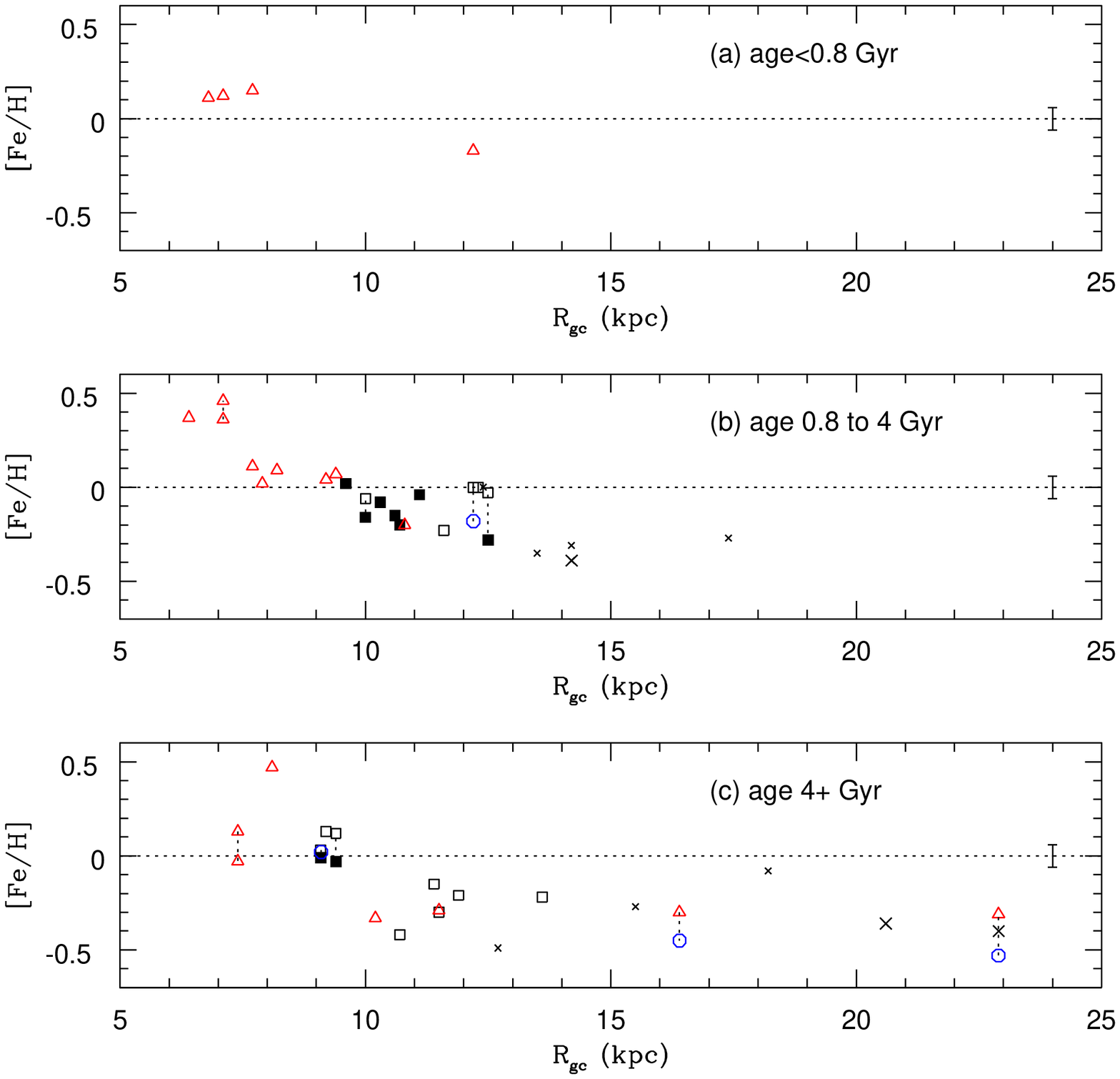}
\caption{Cluster radial metallicity distribution as a function of cluster age.  Symbols same as in Figure~\ref{f:fe_rgc}.  Top panel: clusters younger than 800 Myr; middle panel: clusters 800 Myr to 4 Gyr old; bottom panel: clusters 4 Gyr and older. The typical standard deviation of the weighted mean [Fe/H] for our cluster  sample (squares) is indicated to the right of each panel.}
\label{f:fe_evol}
\end{figure*}

\begin{figure*}
\epsscale{0.85}
\plotone{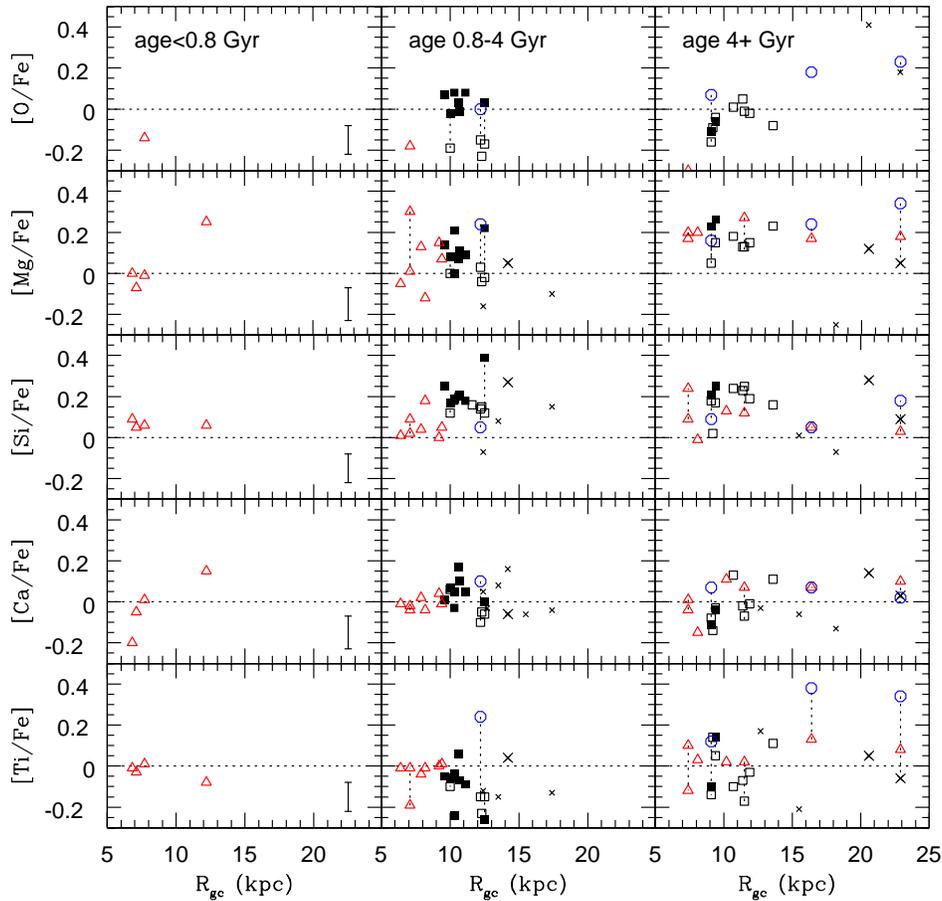}
\caption{The radial distribution of $\alpha$ element [X/Fe] ratios as a function of cluster age.  Symbols same as in Figure~\ref{f:fe_rgc}.  Left column: clusters younger than 800 Myr; middle column: clusters 800 Myr to 4 Gyr old; right column: clusters 4 Gyr and older.  Representative standard deviations of the weighted mean [X/Fe] for our cluster sample are indicated in the lower right corner of the left-most panels.}
\label{f:alpha_evol}
\end{figure*}

\begin{figure*}
\epsscale{0.85}
\plotone{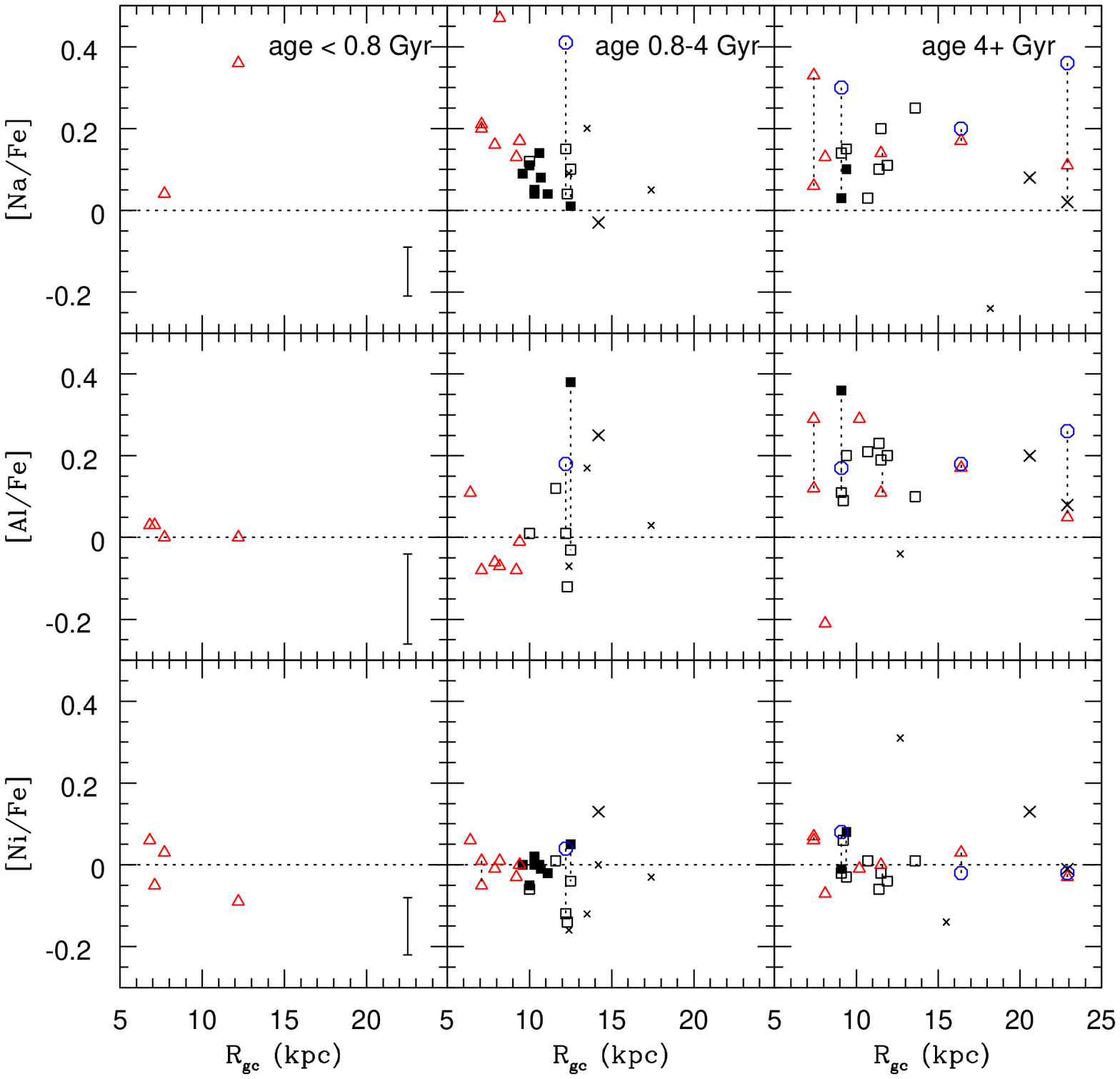}
\caption{Same as Figure~\ref{f:alpha_evol}, for [Na/Fe], [Al/Fe] and [Ni/Fe].}
\label{f:other_evol}
\end{figure*}

The combined cluster samples considered here span nearly the full age range of the disk, so in Figures~\ref{f:fe_evol}--\ref{f:other_evol} we have plotted the radial abundance distributions for three different age groups.  For ease of comparison to, e.g., Magrini et al.\ (2009) and A10, we have divided the cluster sample into those with ages 800 Myr and younger, 800 Myr to 4 Gyr, and 4 Gyr and older.  At most only four clusters are in the youngest age bin, so we can make no statements about the metallicity distributions of the youngest open clusters.  For the older two age bins, one can see that 
Figure~\ref{f:fe_evol} is very similar to that found in A10, although there is a difference in adopted distance scales.  

Metallicities of clusters 0.8--4 Gyr in age follow a negative linear gradient with relatively small dispersion out to \rgc$\sim$14 kpc, where the distribution basically ends.  A10 calculated a slope of $\Delta$[Fe/H]/$\Delta$\rgc\ = $-$0.07 for clusters interior to \rgc\ = 12 kpc (their Figure 19; middle panel).  For the same \rgc\ range, we have found a slightly steeper value for middle-aged clusters in Figure~\ref{f:fe_evol}: $-$0.083$\pm$0.019 to $-$0.086$\pm$0.015, depending on which [Fe/H] measures for NGC 2141 and NGC 6253 are adopted.  In contrast, the inner disk abundance distribution of the oldest clusters is more difficult to interpret given the smaller number of clusters inside 10 kpc, but what clusters there are indicate a steeper slope that flattens out at a smaller galactocentric distance: $\sim$10 kpc.  The slope, $\Delta$[Fe/H]/$\Delta$\rgc\ for the oldest clusters inside \rgc\ = 12 kpc ranges from $-$0.115$\pm$0.045 to $-$0.134$\pm$0.035, depending on which [Fe/H] value is adopted for Cr 261.  These values are slightly shallower than that found by A10: $-$0.15.  While we and A10 have found evidence for a flattening of the gradient with time, similar to that of, e.g., Magrini et al.\ (2009), it must be emphasized that the different \rgc\ ranges spanned by the intermediate- and old-aged clusters may well influence this interpretation.

Another interesting feature for the two older age bins in Figure~\ref{f:fe_evol} is the location of the transition from a gradient to the flat [Fe/H] distribution of the outer disk: for the intermediate-aged clusters, this transition appears to occur near or beyond \rgc$\sim$14 kpc, where the cluster sample basically ends.  In contrast, the older clusters exhibit a transition at a smaller galactocentric distance: $\sim$10 kpc.  Could this be an indication of a change in the chemical enrichment history of the disk?  The uniformity of the older population's abundance patterns may reflect the domination of  well-mixed infalling material on the enrichment history of the early disk, while the inner disk's gradient could be the result of the outward expansion of the younger disk with its enhanced metallicity. 

Figure~\ref{f:alpha_evol} shows the temporal evolution of [$\alpha$/Fe] ratios.  Here again one is struck by the flat distributions of [X/Fe] with \rgc\ for the oldest clusters.  Once more, [O/Fe] stands out with a more intriguing pattern, but this is a product of the [O/Fe] vs.\ [Fe/H] distribution (Figure 7), and the fact that oxygen abundances for many clusters are still missing from the diagram, which makes the picture more incomplete than for the other $\alpha$ elements.  The element radial distributions for the middle-aged clusters appear more varied than for the older clusters.  The reduced dispersion in [Mg/Fe] for the oldest clusters compared to that of the intermediate-aged clusters is noticeable.  This can also be seen in Figure 9, which shows not only the decrease in dispersion of [Mg/Fe] with age, but the tendency for older clusters to typically have enhanced [Mg/Fe] ratios.  How much of this is attributable to selection biases is unclear.  The distributions for the light elements shown in Figure~\ref{f:other_evol} are essentially flat with large scatter, while [Ni/Fe] shows little, if any, variation with \rgc.

The open cluster chemical abundance patterns highlighted above and in the previous sections offer important constraints to theoretical models of the formation and evolution of the Milky Way disk.  It must be kept in mind, however, that the clusters presented here comprise only a small fraction of the total cluster population and that sample biases may influence the perceived abundance distributions in unknown ways.  With that caveat, our current understanding of the open cluster chemical abundance distributions requires a disk chemical evolution model to reproduce:
\begin{itemize}
\item A trend of decreasing [Fe/H] that then flattens out beyond \rgc$\sim$12 kpc.
\item Flat distributions with roughly constant dispersion across the full extent of the disk for other element [X/Fe] ratios.
\item At most only a slight flattening of the [Fe/H] gradient inside \rgc\ $\sim$12 kpc over time, and perhaps a change in the location of the transition from the inner disk gradient to outer disk constant [Fe/H] to larger Galactocentric radii over time. 
\item Little to no evolution of [X/Fe] radial distributions with time, (bearing in mind the different \rgc\ ranges of the two older age bins in Figures~\ref{f:fe_evol}-\ref{f:other_evol}).
\item No obvious trends of [Fe/H] or [X/Fe] with cluster age (though this may change once outer disk cluster [O/Fe] ratios are better constrained).
\item Different dispersions in [Fe/H], [X/Fe] versus age, \rgc\ and [Fe/H].
\end{itemize}

Naturally, many open questions remain.  For example, are the lack of obvious trends for most [X/Fe] and the different abundance dispersions indicative of azimuthal variations in disk abundances, as found in the Cepheid study by Luck et al.\ (2006)?  Possible trends of abundance with \rgc\ may well be smoothed out for a cluster population that spans a wide range of Galactic longitude or different Galactic quadrants.  Unfortunately, the number of clusters for which we have detailed abundance patterns is insufficient to properly investigate this question, and indeed it is possible that the spatial distribution in the disk of the currently-known population of clusters is too scant to answer this question with clusters alone.

Another important question is the effect of radial migration of stars in the disk on the interpretation of radial chemical abundance patterns (see, e.g., Sellwood \& Binney 2002).  It has been implicitly assumed that open clusters are unaffected by radial migration forces, either because most clusters are dissolved before migration can occur, or else some clusters (especially many of the older clusters still around today) have orbits that keep them out of the densest parts of disk for most of their lives.  Yet, recent simulations of radial migration seem to indicate that clusters can change their orbits via interactions with transient spiral density waves, especially those on near-circular orbits confined to the disk (Ro\v{s}kar et al.\ 2008).  A general consequence of radial migration would be a flattening of any existing abundance gradient in the disk, as stars from the metal-poor outer disk migrate inwards and vice versa.  In their recent study of open cluster orbits, Wu et al.\ (2009) plotted cluster [Fe/H] versus apogalactic distance and found it to be very similar to the metallicity distribution found using clusters' current \rgc\ values.  A10 have stated that this implies that clusters are little affected by migration and therefore the observed abundance distributions are indicative of the state of the disk at the time they formed.  However, our understanding is that current radial migration models show that interaction with transient spiral density waves can move objects 1,2,3 kpc from their birth radii while still maintaining their circular orbits, so it is unclear why one would expect that any difference between abundance distributions of clusters at their current \rgc\ values and their apogalactic positions would be an indication of the presence of radial migration.

However, as already stated, if radial migration is an important factor in open cluster kinematics, any metallicity gradient would be slowly washed out over time, with the oldest clusters exhibiting a shallower gradient than younger clusters.  The fact that the older clusters appear to exhibit a slightly steeper gradient than the younger clusters may very well indicate that migration is not important.
Therefore, assuming for now that open clusters are not affected by radial migration as field stars are, a natural question to ask is whether the more metal-poor outer disk clusters have similar abundance patterns to metal poor solar neighborhood field stars?  If they originated from the same part of the disk, then it seems reasonable that they should.  The picture indicated in Figure~\ref{f:bensxfe_fe} is difficult to interpret in this context, since it is uncertain what systematic affects may still be present between the open cluster giant star abundances and the field dwarf star abundances.  A differential abundance analysis of nearby field stars and outer disk open clusters, based on data taken with the same telescope/instrument combination, would be the ideal way to explore this question.

As pointed out by A10, the general radial element abundance distributions -- namely, the flattening out in the outer disk -- can be explained as both a natural consequence of disk formation and evolution (e.g., Carraro et al.\ 2007) or as a consequence of satellite mergers (e.g., Yong et al.\ 2005, Carraro \& Bensby 2009).  Can any of the other abundance patterns discussed here distinguish between these two scenarios?  The homogeneity of outer disk abundances is an important constraint: if it was built by mergers, then the chemical composition of the merging material would need to be rather finely tuned to create such homogeneity, as discussed by Magrini et al.\ (2009).  Indeed, the flat distribution of [X/Fe] across the full extent of the disk, especially for the older clusters (which the outer disk population is a part of) as seen in Figures~\ref{f:alpha_evol} and \ref{f:other_evol}, seem to imply a similar chemical evolutionary history across the extent of the disk.
It goes without saying that a larger cluster sample in the outer disk would be extremely helpful, not only to better constrain the abundance distributions but also the age range of objects in the outer disk.

\section{Summary}
We have presented a detailed chemical abundance study of ten open clusters based on single-order echelle spectra of giant stars obtained with the Hydra multi-object spectrograph on the WIYN 3.5m telescope.  To our knowledge, this is the first detailed abundance study for four clusters in our sample: NGC 1245, NGC 2194, NGC 2355 and NGC 2425.  Using an analysis involving both equivalent width measurement and spectrum synthesis, we have determined abundances for the elements Fe, O, Na, Mg, Si, Ca, Ti, Ni and Zr, and for two clusters in the sample, Al and Cr.  Of these, the Mg and Cr abundances have relatively large uncertainty, as they are based on the measurement of one line only.  We previously carried out an abundance analysis for four clusters in this sample, M 67, NGC 188, NGC 1817 and NGC 2158, based on multi-order KPNO 4m echelle spectra of one to four stars per cluster.  Comparison of the results from the current study to these found general good agreement in atmospheric parameters and chemical abundances for three of the clusters, though with the cluster mean [Fe/H] determined in this study being slightly lower than found from the KPNO 4m data.  As agreement between the two analyses is generally good for individual stars, this difference in metallicity is likely due to the increased number of stars analyzed in the current sample leading to a more robust result.  The mean metallicity for NGC 2158, [Fe/H] = $-$0.28$\pm$0.10 (s.d.) is significantly lower than found in an earlier study ($-$0.03; Jacobson et al.\ 2009), which we attribute to differences in temperature scales. 

This cluster sample was selected to explore the element abundance trends with Galactocentric distance within the range \rgc$\sim$9-13 kpc, in order to better characterize the transition between the inner and outer Milky Way disk.  This sample supplements the twelve clusters from our previous work that occupy this same range in Galactocentric radius, and we have further combined our sample with those of other groups to investigate cluster abundance trends over the full radial extent of the disk.  We have found that distribution of cluster [Fe/H] appears to be smooth through the transition between the inner and outer disk, with a gradual decrease in [Fe/H] with increasing \rgc, at least through \rgc$\sim$13 kpc, although this trend is rather dependent on the abundances of a few key clusters at \rgc$\sim$12 kpc, such as NGC 2158.  The distribution of cluster mean [X/Fe] ratios  for most other elements also appears continuous and flat for the full extent of the Galactic disk, with 0.2--0.4 dex dispersion at least in the inner disk and transition region which are well sampled.  A flat distribution in [X/Fe] across the disk implies that the chemical evolution of the outer disk is similar to that of the inner disk.  For example, the comparable [$\alpha$/Fe] ratios of inner disk and outer disk clusters indicates that the relative contributions of Type Ia and Type II supernovae were similar in the two regions.

The element abundance patterns of the  clusters studied here do not appear to exhibit any trends with cluster age, as has been seen by other studies.  In general, our results are qualitatively similar to those found from larger, heterogeneous cluster samples (Pancino et al.\ 2010), although much more work must be done to understand and minimize systematic differences among different studies.  Until such work is done, any real changes in cluster abundance patterns present in the disk on the level of 0.3 to 0.2 dex or smaller cannot be considered reliable.  In a similar vein, it is as yet still difficult to make a detailed comparison of cluster abundance ratios to those of solar neighborhood field stars.  For now, all one can say with confidence is that cluster [X/Fe] ratios have the same general trends with [Fe/H] as do thin and thick disk stars.

Plots of cluster element abundance ratios versus \rgc\ and [Fe/H] for two different age ranges of clusters (0.8--4 Gyr and 4$+$ Gyr) show some possible differences in slope.  For example, the older clusters show remarkably flat distributions of [X/Fe] with \rgc, while the younger clusters may show a hint of increasing [X/Fe] with \rgc\ for some elements (though the narrower \rgc\ range of this sample should be kept in mind).

As such, the exact nature of the transition between the inner and outer disk remains ambiguous, and the question of whether or not the outer disk was formed as a natural part of disk evolution or from merger events remains unanswered.  Remaining systematic errors between abundance studies of different research groups must be minimized, and the abundance ratios of some key clusters better constrained by more observations, before we can come closer to answering these questions.  The inner disk must not be neglected in this research, either.   We note the recent work of Magrini et al.\ (2010) on three clusters with \rgc$\sim$6--7 kpc that found indications of a steepening of the [Fe/H] distribution towards the Galactic center, as has been found for Cepheids at similar Galactocentric location (e.g., Andrievsky et al.\ 2002b).

That said, we as a research field have come far in a few short years in improving our understanding of the chemical abundance distribution of the Galactic disk.  Continued efforts, both on a cluster by cluster basis as individual groups continue to increase their sample sizes, and on survey-sized scales with the likes of APOGEE\footnote{See http://www.sdss3.org/surveys/apogee.php}  and HERMES\footnote{See http://www.aao.gov.au/HERMES/} and Gaia\footnote{See http://sci.esa.int/science-e/www/area/index.cfm?fareaid=26} promise even further advances in our understanding.

\acknowledgements 
We are grateful to the referee for suggestions that improved the presentation of this research.  
H.R.J. thanks the many telescope operators at WIYN over the years who have assisted in obtaining these observations, as well as T.\ Monroe who obtained many of them on our behalf.  D.\ Harmer and D.\ Willmarth are also gratefully acknowledged for their expertise on all things Hydra-related.  This research has been supported by a National Science Foundation Graduate Research Fellowship and an Indiana Space Grant Graduate Research Fellowship.  H.R.J. also acknowledges support from the National Science Foundation through the NSF Astronomy and Astrophysics Postdoctoral Fellowship under award AST-0901919 during the preparation of this manuscript.  C.A.P.  acknowledges the support of the Daniel Kirkwood Research Fund at Indiana University.  This work has used data products from the Two Micron All Sky Survey, which is a joint project of
the University of Massachusetts and the Infrared Processing and Analysis Center/California Institute 
of Technology, funded by the National Aeronautics and Space Administration and the National Science Foundation, and of the USNOFS Image and Catalogue Archive operated by the United States Naval Observatory, Flagstaff Station.  This study also made extensive use of the WEBDA database, operated at the Institute for Astronomy at the University of Vienna,  the ELODIE archive at Observatoire de Haute-Provence (OHP) and the SIMBAD database,
operated at CDS, Strasbourg, France.  Lastly, the existence of NASA's Astrophysics Data System Bibliographic Services is gratefully acknowledged.

%\clearpage

\appendix
\section{Cluster Parameters}
In this appendix we briefly review each cluster in the current analysis and include a summary of previous studies and their results.  Where relevant, the distance determination based on cluster red clump star luminosity is also described.  
A discussion of metallicity and element abundance determinations for each cluster is found in Appendix B.

\begin{itemize}
\item{\it M 67:}
M 67 ({\it l} = 215.6$^{o}$, {\it b} = $+$31.7$^{o}$) is one of the most well-studied clusters in the Milky Way disk.  Cluster
membership has been reliably determined using both radial velocity (Mathieu et al.\ 1986) and proper motion (Sanders 1977) studies.  The canonical photometry study of M 67 is that of Montgomery, Marschall \& Janes (1993).  They found E(B$-$V)=0.05$\pm$0.01 and (m$-$M)$_{V}$ = 9.6, though they remarked on the poor match between the data and theoretical isochrones.  Twarog et al.\ (1997) used their photometry and found (m$-$M)$_{V}$ = 9.8$\pm$0.2, assuming E(B$-$V)=0.04.  M 67 was one of the clusters used for calibration in GS02 and CLD05, so the distance determination using red clump stars indicates the robustness of the method.  Confirmed cluster member red clump giant stars in M 67 have been identified in both photometric and spectroscopic studies of the cluster.  Using 2MASS photometry of seven red clump stars, the average clump K-band magnitude and J$-$K color are 7.959$\pm$0.023, 0.638$\pm$0.023.  Using the CLD05 relation, E(J$-$K) = 0.022$\pm$0.029, E(B$-$V) = 0.041$\pm$0.029, and A$_{\rm K}$ = 0.014.  This results in a distance modulus (K$-$M$_{\rm K}$)$_{0}$ = 9.605$\pm$0.064, which corresponds to a distance d = 834$\pm$25 pc, and \rgc\ = 9.09$\pm$0.02 kpc, z = 438$\pm$13 pc.  These values are in excellent agreement with those of, e.g., Twarog et al. (1997; d = 860$\pm$50 pc).
On the scale of Salaris et al.\ (2004), M 67 has an age of 4.3$\pm$0.5 Gyr.
Stellar ID's come from WEBDA (based on the identifications of Fagerholm (1906)), while photometry and alternate identifications come from Montgomery, Marschall \& Janes (1993).   
Coordinates given in Table 3 are those of Fan et al.\ (1996).
We have used E(B$-$V) = 0.04$\pm$0.02 and (m$-$M)$_{0}$ = 9.68$\pm$0.20 (Twarog et al.\ 1997) for determination of stellar atmospheric parameters.

\item{\it NGC 188:}
Along with M 67, NGC 188 ({\it l} = 122.8$^{o}$, {\it b} = $+$22.5$^{o}$) is an extremely well-studied open cluster.  It is also one of the older open clusters, with an age of 6.3$\pm$0.8 Gyr (Salaris et al.\ 2004).  
We refer the reader to Fornal et al.\ (2007) for a summary of previous studies in the literature.  As in our previous study of NGC 188 stars (Friel et al.\ 2010), we chose not to determine the cluster's fundamental parameters based on red clump morphology, given the difficulty in identifying red clump stars.
Instead, we have adopted the distance and reddening found by Sarajedini et al.\ (1999): E(B$-$V) = 0.09$\pm$0.02, d = 1.70$\pm$0.07 kpc, \rgc = 9.40$\pm$0.09 kpc, z = $+$650$\pm$30 pc.  These are in excellent agreement with, e.g., the findings of Twarog et al.\ (1997).  Incidentally, Van Helshoecht \& Groenewegen (2007) included NGC 188 in their investigation of open cluster K-band red clump luminosity.  The stars they identified as red clump stars have M$_{\rm K}$ = $-$1.40 in the 2MASS magnitude system, which, working backwards with their choice of cluster reddening and distance modulus, indicates a mean clump K = 9.67.  This value agrees with the K magnitudes of stars tentatively identified as members of the red clump by Sarajedini et al.\ (1999), as described in Friel et al.\ (2010).  The resulting distance modulus, using M$_{\rm K,0}$ = $-$1.66, is 11.33 magnitudes, or $\sim$0.2 magnitudes larger than found by other photometric studies.  
In Table 3, identifications are from WEBDA, while alternate ID's, J2000 coordinates and photometry are from Platais et al.\ (2003).

\item{\it NGC 1245:}
The first photometric studies of NGC 1245 ({\it l} = 146.6$^{o}$, {\it b} = $-$8.9$^{o}$) were carried out in the 1960's (Hoag et al.\ 1961, Chincarini 1964).  It is a very rich cluster, 1.1$\pm$0.2 Gyr old (Salaris et al.\ 2004) and with a populous red giant clump.  It has been the subject of several CCD photometric studies in the past two decades, but at least one appears to be troubled with incorrect color calibrations (Subramaniam 2003).  Carraro \& Patat (1994) found evidence for differential extinction across the field of the cluster, based on star counts in different regions of the field; Wee \& Lee (1996) also found differential reddening, but found it to increase in the opposite direction.  Burke et al.\ (2004) presented BVI photometry of a very large area around the cluster and found no evidence of variable extinction.  Values for interstellar reddening range from E(B$-$V) = 0.21 (Burke et al.\ 2004) to 0.28 (Wee \& Lee 1996).  Distance moduli vary from (m$-$M)$_{0}$ = 12.0 (Wee \& Lee 1996) to 12.4 (Carraro \& Patat 1994, Subramaniam 2003, Burke et al.\ 2004).  We identified red clump stars by eye in the cluster 2MASS JK CMD, which features a very broad, crowded red clump at (K, J$-$K) $\sim$ 11.1$\pm$0.2, 0.64$\pm$0.10.  Assuming solar metallicity for the cluster, E(J$-$K) = 0.024$\pm$0.102, and E(B$-$V) = 0.045$\pm$0.102, A$_{\rm K}$ = 0.015.  These values are much lower than found by the studies mentioned above.  Nonetheless, the resulting distance modulus, (K$-$M$_{\rm K}$)$_{0}$ = 12.4$\pm$0.2, is consistent with literature values.  This places NGC 1245 3.0$\pm$0.3 kpc from the Sun, with \rgc\ = 11.1$\pm$0.3 kpc, and z = $-$460$\pm$50 pc.  
Stellar identifications come from Chincarini (1964); alternate ID's come from Wee \& Lee (1996).  J2000 coordinates have been taken from the USNO-B1.0 catalog\footnote{See http://www.nofs.navy.mil/data/FchPix/}.  The optical  
photometry listed in Table 3 come from Subramaniam (2003), save for star 382, for which we give the photoelectric photometry from Jennens \& Helfer (1975).  Additional likely evolved star targets  were selected from the 2MASS JK CMD of the cluster.  These targets have been given arbitrary identification numbers starting with 900.  
Given the possible large zero-point offsets for the Subramaniam photometry (see Burke et al.\ 2004 for discussion), we determined stellar atmospheric parameters using J$-$K colors {\it only}.  
For these calculations, we have adopted the distance and reddening values of Burke et al.\ (2004): E(B$-$V) = 0.21$\pm$0.03, (m$-$M)$_{0}$ = 12.27$\pm$0.02.

\item{\it NGC 1817:}
A summary of previous studies of NGC 1817 ({\it l} = 186.1$^{o}$, {\it b} = $-$13.1$^{o}$) was given in our first study of stars in this cluster (Jacobson et al.\ 2009).  There, we did not determine its distance and reddening using its red clump morphology since it was used as a calibration cluster by GS02.  Instead, we adopted the findings of Balaguer-N\'{u}\~{n}ez et al.\ (2004b), who found E(B$-$V) = 0.27, (m$-$M)$_{0}$ = 10.9$\pm$0.6, d = 1.5$\pm$0.5 kpc, \rgc = 10.0$\pm$0.4 kpc, z = $-$340$\pm$100 pc.  However, as for M 67, we have determined its fundamental parameters using its red clump photometry in order to compare with other findings in the literature.  In actuality, identification of red clump stars in the 2MASS CMD is difficult, and restricting the analysis to radial velocity and/or proper motion members does not make identification easier.  This is due to the extended appearance of the red clump, even in optical CMDs, as noted by Mermilliod et al.\ (2003).  Van Helshoecht \& Groenewegen (2007) cautioned that inclusion of noncluster members can alter the red clump morphology and the determined distance and reddening to clusters; NGC 1817 serves as a useful example of this, and even of how the inclusion/exclusion of member stars can alter the results when the red clump is especially extended.  For example, depending on whether a red star clump is identified in the cluster 2MASS CMD with no membership information, with membership information, or with cross-identification of clump stars in the optical CMDs of Mermilliod et al.\ (2003) or Balaguer-N\'{u}\~{n}ez et al.\ (2004b), the mean red clump K-band magnitude and J$-$K color can differ by more than 0.3 and 0.04 magnitudes, respectively, with dispersions of up to 0.2 and 0.1 magnitudes.  The E(J$-$K) values range from 0.05 to 0.09,  indicating E(B$-$V) = 0.15--0.18 -- much lower values than found by other studies in the literature.  Corresponding absolute distance moduli range  from 11.1 to 11.5, which are also larger than found by other studies.  As a result of this, we have chosen to use the reddening and distance to NGC 1817 found by Balaguer-N\'{u}\~{n}ez et al.\ (2004b), as we did in Jacobson et al.\ (2009).  Star identifications for this  1.1$\pm$0.2 Gyr old cluster (Salaris et al.\ 2004) have been taken from WEBDA, and are those of Cuffy (1938).  Alternative ID's in Table 3 come from Balaguer-N\'{u}\~{n}ez et al.\ (2004b).  Note that the photometry listed in Table 3 are V{\it by} photometry from Balaguer-N\'{u}\~{n}ez et al.\ (2004b), and the coordinates are from Balaguer-N\'{u}\~{n}ez et al.\ (2004a).

\item{\it NGC 2158:}
A summary of previous studies of NGC 2158 ($\it l$ = 186$^{o}.$6 , $\it b$ = $+$1$^{o}.$8) was presented in Jacobson et al.\ (2009).  As described there, 
GS02 determined fundamental parameters for NGC 2158 based on their calibration of the red clump star {\it K} magnitude as a distance indicator.  They found E(B$-$V)=0.430$\pm$0.013 and (m$-$M)$_{V}$=14.38$\pm$0.09 (corresponding to a distance of $\sim$4 kpc from the Sun).  Using their values for the clump's $<$M$_{K}$$>$ and $<$(J$-$K)$>$, and adopting [Fe/H] = $-$0.30 for the cluster, we redetermined the distance modulus and reddening for this study: E(J$-$K) = 0.225$\pm$0.033, E(B$-$V) = 0.417$\pm$0.033, (K$-$M$_{\rm K}$)$_{0}$ = 13.03$\pm$0.06.  These values are in excellent agreement with GS02, as they should be.  These values correspond to d = 4036$\pm$125 pc, \rgc = 12.5$\pm$0.1 kpc, and z = 127$\pm$4 kpc.  
Stellar identifications come from WEBDA, and alternate ID's and stellar photometry from Carraro et al.\ (2002), save for star 4230.  Photometry for this star has been taken from the photographic study of Arp \& Cuffey (1962; star ``c" in their catalog).  Stellar coordinates are from the USNO-B1.0 catalog.  NGC 2158 has an age of 1.9$\pm$0.5 Gyr on the scale of Salaris et al.\ (2004).  We have adopted the distance and reddening of GS02 for NGC 2158 for the determination of stellar atmospheric parameters.

\item{\it NGC 2194:}
The CCD photometry studies of NGC 2194 ($\it l$ = 197$^{o}.$3 , $\it b$ = $-$2$^{o}.$3) include Sanner et al.\ (2000), Piatti et al.\ (2003) and Kyeong et al.\ (2005).  The main sequence found by each of these studies is very broad, indicating a high degree of variability in interstellar extinction, the presence of a prominent binary main sequence, large uncertainties in the  photometry, or some combination of these effects.  All these studies agree on its young age, with values ranging from 400--630 Myr, and the cluster exhibits a well-populated red clump and several blue straggler candidates.  
Salaris et al.\ (2004) found a slightly older age of 900$\pm$200 Myr, based on its CMD morphology.
Reddening estimates from these three studies range E(B$-$V) = 0.40--0.47, while distance moduli vary (m$-$M)$_{0}$=12.1--12.5.   
We determined the reddening and distance to the cluster based on the JK photometry of its red clump stars.  Red clump stars were identified by eye in the 2MASS CMD, as well as by cross-checking with red clump stars identified in the optical photometry studies of Piatti et al.\ (2003) and Kyeong et al.\ (2005).  Kyeong et al.\ also obtained JHK photometry of the cluster, and our determined red clump K-band magnitude and (J$-$K) color,  10.2$\pm$0.1, 0.78$\pm$0.03 is in good agreement with their values (differences in filter systems considered).  Using these values and assuming [Fe/H] = 0, we found E(J$-$K) = 0.165$\pm$0.037, E(B$-$V) = 0.306$\pm$0.037, A$_{\rm K}$ = 0.103.  The absolute distance modulus, (K$-$M$_{\rm K}$)$_{0}$ = 11.4$\pm$0.2, which is much smaller than the values of 
Sanner et al., Piatti et al., \& Kyeong et al.  This difference can largely be ascribed to our use of the larger age of Salaris et al.\ (2004) to determine the mean clump luminosity: M$_{\rm K}$ = $-$1.3$\pm$0.1, as opposed to $-$1.6 used by Kyeong et al.  Our distance modulus for NGC 2194 corresponds to d = 1903$\pm$120 pc, \rgc = 10.3$\pm$0.2 kpc, and z = $-$80$\pm$5 pc.  
Stellar identifications and J2000 coordinates come from Sanner et al.\ (2000) and are also those used in WEBDA.  Also shown in Table 3 are the ID's from Cuffey (1943), and the stellar photometry from Sanner et al.  The reddening and distance modulus of Sanner et al.\ were also used to determine stellar atmospheric parameters.

\item{\it NGC 2355:}
CCD photometry studies of NGC 2355 ($\it l$ = 203$^{o}.$4 , $\it b$ = $+$11$^{o}.8$) are those of Kaluzny \& Mazur(1991b), Phelps et al.\ (1994), and most recently, Ann et al.\ (1999).  Reddening values and distance moduli from this small number of studies are quite different, with E(B$-$V) values ranging from 0.12 to 0.25, and (m$-$M)$_{0}$ ranging from 12.1 to 11.4 (both Kaluzny \& Mazur 1991, Ann et al.\ 1999, respectively).  This intermediate-aged cluster (0.8 Gyr; Salaris et al.\ 2004) has an identifiable red clump that we used to determine its distance and reddening.  Soubiran et al.\ (2000) presented a proper motion study of the cluster, and also determined radial velocities, atmospheric parameters, and metallicities for several stars in the field by matching R$\sim$40,000, S/N$\sim$15, spectra to a library of template spectra.  Combining their results with the photometry of Kaluzny \& Mazur (1991) and 2MASS (Cutri et al.\ 2003), they determined NGC 2355 to have E(B$-$V)=0.16 and (m$-$M)$_{0}$=11.06$\pm$0.10.
The JK CMD of the cluster field shows a lot of field star contamination, so even with the identification of radial velocity members, it was difficult to identify the exact location of the red clump.  We have identified the red clump as having (K, J$-$K) = 10.1$\pm$0.1, 0.62$\pm$0.03.  Assuming [Fe/H] = 0 for the cluster, E(J$-$K) = 0.022$\pm$0.044, and E(B$-$V) = 0.041$\pm$0.044, A$_{\rm K}$ = 0.014.  As found for other clusters, this reddening value is much lower than that found by other studies.  That said, the absolute distance modulus, (K$-$M$_{\rm K}$)$_{0}$ = 11.43$\pm$0.14, is consistent with that found by, e.g., Ann et al.\ (1999).  This places NGC 2355 1928$\pm$130 pc from the Sun, at \rgc = 10.3$\pm$0.1 kpc, z = 400$\pm$25 pc.

Also, Soubiran et al. remarked upon the presence of a very bright giant star, two magnitudes brighter than the red clump in V-band, but which has the same color and temperature of the clump stars and is a radial velocity member.  This star (\# 398) is also in our sample, and we confirm its membership.  A high-resolution spectrum of this star is publicly available from the ELODIE archive\footnote{See http://atlas.obs-hp.fr/elodie/intro.html}, and is of high enough S/N to carry out a detailed abundance analysis.  We used this spectrum in addition to our own in the analysis of this cluster.  Stellar identifications come from Ann et al.\ (1999; these are the ID's adopted in WEBDA).  The identifications of Kaluzny \& Mazur (1991) are also given, along with their BV photometry for all stars save 144, 203 and 668, for which we give the photometry of Ann et al.\ (1999).  Stellar coordinates come from Lasker et al.\ (1990).  
For calculation of stellar atmospheric parameters, we adopted the reddening and distance modulus of Soubiran et al.

\item{\it NGC 2420:}
NGC 2420 is a canonical metal-poor cluster that resides in the outer disk ($\it l$ = 198$^{o}.$1, $\it b$ = $+$19$^{o}.$6).  On the age-scale of Salaris et al.\ (2004), it is 2.2$\pm$0.3 Gyr old.  It has been the subject of numerous photometry studies over the past four decades or more, all of which indicate low values of extinction towards the cluster: E(B$-$V)$\approx$0.04 (e.g., West 1967; McClure et al.\ 1974; Anthony-Twarog et al.\ 2006; Sharma et al.\ 2006).  Distance moduli range from (m$-$M)$_{0}$ = 11.4--11.9 (McClure et al. 1974; Twarog et al.\ 1997), with the majority of studies finding the latter value.  NGC 2420 is used by GS02 as calibration of their relation, so here again our analysis using the red clump morphology serves as validation of the relation.  Red clump stars were identified in the V{\it by} CMD of Anthony-Twarog et al.\ (2006).  The mean (K, J$-$K) of 12 clump stars is 10.330$\pm$0.241, 0.615$\pm$0.021.  Adopting [Fe/H] = $-$0.20, E(J$-$K) = 0.038$\pm$0.038, E(B$-$V) = 0.071$\pm$0.038, A$_{\rm K}$ = 0.024.  This reddening is slightly larger than that found by the studies mentioned above.  (K$-$M$_{\rm K}$)$_{0}$ = 11.97$\pm$0.04, which is also slightly larger than that found by previous studies.  This indicates a distance of 2473$\pm$285 pc from the Sun, or \rgc = 10.7$\pm$0.3 kpc, z = 830$\pm$100 pc for NGC 2420.
Stellar identifications are those used in WEBDA, and come from Cannon \& Lloyd (1970).  Alternative ID's are from West (1967) and we also used the photometry of Anthony-Twarog et al.\ (1990; Table 3) and Anthony-Twarog et al.\ (2006).  For stars 131 and 255, not in the catalog of Anthony-Twarog et al.\ (1990), we give the BV photometry of Sharma et al. (2006).  J2000 coordinates are from the USNO-B1.0 catalog.  
Additional likely evolved star targets  were selected from the 2MASS JK CMD of the cluster.  These targets have been given arbitrary identification numbers starting with 8000.  The reddening value E(B$-$V) = 0.050$\pm$0.004 (Anthony-Twarog et al.\  2006) and distance modulus (m$-$M)$_{0}$ = 11.9$\pm$0.2 (Twarog et al.\ 1997) were used to calculate stellar temperatures and gravities.

\item{\it NGC 2425:}
NGC 2425 is located at  $\it l$ = 231$^{o}.$5 , $\it b$ = $+$3$^{o}.$3.  It is a relatively little-studied, old cluster, with the first photometric studies done in 2006.  The UBV(RI)$_{C}$ photometry of Moitinho et al.\ (2006) extends down to V$\sim$21 and revealed a clear red clump and sparse giant branch.  The BVI photometry of Pietrukowicz et al.\ (2006) extends down to a similar magnitude.  These two studies found E(B$-$V)$\sim$0.21-0.29 and (m$-$M)$_{V}$=13.2--13.6, and age 2.2--2.5 Gyr based on ZAMS-fitting and matches to theoretical isochrones.  Stellar identifications are those used in WEBDA, and come from Pietrukowicz et al.\ (2006).  We determined the distance and reddening to the cluster based on the 2MASS photometry of its red clump stars.  The red clump was identified in the cluster 2MASS CMD, with the aid of radial velocity information to remove non-members.  We found (K, J$-$K) = 10.87$\pm$0.04, 0.76$\pm$0.02 for the clump.  Adopting [Fe/H] = $-$0.10, E(J$-$K) = 0.16$\pm$0.04, E(B$-$V) = 0.30$\pm$0.04, A$_{\rm K}$ = 0.10.  (K$-$M$_{\rm K}$)$_{0}$ = 12.43$\pm$0.07, which is consistent with literature values.  These place NGC 2425 3062$\pm$100 pc from the Sun, at \rgc = 10.7$\pm$0.1 kpc, z = 180$\pm$10 pc.
Photometry comes from Pietrukowicz et al.\ (2006), and alternate ID's from Moitinho et al. (2006).  Stellar coordinates were taken from the USNO-B1.0 catalog.

\item{\it NGC 7789:}
NGC 7789 ($\it l$ = 115$^{o}$.5 , $\it b$ = $-$5$^{o}$.4) is another well-studied, rich open cluster.  The first photometric study of it was performed by Burbidge \& Sandage (1958) who obtained UBV photographic and photoelectric photometry down to V$\sim$16 and determined E(B$-$V) = 0.28, (m$-$M)$_{0}$ = 11.36$\pm$0.20 for the cluster.  Photometry studies since then have found interstellar extinction values ranging from   E(B$-$V) = 0.24$\pm$0.01 (Janes 1977) to E(B$-$V) = 0.28$\pm$0.02 (Gim et al.\ 1998b), with some studies finding E(B$-$V)$\approx$0.27 (e.g., Twarog et al.\ 1997, Breger \& Wheeler 1980).  Distance moduli generally range from (m$-$M)$_{V}$ = 11.5--12.45 in these studies.  NGC 7789 has also been the subject of detailed proper motion (McNamara \& Solomon 1981) and radial velocity (Gim et al.\ 1998a) studies.  
NGC 7789 has a very tight, well-populated red clump that we used to determine its distance and reddening.  The clump's position is very clear in the 2MASS CMD, though we also made use of the clump star identifications of Gim et al.\ (1998b).  We have found (K, J$-$K) = 10.1$\pm$0.2, 0.72$\pm$0.10.  Assuming solar metallicity for the cluster, E(J$-$K) = 0.10$\pm$0.10, E(B$-$V) = 0.19$\pm$0.10, and A$_{\rm K}$ = 0.063.  This reddening value is $\sim$0.1 magnitude lower than that found by other studies.  However, the resulting distance modulus, (K$-$M$_{\rm K}$)$_{0}$ = 11.70$\pm$0.21, is in the 
mid-range of values found in the literature, and agrees within the errors, with that of Twarog et al.\ (1997).  This places NGC 7789 2185$\pm$210 pc from the Sun, with \rgc = 9.6$\pm$0.2 kpc, and z = $-$206$\pm$20 pc.  
Stellar identifications are from WEBDA (originally from Gim et al. 1998b), and the photometry is taken from Mochejska \& Kaluzny (1999). The identifications and 1950B coordinates from K\"{u}stner (1923) are also given for each star.
\end{itemize}

\section{Comparison to Literature Abundance Results}
In this appendix we compare our abundance results to those found in the literature.  As already discussed, some clusters (e.g., M 67) have been subject to numerous detailed abundance studies, and where such studies exist, we generally confine the discussion to them.  However, some clusters have been subject to photometric studies only, in which case we compare our [Fe/H] values with photometric metallicities.
\begin{itemize}
\item {\it M 67}:  We have previously made a careful comparison of the results of three different high resolution spectroscopic studies of M 67; we refer the reader to Friel et al.\ (2010) for details.  The mean metallicity for M 67 determined from the Hydra spectra, [Fe/H] = $-$0.01$\pm$0.05 is in excellent agreement with other studies of this solar metallicity cluster.  Agreement among different studies for the abundances of other elements is not as good, however.  For example, Tautvai\v{s}ien\.{e} et al.(2000) and Yong et al.\ (2005) found larger enhancements of [Na/Fe] for this cluster than did we or Pancino et al.\ (2010).  Our [Ca/Fe] ratio is intermediate between those of Pancino et al., Yong et al.\ and Tautvai\v{s}ien\.{e} et al., while our [Ti/Fe] and [O/Fe] ratios are lower than those of all three studies.  As these different studies generally rely on the same samples of M 67 giants, and the agreement in atmospheric parameters is reasonable, these abundance differences are likely due to details in the different analyses.

\item {\it NGC 188}:  Though many spectroscopic studies of NGC 188 exist in the literature, relatively few of them focus on the determination of Fe or elements heavier than the light elements.  Hobbs et al.\ (1990) found [Fe/H] = $-$0.12$\pm$0.16 based on a study of seven stars in the cluster.  Randich et al.\ (2003) found [Fe/H] = $+$0.01$\pm$0.08 based on co-added spectra of main sequence stars.  These values are in good agreement with our value of [Fe/H] = $-$0.03$\pm$0.04.

\item {\it NGC 1245}: 
To our knowledge, this is the first detailed chemical abundance study of NGC 1245; previous metallicity estimates have been determined from photometry.  NGC 1245 has [Fe/H] = $+$0.14 in the Lyng\aa's open cluster catalog (Lyng\aa\ 1987).  Based on Washington photometry, Wee \& Lee (1996) found [Fe/H] = $-$0.04$\pm$0.05, in excellent agreement with our value of [Fe/H] = $-$0.04$\pm$0.05, as is the value found by Burke et al.\ using isochrone-fitting (2004: $-$0.05$\pm$0.03). 

\item {\it NGC 1817}: Parisi et al.\ (2005) determined a metallicity for NGC 1817 based on Washington photometry of 15 cluster stars.  They found [Fe/H] = $-$0.33$\pm$0.08.  This is in decent agreement with our value of [Fe/H] = $-$0.16$\pm$0.03 based on 28 stars.  In our analysis of KPNO 4m echelle spectra of two stars in this cluster (Jacobson et al.\ 2009), we remarked that the E(B$-$V) value adopted by Parisi et al.\ (2005) was slightly lower than that found by Balaguer-N\'{u}\~{n}ez et al.(2004b), which we have adopted here.  Use of E(B$-$V) = 0.27 instead of E(B$-$V) = 0.23 would increase the metallicity found from the Washington photometric indices by $\sim$0.15 dex, to [Fe/H] $\approx$ $-$0.16 dex, in excellent agreement with our value. 

\item {\it NGC 2158}: To our knowledge, the only high resolution spectroscopic abundance study of NGC 2158 in the literature is our work based on one star in the cluster (Jacobson et al.\ 2009).  As we've already compared our results to this work, we turn briefly to other photometric and/or low resolution spectroscopic studies.  Janes (1979) found [Fe/H] = $-$0.64$\pm$0.24 based on DDO photometry; Twarog et al.\ (1997) found [Fe/H] = $-$0.238$\pm$0.064 (s.d.) based on a recalibration of Janes's metallicity scale.  From Washington photometry of stars in the field, Geisler (1987) found [A/H] = $-$0.88$\pm$0.10 or $-$0.30$\pm$0.15 assuming E(B$-$V) = 0.40 or 0.55, respectively.  Based on low-resolution spectroscopy of seven stars, Friel et al.\ (2002) found [Fe/H] = $-$0.29$\pm$0.09.
Our metallicity of [Fe/H] = $-$0.28$\pm$0.05 is in excellent agreement with this value.

\item {\it NGC 2194}:  NGC 2194 is another cluster for which only photometric abundances can be found in the literature.  Sanner et al.\ (2000) found solar metallicity isochrones provided the best match to the observed cluster CMD, in agreement with our determined [Fe/H] = $-$0.08$\pm$0.08 for this cluster.  However, Piatti et al.\ (2003) found [Fe/H] = $-$0.27$\pm$0.03 based on Washington photometry.  Some of this difference may be explained by their choice of reddening to the cluster: they found E(V$-$I) = 0.75, which corresponds to E(B$-$V) = 0.55.  This is much larger than the value found in our study.  Like NGC 2158, uncertainties in the (large) extinction to this cluster can adversely effect the photometric temperature scale we adopted to determine the cluster abundances.  However, a larger reddening would result in hotter temperatures, which would increase stellar metallicity.  Therefore, selection of a larger reddening value would not bring our results into better agreement with Piatti et al.\ (2003).  The reddening to this cluster should be better constrained, or else stellar temperatures should be spectroscopically-determined in order to untangle reddening and metallicity of this cluster.

\item {\it NGC 2355}:  A handful of photometry studies of NGC 2355 exist in the literature.  Metallicity estimates range from [Fe/H] = $+$0.13 (Kaluzny \& Mazur 1991) to [Fe/H] = $-$0.32 (Ann et al.\ 1999).  Soubiran et al.\ (2000) determined a mean cluster metallicity by comparison of low S/N, high resolution ELODIE spectra of cluster members to a library of stars with well-defined parameters (the TGMET library; Soubiran et al.\ 1998).  They found a weighted mean [Fe/H] = $-$0.07$\pm$0.11 based on 17 stars, which is in excellent agreement with our value of [Fe/H] = $-$0.08$\pm$0.08.

\item {\it NGC 2420}:  At least three high resolution spectroscopic studies of NGC 2420 have been performed previously.  Smith \& Suntzeff (1987) found [Fe/H] $\sim$ $-$0.45$\pm$0.08 (their analysis was relative to a Hyades giant star rather than the Sun).  Based on high resolution Lick spectra of three stars, Bosler (2004) found an average [Fe/H] = $-$0.33$\pm$0.19.  Our analysis of these same spectra found the same result (Jacobson et al.\ 2009).  Most recently, Pancino et al.\ (2010) found [Fe/H] = $-$0.05$\pm$0.10 (s.d.) based on spectra of three evolved stars.  Our value, [Fe/H] = $-$0.20$\pm$0.06 lies in between all these values, most closely agreeing with that of Bosler (2004).  For the other elements, our [X/Fe] ratios for Ca, Ti and Ni agree with those of Smith \& Suntzeff (1987) within 0.15 dex, which is decent considering the lower resolution and even smaller wavelength range of their spectra compared to ours.  Compared to the results of Pancino et al.\ (2010), our values for [O/Fe], [Ca/Fe], [Si/Fe] and [Na/Fe] differ by $\sim$0.2 dex, with our values being larger for all but oxygen.  Abundance ratios for Mg, Ti and Ni agree within 0.03 dex.

\item {\it NGC 2425}:  Photometric metallicities for NGC 2425 have been determined by Moitinho et al.\ (2006) and Pietrukowicz et al.\ (2006).  Based on fits of theoretical isochrones to optical CMD's, Moitinho et al.\ found solar metallicity, while Pietrukowicz et al.\ found good matches using both solar metallicity isochrones and isochrones with Z = 0.008.  Our value of [Fe/H] = $-$0.15$\pm$0.09 is in good agreement with these results.

\item {\it NGC 7789}: Tautvai\v{s}ien\.{e} et al.\ (2005) carried out a detailed abundance study of 9 evolved stars in NGC 7789.  They found [Fe/H] = $-$0.04$\pm$0.05, which is in good agreement with our value of [Fe/H] = $+$0.02$\pm$0.04.  Abundances of non-Fe elements all agree within 0.11 dex, save for Si and O (our values are 0.15-0.2 dex larger).  Pancino et al.\ (2010) determined abundances for three giants in NGC 7789, and their determined mean metallicity is also in excellent agreement with our value: [Fe/H] = $+$0.04$\pm$0.10.  [X/Fe] ratios agree less well, with differences between our and their values for Ca, Mg, Na and Si differing by $\sim$0.15 dex or more.  Our abundance ratios tend to be larger for all elements in common save for Mg and O.  
\end{itemize}

\clearpage
\begin{turnpage}
% Values checked 5 May 2011
%\begin{landscape}
\begin{deluxetable}{lrcccccccccccccccc}
%\setlength{\tabcolsep}{0.02in} 
%\rotate
\tabletypesize{\scriptsize}
\tablecolumns{18}
\tablenum{12}
%\tablewidth{0pt}
\tablecaption{Mean Cluster [X/H] Ratios\label{cluster_wxh}}
\tablehead{ \colhead{Cluster} &
\colhead{\# Stars} & \colhead{[Fe/H]} & \colhead{$\sigma$[Fe/H]} &
 \colhead{[Na/H]} & \colhead{$\sigma$[Na/H]} & \colhead{[Mg/H]} & \colhead{$\sigma$[Mg/H]} & \colhead{[Si/H]} & \colhead{$\sigma$[Si/H]} & 
\colhead{[Ca/H]} & \colhead{$\sigma$[Ca/H]} & \colhead{[Ti/H]} & \colhead{$\sigma$[Ti/H]} &
  \colhead{[Ni/H]} & \colhead{$\sigma$[Ni/H]} & \colhead{[Zr/H]} & \colhead{$\sigma$[Zr/H]} %& \colhead{[Si/H]} & \colhead{$\sigma$[Si/H]}
 }
\startdata
M67   &   19 & $-$0.01 & 0.05 & $-$0.02 & 0.02 & $+$0.22 & 0.05 & $+$0.19 & 0.02 & $-$0.11 & 0.04 & $-$0.07 & 0.02 & $-$0.03 & 0.03 & $+$0.00 & 0.01\\
N188  &   27 & $-$0.03 & 0.04 & $+$0.14 & 0.01 & $+$0.24 & 0.04 & $+$0.24 & 0.01 & $-$0.08 & 0.03 & $+$0.15 & 0.01 & $+$0.07 & 0.03 & $+$0.02 & 0.01\\
N1245 &   13 & $-$0.04 & 0.05 & $+$0.04 & 0.01 & $+$0.04 & 0.06 & $+$0.15 & 0.04 & $-$0.02 & 0.04 & $-$0.14 & 0.03 & $-$0.04 & 0.04 & $+$0.13 & 0.01\\
N1817 &   28 & $-$0.16 & 0.03 & $-$0.02 & 0.01 & $-$0.09 & 0.04 & $+$0.06 & 0.02 & $+$0.02 & 0.02 & $-$0.29 & 0.01 & $-$0.17 & 0.01 & $+$0.12 & 0.01\\
N2158 &   15 & $-$0.28 & 0.05 & $-$0.31 & 0.01 & $-$0.07 & 0.05 & $+$0.11 & 0.02 & $-$0.31 & 0.02 & $-$0.65 & 0.01 & $-$0.24 & 0.03 & $+$0.00 & 0.01\\ 
N2194 &    6 & $-$0.08 & 0.08 & $-$0.05 & 0.02 & $-$0.07 & 0.08 & $+$0.09 & 0.06 & $-$0.12 & 0.04 & $-$0.35 & 0.02 & $-$0.10 & 0.06 & $+$0.13 & 0.05\\
N2355 &    5 & $-$0.08 & 0.08 & $+$0.27 & 0.01 & $+$0.15 & 0.09 & $+$0.14 & 0.04 & $+$0.17 & 0.06 & $-$0.24 & 0.03 & $-$0.06 & 0.05 & $+$0.20 & 0.09\\ 
N2420 &    9 & $-$0.20 & 0.06 & $-$0.12 & 0.01 & $-$0.10 & 0.07 & $+$0.04 & 0.02 & $-$0.08 & 0.03 & $-$0.26 & 0.04 & $-$0.20 & 0.02 & $+$0.02 & 0.04\\
N2425 &    4 & $-$0.15 & 0.09 & $-$0.07 & 0.02 & $-$0.08 & 0.10 & $+$0.03 & 0.04 & $+$0.07 & 0.07 & $-$0.13 & 0.03 & $-$0.13 & 0.05 & $+$0.33 & 0.01\\
N7789 &   28 & $+$0.02 & 0.04 & $-$0.06 & 0.02 & $+$0.18 & 0.04 & $+$0.28 & 0.01 & $+$0.07 & 0.02 & $+$0.02 & 0.01 & $+$0.05 & 0.03 & $+$0.31 & 0.01\\ 
\enddata
\end{deluxetable}
%\end{landscape}
%12
% Values checked 5 May 2011
%\begin{landscape}
\begin{deluxetable}{lrcccccccccccccc}
\tabletypesize{\scriptsize}
\tablecolumns{18}
%\tablewidth{0pt}
\tablenum{13}
\tablecaption{Mean Cluster [X/Fe] Ratios\label{cluster_wxfe}}
\tablehead{ \colhead{Cluster} &
\colhead{\# Stars} & %\colhead{[Fe/H]} & \colhead{$\sigma$[Fe/H]} &
 \colhead{[Na/Fe]} & \colhead{$\sigma$[Na/Fe]} & \colhead{[Mg/Fe]} & \colhead{$\sigma$[Mg/Fe]} & \colhead{[Si/Fe]} & \colhead{$\sigma$[Si/Fe]} & 
\colhead{[Ca/Fe]} & \colhead{$\sigma$[Ca/Fe]} & \colhead{[Ti/Fe]} & \colhead{$\sigma$[Ti/Fe]} &
  \colhead{[Ni/Fe]} & \colhead{$\sigma$[Ni/Fe]} & \colhead{[Zr/Fe]} & \colhead{$\sigma$[Zr/Fe]} %& \colhead{[Si/H]} & \colhead{$\sigma$[Si/H]}
 }
\startdata
M67   &   19 & $+$0.03 & 0.06 & $+$0.23 & 0.07 & $+$0.21 & 0.05 & $-$0.11 & 0.07 & $-$0.10 & 0.05 & $-$0.01 & 0.06 & $-$0.03 & 0.06 \\
N188  &   27 & $+$0.10 & 0.05 & $+$0.26 & 0.05 & $+$0.25 & 0.05 & $-$0.04 & 0.06 & $+$0.14 & 0.05 & $+$0.08 & 0.05 & $+$0.10 & 0.05 \\
N1245 &   13 & $+$0.04 & 0.06 & $+$0.09 & 0.08 & $+$0.18 & 0.07 & $+$0.05 & 0.07 & $-$0.09 & 0.06 & $-$0.02 & 0.07 & $+$0.33 & 0.06 \\
N1817 &   28 & $+$0.11 & 0.04 & $+$0.08 & 0.05 & $+$0.17 & 0.04 & $+$0.07 & 0.04 & $-$0.06 & 0.04 & $-$0.05 & 0.04 & $+$0.34 & 0.04 \\
N2158 &   15 & $+$0.01 & 0.05 & $+$0.22 & 0.07 & $+$0.39 & 0.05 & $+$0.00 & 0.06 & $-$0.26 & 0.05 & $+$0.05 & 0.06 & $+$0.00 & 0.06 \\ 
N2194 &    6 & $+$0.04 & 0.09 & $+$0.00 & 0.12 & $+$0.18 & 0.11 & $-$0.03 & 0.12 & $-$0.24 & 0.11 & $-$0.03 & 0.10 & $+$0.22 & 0.12 \\
N2355 &    5 & $+$0.05 & 0.08 & $+$0.21 & 0.12 & $+$0.19 & 0.10 & $+$0.05 & 0.13 & $-$0.04 & 0.09 & $+$0.02 & 0.10 & $+$0.24 & 0.12 \\ 
N2420 &    9 & $+$0.08 & 0.06 & $+$0.11 & 0.09 & $+$0.21 & 0.07 & $+$0.10 & 0.07 & $-$0.07 & 0.08 & $-$0.01 & 0.07 & $+$0.24 & 0.08 \\
N2425 &    4 & $+$0.14 & 0.10 & $+$0.07 & 0.14 & $+$0.20 & 0.11 & $+$0.17 & 0.12 & $+$0.06 & 0.11 & $+$0.00 & 0.11 & $+$0.33 & 0.10 \\
N7789 &   28 & $+$0.09 & 0.05 & $+$0.14 & 0.05 & $+$0.25 & 0.05 & $+$0.01 & 0.05 & $-$0.05 & 0.04 & $+$0.00 & 0.05 & $+$0.18 & 0.05 \\ 
\enddata
\end{deluxetable}
%\end{landscape}%13
\clearpage
%\begin{landscape}
 \begin{deluxetable}{lccccccccccccc}
\tabletypesize{\scriptsize}
\tablecolumns{12}
\tablenum{18}
\tablewidth{0pt}
\tablecaption{Previously Studied Clusters\label{prev_ocs}}
\tablehead{ \colhead{Cluster} & \colhead{R$_{gc}$ (kpc)} & \colhead{Age (Gyr)} & \colhead{\# Stars} &
 \colhead{[Fe/H]} & \colhead{[Na/Fe]} & \colhead{[Al/Fe]} & \colhead{[Mg/Fe]} & 
 \colhead{[Si/Fe]} & \colhead{[Ca/Fe]} & \colhead{[Ti/Fe]} & %\colhead{[Ni/Fe]} & 
 \colhead{[O/Fe]} %& \colhead{[$\alpha$/Fe]}
 }
\startdata
Be 17\tablenotemark{1} & 11.4 & 10.1 & 3 & $-$0.15$\pm$0.09 & $+$0.10$\pm$0.13 & $+$0.23$\pm$0.09 & $+$0.13$\pm$0.12 & $+$0.23$\pm$0.06 & $-$0.02$\pm$0.06 & $-$0.07$\pm$0.07  & $+$0.05$\pm$0.06 \\%& $+$0.07$\pm$0.14\\
Be 32\tablenotemark{4} & 11.5 & 5.9 & 2 & $-$0.30$\pm$0.02 & $+$0.20$\pm$0.01 & $+$0.19$\pm$0.06 & $+$0.13$\pm$0.01 & $+$0.25$\pm$0.05 & $-$0.07$\pm$0.01 & $-$0.17$\pm$0.01  & $-$0.01$\pm$0.03 \\%& $+$0.03$\pm$0.20\\
Be 39\tablenotemark{4} & 11.9 & 7.0 & 4 & $-$0.21$\pm$0.01 & $+$0.11$\pm$0.04 & $+$0.20$\pm$0.03 & $+$0.15$\pm$0.03 & $+$0.19$\pm$0.04 & $-$0.01$\pm$0.07 & $-$0.03$\pm$0.09  & $-$0.02$\pm$0.05 \\%& $+$0.07$\pm$0.11\\
M 67\tablenotemark{4} & 9.1 & 4.3  & 3 & $+$0.03$\pm$0.07 & $+$0.14$\pm$0.09 & $+$0.11$\pm$0.07 & $+$0.05$\pm$0.03 & $+$0.18$\pm$0.04 & $-$0.08$\pm$0.01 & $-$0.14$\pm$0.05  & $-$0.16$\pm$0.05 \\%& $+$0.00$\pm$0.14\\
NGC 188\tablenotemark{4} & 9.4 & 6.3 & 4 & $+$0.12$\pm$0.02 & $+$0.15$\pm$0.03 & $+$0.20$\pm$0.03 & $+$0.15$\pm$0.06 & $+$0.17$\pm$0.08 & $-$0.03$\pm$0.06 & $+$0.05$\pm$0.12  & $-$0.04$\pm$0.10 \\%& $+$0.07$\pm$0.09\\
NGC 1193\tablenotemark{4} & 13.6 & 4.2 & 1 & $-$0.22$\pm$0.14 & $+$0.25$\pm$0.14 & $+$0.10$\pm$0.06 & $+$0.23$\pm$0.18 & $+$0.16$\pm$0.09 & $+$0.11$\pm$0.15 & $+$0.11$\pm$0.12  & $-$0.08$\pm$0.10 \\%& $+$0.15$\pm$0.06\\
NGC 1817\tablenotemark{3} & 10.0 & 1.1 & 2 & $-$0.07$\pm$0.04 & $+$0.12$\pm$0.01 & $+$0.01$\pm$0.01 & $+$0.01$\pm$0.06 & $+$0.12$\pm$0.01 & $+$0.06$\pm$0.01 & $-$0.10$\pm$0.06  & $-$0.19$\pm$0.11 \\%& $+$0.02$\pm$0.09\\
NGC 1883\tablenotemark{3} & 12.3 & 0.7 & 2 & $-$0.01$\pm$0.01 & $+$0.04$\pm$0.06 & $-$0.12$\pm$0.04 & $-$0.04$\pm$0.01 & $+$0.15$\pm$0.03 & $-$0.05$\pm$0.01 & $-$0.23$\pm$0.08  & $-$0.22$\pm$0.03 \\%& $-$0.04$\pm$0.16\\
NGC 2141\tablenotemark{3} & 12.2 & 2.4 & 1 & $+$0.00$\pm$0.16 & $+$0.15$\pm$0.15 & $+$0.01$\pm$0.04 & $+$0.03$\pm$0.10 & $+$0.14$\pm$0.14 & $-$0.11$\pm$0.11 & $-$0.15$\pm$0.12  & $-$0.15$\pm$0.10 \\%& $-$0.02$\pm$0.13\\
NGC 2158\tablenotemark{3} & 12.5 & 1.9 & 1 & $-$0.03$\pm$0.14 & $+$0.10$\pm$0.15 & $-$0.03$\pm$0.10 & $-$0.02$\pm$0.11 & $+$0.12$\pm$0.21 & $-$0.06$\pm$0.08 & $-$0.15$\pm$0.10  & $-$0.17$\pm$0.10 \\%& $-$0.03$\pm$0.11\\
NGC 2204\tablenotemark{5} & 11.6 & 2.0 & 13 & $-$0.23$\pm$0.04 & \nodata & $+$0.12$\pm$0.06 & \nodata & $+$0.16$\pm$0.06 & \nodata & $-$0.40$\pm$0.06 & \nodata \\
NGC 2243\tablenotemark{5} & 10.7 & 4.7 & 10 & $-$0.42$\pm$0.05 & $+$0.03$\pm$0.06 & $+$0.21$\pm$0.07 & $+$0.18$\pm$0.09 & $+$0.24$\pm$0.08 & $+$0.13$\pm$0.07 & $-$0.10$\pm$0.07 & $+$0.01$\pm$0.16\\
NGC 7142\tablenotemark{2} & 9.2 & 4.0 & 4 & $+$0.13$\pm$0.02 & $+$0.24$\pm$0.07 & $+$0.09$\pm$0.10 & $-$0.12$\pm$0.15 & $+$0.02$\pm$0.07 & $-$0.14$\pm$0.12 & $+$0.14$\pm$0.09  & $-$0.09$\pm$0.03 \\%& $-$0.03$\pm$0.13\\
\enddata
\tablenotetext{1}{Friel et al.\ (2005)}
\tablenotetext{2}{Jacobson et al.\ (2008)}
\tablenotetext{3}{Jacobson et al.\ (2009)}
\tablenotetext{4}{Friel et al.\ (2010)}
\tablenotetext{5}{Jacobson et al.\ (2011)}
\end{deluxetable}
%\end{landscape}
\clearpage
\end{turnpage}
\clearpage

\begin{thebibliography}{}
\bibitem[Allen(1976)]{aaq} Allen, C. W. 1976, Astrophysical Quantities (4th ed.; London: Athlone Press)
\bibitem[Alonso et al.(1999)]{alonso99b} Alonso, A., Arribas, S. \& Martinez-Roger, C.
1999, \aaps, 140, 261
\bibitem[Anders \& Grevesse(1989)]{ag89} Anders, E. \& Grevesse, N. 1989, GeCoA, 53, 197
\bibitem[Andrievsky et al.(2002a)]{and02a} Andrievsky, S. M., Bersier, D., Kovtyukh, V. V., Luck, R. E., Maciel, W. J., Lepine, J. R. D., \& Beletsky, Y. V. 2002a, \aap, 384, 140
\bibitem[Andrievsky et al.(2002b)]{and02b} Andrievsky, S.M., Kovtyukh, V.V., Luck, R.E., Lepine, J.R.D., Maciel, W.J., \& Beletsky, Y.V. 2002b, \aap, 392, 491
\bibitem[Andreuzzi et al.(2010)]{and10} Andreuzzi, G., Bragaglia, A., Tosi, M. \& Marconi, G. 2011, \mnras, 412, 1265 (A10)
\bibitem[Ann et al.(1999)]{ann_n2355} Ann, H. B., Lee, M. G., Chun, M. Y., et al. 1999, JKAS, 32, 7
\bibitem[Anthony-Twarog et al.(1990)]{akst90_n2420} Anthony-Twarog, B. J., Kaluzny, J., Shara, M. M. \& Twarog, B. A. 1990, \aj, 99, 1504
\bibitem[Anthony-Twarog et al.(2006)]{at2420} Anthony-Twarog, B. J., Tanner, D., Cracraft, M., \& Twarog, B. A. 2006, \aj, 131, 461
\bibitem[Arp \& Cuffey(1962)]{ac62} Arp, H. C. \& Cuffey, J. 1962, \apj, 136, 51
\bibitem[Balaguer-N\'{u}\~{n}ez et al.(2004b)]{bjemb} Balaguer-N\'{u}\~{n}ez, L., Jordi, C., Galad\'{i}-Enr\'{i}quez, \& Masana, E. 2004b, \aap, 426, 827
\bibitem[Balaguer-N\'{u}\~{n}ez et al.(2004a)]{bjema} Balaguer-N\'{u}\~{n}ez, L., Jordi, C., Galad\'{i}-Enr\'{i}quez, \& Zhao, J. L. 2004a, \aap, 426, 819
\bibitem[Balaguer-N\'{u}\~{n}ez et al.(1998)]{btz} Balaguer-N\'{u}\~{n}ez, L., Tian, K. P., \& Zhao, J. L. 1998, \aaps, 133, 387
\bibitem[Bell et al.(1976)]{bell} Bell, R., Eriksson, K., Gustafsson, B. \& Nordlund, A.  1976, \aaps, 23, 37
\bibitem[Bensby et al.(2003)]{b03} Bensby, T., Feltzing, S. \& Lundstr\"{o}m, I. 2003, \aap, 410, 527
\bibitem[Bensby et al.(2005)]{bensby} Bensby, T., Feltzing, S., Lundstr\"{o}m, I. \& Ilyin, I. 2005, \aap, 433, 185
\bibitem[Bosler, T.(2004)]{tlb} Bosler, T. L. 2004, Thesis, University of California, Irvine.
\bibitem[Bragaglia et al.(2001)]{n6819} Bragaglia, A., et al. 2001, \aj, 121, 327
\bibitem[Bragaglia et al.(2008)]{b08} Bragaglia, A., Sestito, P., Villanova, S., Carretta, E., Randich, S., \& Tosi, M. 2008, \aap, 480, 79 %(SB)
\bibitem[Brown et al.(1996)]{brown} Brown, J. A., Wallerstein, G., Geisler, D. \& Oke, J. B. 1996, \aj, 112, 1551
\bibitem[Breger \& Wheeler(1980)]{bw80} Breger, M. \& Wheeler, J. C. 1980, \pasp, 92, 514
\bibitem[Burbidge \& Sandage(1958)]{bs58} Burbidge, E. M. \& Sandage, A. 1958, \apj, 128, 174
\bibitem[Burke et al.(2004)]{burke04} Burke, C. J., Gaudi, B. S. DePoy, D. L., Pogge, R. W., \& 
Pinsonneault, M. H. 2004, \aj, 127, 2382
\bibitem[Burke et al.(2005)]{burke05} Burke, C. J., Gaudi, B. S., DePoy, D. L. \& Pogge, R. W. 2006, \aj, 132, 210
\bibitem[Cannon \& Lloyd(1970)]{cl70} Cannon, R. D. \& Lloyd, C. 1970, \mnras, 150, 279
\bibitem[Carney et al.(2005)]{carn05a} Carney, B. W., Lee, J.-W., \& Dodson, B. 2005, \aj, 129, 656 (CLD05)
\bibitem[Carraro \& Bensby(2009)]{cb09} Carraro, G. \& Bensby, T. 2009, \mnras, 397, L106
\bibitem[Carraro et al.(2004)]{carr04} Carraro, G., Bresolin, F., Villanova, S., Matteucci, F., Patat, F. \& Romaniello, M. 2004, \aj, 128, 1676
\bibitem[Carraro et al.(2007)]{carr07} Carraro, G., Geisler, D., Villanova, S., Frinchaboy, P. M., \& Majewski, S. R. 2007, \aap, 476, 217
\bibitem[Carraro et al.(2002)]{ca02} Carraro, G., Girardi, L., \& Marigo, P. 2002, \mnras, 332, 705 %(Ca02)
\bibitem[Carraro et al.(2006)]{car06} Carraro, G., Villanova, S., Demarque, P., McSwain, M. V., Piotto, G., \& Bedin, L. R. 2006, \apj, 643, 1151
\bibitem[Carraro \& Patat(1994)]{cp94} Carraro, G. \& Patat, F. 1994, \aap, 289, 397
\bibitem[Carretta et al.(2007)]{car07} Carretta, E., Bragaglia, A., \& Gratton, R. G. 2007, \aap, 473, 129
\bibitem[Carretta et al.(2004)]{car04} Carretta, E., Bragaglia, A., Gratton, R. G., \& Tosi, M. 2004, \aap, 422, 951
\bibitem[Carretta et al.(2005)]{car05} Carretta, E., Bragaglia, A., Gratton, R. G., \& Tosi, M. 2005, \aap, 441, 131
\bibitem[Chiappini et al.(2001)]{chiappini} Chiappini, C., Matteucci, F. \& Romano, D. 2001, \apj, 554, 1044
\bibitem[Chincarini(1964)]{chin64} Chincarini, G. 1964, Mem. S. A. It., 35, 2
\bibitem[Cuffey(1938)]{cuffey} Cuffey, J. 1938, Ann. Harvard Obs., 106, 39
\bibitem[Cuffey(1943)]{cuffey_n2194} Cuffey, J. 1943, \apj, 97, 93
\bibitem[Cutri et al.(2003)]{2mass} Cutri, R. M. et al. 2003, The IRSA 2MASS All-Sky Point Source Catalog (Pasadena: NASA/IPAC)
\bibitem[Fagerholm(1906)]{m67ID} Fagerholm, E. 1906, Ph.D. thesis, Uppsala Univ.
\bibitem[Fan et al.(1996)]{fanm67} Fan, X., Burstein, D., Chen, J. -S., Zhu, J., Jlang, Z., Wu, H., et al., 1996, \aj, 112, 628
\bibitem[Fornal et al.(2007)]{n188fornal} Fornal, B., Tucker, D. L., Smith, J. A., Allam, S. S., Rider, C. J., \& Sung, H. 2007, \aj, 133, 1409
\bibitem[Friel(2006)]{edf06} Friel, E. D. 2006, in ``Chemical Abundances and Mixing in Stars in the Milky Way and its Satellites", ed. S. Randich \& L. Pasquini, ESO Astrophysics Symposia (Heidelberg: Springer Berlin), Vol. 24, p. 3 
\bibitem[Friel et al.(2003)]{cr261} Friel, E. D., Jacobson, H. R., Barrett, E., Fullton, L., Balachandran, S., \& Pilachowski, C. A. 2003, \aj, 126, 2372 %(Paper I)
\bibitem[Friel et al.(2005)]{be17} Friel, E. D., Jacobson, H. R., \& Pilachowski, C. A. 2005, \aj, 129, 2725
\bibitem[Friel et al.(2010)]{pV} Friel, E. D., Jacobson, H. R., \& Pilachowski, C. A. 2010, \aj, 139, 1942
\bibitem[Friel et al.(2002)]{f02} Friel, E. D., Janes, K. A., Tavarez, M., Scott, J., Katsanis, R., Lotz, J., Hong, L., \& Miller, N. 2002, \aj, 124, 2693
\bibitem[Fulbright et al.(2006)]{fmr06} Fulbright, J. P., McWilliam, A., \& Rich, R. M. 2006, 
\apj, 636, 821
\bibitem[Fulbright et al.(2007)]{fmr07} Fulbright, J. P., McWilliam, A. \& Rich, R. M. 2007, \apj, 661, 1152
\bibitem[Geisler, D.(1987)]{geis87} Geisler, D. 1987, \aj, 94, 84
\bibitem[Geisler, D.(1988)]{geisl88} Geisler, D. 1988, \pasp, 100, 338
\bibitem[Geller et al.(2008)]{gel} Geller, A., Mathieu, R. D., Harris, H. C. \& McClure, R. D. 2008, \aj, 135, 2264
\bibitem[Gim et al.(1998a)]{gima} Gim, M., Hesser, J. E., McClure, R. D. \& Stetson, P. B. 1998(a), \pasp, 110, 1172
\bibitem[Gim et al.(1998b)]{gimb} Gim, M., VandenBerg, D. A., Stetson, P. B., Hesser, J. E. \& Zurek, D. R. 1998(b), \pasp, 110, 1318
\bibitem[Gratton et al.(1996)]{grat96} Gratton, R. G., Carretta, E. \& Castelli, F. 1996, \aap, 314, 191
\bibitem[Grochalski \& Sarajedini(2002)]{GS02} Grochalski, A. J., \& Sarajedini, A. 2002, \aj, 123, 1603 (GS02)
\bibitem[Gustafsson et al.(2008)]{sphMARCS} Gustafsson, B., Edvardsson, B., Eriksson, K., Jorgensen, U. G., Nordlund, \AA., \& Plez, B. 2008, \aap, 486, 951
\bibitem[Hinkle et al.(2000)]{arcsp} Hinkle, K., Wallace, L., Valenti,
J., \& Harmer, D. 2000, Visible and Near Infrared Atlas of the Arcturus
Spectrum, (San Francisco:ASP)
\bibitem[Hoag et al.(1961)]{hoag} Hoag, A. A., Johnson, H. L., Iriarte, B., Mitchell, R. I., Hallam, K. L., \& Sharpless, S. 1961, Publ. U. S. Naval Obs., 17, 345
\bibitem[Hobbs et al.(1990)]{n188hobbs} Hobbs, L. M., Thorburn, J. A., \& Rodriguez-Bell, T. 1990, \aj, 100, 710
\bibitem[Hou et al.(2000)]{hou} Hou, J. L.,, Prantzos, N. \& Boissier, S. 2000, \aap, 362, 921
\bibitem[Jacobson et al.(2008)]{pIII} Jacobson, H. R., Friel, E. D., \& Pilachowski, C. A. 2008, \aj, 135, 2341 %(Paper III)
\bibitem[Jacobson et al.(2009)]{pIV} Jacobson, H. R., Friel, E. D., \& Pilachowski, C. A. 2009, \aj, 137, 4753 %(Paper IV)
\bibitem[Jacobson et al.(2011)]{ctio} Jacobson, H. R., Friel, E. D., \& Pilachowski, C. A. 2011, \aj, 141 58 %(CTIO)
\bibitem[Janes(1977)]{ja77} Janes, K. A. 1977, \aj, 82, 35
\bibitem[Janes(1979)]{ja79} Janes, K. A. 1979, \apjs, 39, 135
\bibitem[Jennens \& Helfer(1995)]{jh75} Jennens, P. A. \& Helfer, H. L. 1975, \mnras, 172, 681
\bibitem[Johansson et al.(2003)]{johansson} Johansson, S., Litz\'{e}n, U., Lundberg, H., \& Zhang, Z. 2003, \apj, 584, L107
\bibitem[Johnson et al.(2008)]{cijohn} Johnson, C. I., Pilachowski, C. A., Simmerer, J., \& Schwenk, D. 2008, \apj, 681, 1505
\bibitem[Kaluzny \& Mazur(1991)]{km91b_n2355} Kaluzny, J. \& Mazur, B. 1991, AcA, 41, 279
\bibitem[K\"{u}stner(1923)]{kustner_n7789} K\"{u}stner, F. 1923, VeBon, 18, 1
\bibitem[Kyeong et al.(2005)]{kyeong_n2194} Kyeong, J., Byun, Y.-K. \& Sung, E.-C. 2005, 38, 415
\bibitem[Lasker et al.(1990)]{lasker} Lasker, B. M., Russel, J. N., Jenkner, H., Sturch, C. R., McLean, B. J., \& Shara, M. M. 1990, \aj, 99, 2019
\bibitem[Luck et al.(2006)]{luck06} Luck, R. L., Kovtyukh, V. V. \& Andrievsky, S. M. 2006, \aj, 132, 902
\bibitem[Lyng\aa(1987)]{lynga} Lyng\aa, G. 1987, Catalogue of Open Cluster Data (5th ed.; Lund: Lund Obs.)
\bibitem[Magain(1984)]{magain} Magain, P. 1984, \aap, 132, 208
\bibitem[Magrini et al.(2010)]{mag10} Magrini, L., Randich, S., Zoccali, M., Jilkova, L., Carraro, G., Galli, D., Maiorca, E. \& Busso, M. 2010, \aap, 523A, 11
\bibitem[Magrini et al.(2009)]{magrini} Magrini, L., Sestito, P., Randich, S. \& Galli, D. 2009, \aap, 494, 95
\bibitem[Marino et al.(2008)]{marino} Marino, A. F., Villanova, S., Piotto, G., Milone, A. P., Momany, Y., Bedin, L. R. \& Medling, A. M. 2008, \aap, 490, 625
\bibitem[Mashonkina et al.(2000)]{mash00} Mashonkina, L. I., Shimanskii, V. V., \& Sakhibullin, N. A. 2000, ARep, 44, 790
\bibitem[Mathieu(2000)]{wocs} Mathieu, R. D. 2000, ASP, 198, 517
\bibitem[Mathieu et al.(1986)]{mathieu} Mathieu, R. D., Latham, D. W., Griffen, R. F., \& Gunn, J. E. 1986, \aj, 92, 1100
\bibitem[McClure et al.(1974)]{mfg74_n2420} McClure, R. D., Forrester, W. T. \& Gibson, J. 1974, \apj, 189, 409
\bibitem[McNamara \& Solomon(1981)]{ms81} McNamara, B. J. \& Solomon, S. 1981, \aaps, 43, 337
\bibitem[Mermilliod et al.(2003)]{mem03} Mermilliod, J. -C., Latham, D. W., Glushkova, E. V., Ibrahimov, M. A., Batirshinova, V. M., Stefanik, R. P., \& James, D. J. 2003, \aap, 399, 105
\bibitem[Mermilliod \& Mayor(2007)]{mm07} Mermilliod, J. -C. \& Mayor, M. 2007, \aap, 470, 919
\bibitem[Mermilliod et al.(2008)]{mem08} Mermilliod, J. -C., Mayor, M. \& Udry, S. 2008, \aap, 485, 303
\bibitem[Mikolaitis et al.(2010)]{miko} Mikolaitis, \v{S}., Tautvai\v{s}ien\.{e}, G., Gratton, R., Bragaglia, A. \& Carretta, E. 2010, \mnras, 407, 1866
\bibitem[Minniti(1995)]{m95} Minniti, D. 1995, \aaps, 113, 299
\bibitem[Mishenina et al.(2006)]{mish06} Mishenina, T. V., Bienayme, O., Gorbaneva, T. I., Charbonnel, C., Soubrian, C., Korotin, S. A., \& Kovtyukh, V. V. 2006, \aap, 456, 1109
\bibitem[Mochejska \& Kaluzny(1999)]{mk99} Mochejska, B. J. \& Kaluzny, J. 1999, AcA, 49, 351
\bibitem[Moitinho et al.(2006)]{mcbv06_n2425} Moitinho, A., Carraro, G., Baume, G. \& V\'{a}zquez, R. A. 2006, \aap, 445, 493
\bibitem[Montgomery et al.(1993)]{mmj93} Montgomery, K. A., Marschall, L. A., \& Janes, K. A. 1993, \aj, 106, 181
\bibitem[Moultaka et al.(2004)]{elodie} Moultaka, J., Ilovaisky, S. A., Prugniel, P. \&  Soubiran, C. 2004, \pasp, 116, 693
\bibitem[Pace et al.(2010)]{pace10} Pace, G., Danziger, J., Carraro, G., Melendez, J., Fran\c{c}ois, P., Matteucci, F. \& Santos, N. C. 2010, \aap, 515, 28
\bibitem[Pancino et al.(2010)]{pancino} Pancino, E., Carrera, R., Rossetti, E., \& Gallart, C. 2010, \aap, 511, 56
\bibitem[Parisi et al.(2005)]{mcp} Parisi, M. C., Clari\'{a}, J. J., Piatti, A. E., \& Geisler, D. 2005, \mnras, 363, 1247
\bibitem[Pepper \& Burke(2006)]{pepper06} Pepper, J. \& Burke, C. J. 2006, \aj, 132, 1177
\bibitem[Peterson et al.(1993)]{arc} Peterson, R. C., Dalle Ore, C. M., \& Kurucz, R.L. 1993, \apj, 404, 333
\bibitem[Phelps et al.(1994)]{phelps94} Phelps, R. L., Janes, K. A. \& Montgomery, K. A. 1994, \aj, 107, 1079
\bibitem[Piatti et al.(2003)]{piatti_n2194} Piatti, A. E., Clari\'{a}, J. J. \& Ahumada, A. V. 2003, \mnras, 340, 1249
\bibitem[Pietrukowicz et al.(2006)]{pkk06_n2425} Pietrukowicz, P., Kaluzny, J. \& Krzeminski, W. 2006, \mnras, 365, 110
\bibitem[Platais et al.(2003)]{platais} Platais, I., , Kozhurina-Platais, V., Mathieu, R. D., Girard, T. M., 
\& van Altena, W. F. 2003, \aj, 126, 2922
\bibitem[Ram\'{i}rez \& Cohen(2003)]{rc03} Ram\'{i}rez, S. V. \& Cohen, J. G. 2003, \aj, 125, 224
\bibitem[Randich et al.(2003)]{randich188} Randich, S., Sestito, P., \& Pallavincini, R. 2003, \aap, 399, 13
\bibitem[Ro\v{s}kar et al.(2008)]{roskar} Ro\v{s}kar, R., Debattista, V. P., Quinn, T. R., Stinson, G. S. \& Wadsley, J. 2008, \apj, 684, L79
\bibitem[Salaris et al.(2004)]{salaris} Salaris, M., Weiss, A., \& Percival, S. M. 2004, \aap, 414, 163
\bibitem[Sanders(1977)]{san} Sanders, W. L. 1977, \aaps, 27, 89
\bibitem[Sanner et al.(2000)]{sanner_n2194} Sanner, J., Altmann, M., Brunzendorf, J. \& Geffert, M. 2000, \aap, 357, 471
\bibitem[Sarajedini et al.(1999)]{n188ata} Sarajedini,, A., von Hippel, T., Kozhurina-Platais, V., \& Demarque, P. 1999, \aj, 118, 2894
\bibitem[Schuler et al.(2009)]{schu_hyades} Schuler, S. C., King, J. R. \& The, L. -S. 2009, \apj, 701, 837
\bibitem[Sellwood \& Binney(2002)]{sb02} Sellwood, J. A. \& Binney, J. J. 2002, \mnras, 336, 785
\bibitem[Sestito et al.(2006)]{ses06} Sestito, P., Bragaglia, A., Randich, S., Carretta, E., Prisinzano, L., \& Tosi, M. 2006, \aap, 458, 121%(SB)
\bibitem[Sestito et al.(2008)]{ses08} Sestito, P., Bragaglia, A., Randich, S., Pallavicini, R., Andrievsky, S. M., \& Korotin, S. A. 2008, \aap, 488, 943
\bibitem[Sestito et al.(2007)]{ses07} Sestito, P., Randich, S., \& Bragaglia, A. 2007, \aap, 465, 185 
\bibitem[Sharma et al.(2006)]{spomts06_n2420} Sharma, S., Pandey, A. K., Ogura, K., Mito, H., Tarusawa, K., 
\& Sagar, R. 2006, \aj, 132, 1669
\bibitem[Smith \& Suntzeff(1987)]{ss2420} Smith, V. V. \& Suntzeff, N. B. 1987, \aj, 93, 359
\bibitem[Sneden(1973)]{sneden73} Sneden, C. 1973, Thesis, University of Texas at Austin.
\bibitem[Soubiran et al.(1998)]{soub} Soubiran, C., Katz, D. \& Cayrel, R. 1998, \aaps, 133, 221 
\bibitem[Soubiran et al.(2000)]{soc00_n2355} Soubiran, C., Odenkirchen, M. \& Le Campion, J.-F. 2000, \aap, 357, 484
\bibitem[Stello et al.(2007)]{stello} Stello, D., Bruntt, H., Kjeldsen, H., et al. 2007, \mnras, 377, 584
\bibitem[Subramaniam(2003)]{subramaniam} Subramaniam, A. 2003, Bull. Astron. Soc. India, 31, 49
\bibitem[Tautvai\v{s}ien\.{e} et al.(2010)]{taut10} Tautvai\v{s}ien\.{e}, G., Edvardsson, B., Puzeras, E., Barisevi\v{c}ius, G., \& Ilyin, I. 2010, \mnras, 409, 1213
\bibitem[Tautvai\v{s}ien\.{e} et al.(2005)]{taut05} Tautvai\v{s}ien\.{e}, G;, Edvardsson, B., Puzeras, E. \& Ilyin, I. 2005, \aap, 431, 933
\bibitem[Tautvai\v{s}ien\.{e} et al.(2000)]{taut00} Tautvai\v{s}ien\.{e}, G., Edvardsson, B., 
Touminen, I., \& Ilyin, I. 2000, \aap, 360, 499 (T00)
\bibitem[Taylor(1982)]{jrt} Taylor, J. R. 1982, An Introduction to Error Analysis (2nd ed.; Mill Valley: University Science Books)
\bibitem[Twarog et al.(1997)]{taat} Twarog, B. A., Ashman, K. M., \& Anthony-Twarog, B. J. 1997, \aj, 114, 2556
\bibitem[van Dokkum(2001)]{lacos_spec} van Dokkum, P. 2001, \pasp, 113, 1420
\bibitem[Van Helshoecht \& Groenewegen(2007)]{vhg07} Van Helshoecht, V. \& Groenewegen, M. A. T. 2007, \aap, 463, 559
\bibitem[Villanova et al.(2005)]{vill05} Villanova, S., Carraro, G., Bresolin, F. \& Patat, F. 2005, \aj, 130, 652
\bibitem[Wee \& Lee(1996)]{wl96} Wee, S.-O. \& Lee, M. G. 1996, J. Korean Astron. Soc. 29, 181
\bibitem[West(1967)]{west67} West, F. R. 1967, \apjs, 14, 384
\bibitem[Wu et al.(2009)]{wu} Wu, Z. -Y., Zhou, X., Ma, J. \& Du, C. -H. 2009, \mnras, 399, 2146
\bibitem[Yong et al.(2005)]{yong05} Yong, D., Carney, B. W., \& Teixera de Almeida, M. L. 2005, \aj, 130, 597
\end{thebibliography}
\end{document}